\documentclass[%
aps,
pra,
amsmath,amssymb,
reprint,%
]{revtex4-2}

\usepackage{graphicx}
\usepackage{dcolumn}
\usepackage{bm}
\usepackage{graphicx}
\usepackage{multirow}
\usepackage[table]{xcolor}
\usepackage{tabularx}
\usepackage[normalem]{ulem}
\usepackage{booktabs}
\usepackage{gensymb}

\usepackage[utf8]{inputenc}
\usepackage[T1]{fontenc}
\usepackage{mathptmx}
\usepackage{etoolbox}
\usepackage[strings]{underscore}
\usepackage{fixltx2e,amsmath}
\MakeRobust{\eqref}
\DeclareMathAlphabet{\mathdutchcal}{U}{dutchcal}{m}{n}

\makeatletter
\def\@email#1#2{%
	\endgroup
	\patchcmd{\titleblock@produce}
	{\frontmatter@RRAPformat}
	{\frontmatter@RRAPformat{\produce@RRAP{*#1\href{mailto:#2}{#2}}}\frontmatter@RRAPformat}
	{}{}
}%

\newcommand*{\ket}[1]{\lvert#1\rangle}

\newcommand*{\Braket}[3]{\langle#1\vert#2\vert#3\rangle}

\makeatother
\begin{document}
	
	\preprint{AIP/123-QED}
	
	\title{Quantum annealing with pairs of $^2\Sigma$ molecules as qubits}
	\author{K. Asnaashari}
	\author{R. V. Krems}
	\affiliation{ 
		Department of Chemistry, University of British Columbia, \\Vancouver, B.C. V6T 1Z1, Canada \\
		Stewart Blusson Quantum Matter Institute, \\
		Vancouver, B.C. V6T 1Z4, Canada
	}
	
	\date{\today}
	
	\begin{abstract}
		%	We propose a quantum annealing algorithm based on $^2\Sigma$ molecules in an optical lattice. 
		The rotational and fine structure of open-shell molecules in a $\Sigma$ electronic state gives rise to crossings between
		Zeeman states of different parity. These crossings become avoided in the presence of an electric field. We propose
		an algorithm that encodes Ising models into qubits defined by pairs of $^2\Sigma$ molecules sharing an excitation near
		these avoided crossings. This can be used to realize a transverse field Ising model tunable by an external electric
		or magnetic field, suitable for quantum annealing applications. We perform dynamical calculations for several
		examples with one- and two-dimensional connectivities. Our results demonstrate that the probability of obtaining
		valid annealing solutions is high and can be optimized by varying the annealing times.
	\end{abstract}
	
	\maketitle
	
	\section{Introduction}
	Quantum annealing (QA) has been considered as a quantum algorithm for NP-hard optimization problems \cite{kadowaki, qa, search, qa1}. A QA algorithm encodes an optimization problem in the ground-state configuration of an Ising model. Finding the ground state of two-dimensional (2D) and three-dimensional (3D) Ising models has been proven to be NP-hard \cite{spin-glass} and universal, meaning such problems can be used to simulate any classical and quantum spin model with the overhead in the number of spins and interactions being at most polynomial \cite{universal}. Many different hard problems in apparently unrelated fields can be encoded in the ground state of an Ising model \cite{qa1, qa2, qa3, qa4, qa5, qa6, qa7, qa8, qa9, qa10, qa11, qa12, qa13, qa14, qa15, qa16, qa17, qa18}. While it has not been proven that QA can reduce these problems to polynomial complexity \cite{preskill}, it can significantly speed up certain NP optimization problems \cite{mcgeoch}. This continues to stimulate the search for scalable and robust QA architectures with flexible qubit connectivity. Most of the QA devices demonstrated to date are based on superconducting qubits (SQs), as exemplified by the work at D-Wave Systems that has now achieved QA with  $>5000$ SQs  \cite{advantage}.  
	In the present paper, we  propose an architecture for QA based on open-shell molecular radicals placed in superimposed electric and magnetic fields.

	Ultracold molecules trapped in optical lattices have been considered as a platform for quantum simulation of quantum magnetism and quantum spin models \cite{toolbox, superfluid, mag, synthetic, dd-interact, three-body, many-sim, far-equib, dipolar-dyn, mag-lattice, kitaev, control, simulation, b-h, holstein, twointeract, excitons, many-quant, 2d-phase, spin-1, symm-top, 1d-lattice, tweezer}.
	Quantum spin-1/2 models can be realized with ultracold polar molecules by encoding spins in rotational states. The dipole-dipole interactions lead to a general $XXZ$ Hamiltonian with the effective coupling constants depending on the molecules, choice of states, lattice parameters, and the magnitude of applied electric, magnetic, and microwave (MW) fields \cite{toolbox, superfluid, mag, synthetic, dd-interact, three-body, many-sim, far-equib, dipolar-dyn, mag-lattice}. Since the $XXZ$ model can be reduced to the Ising, $XY$, and Heisenberg models, all three cases can possibly be simulated using polar molecules.
	Ultracold molecules have also been considered for quantum information processing \cite{tweezer, demille, robust, gate, rotational-gates, logic-gates, ion, qudit, storage, sigma2, sigma2-gates}, mainly in the context of gate-based models. 
	Robust entangling gates (controlled-NOT and Toffoli) have been implemented using polar $^1\Sigma$ \cite{demille, robust, gate, rotational-gates, logic-gates} and $^2\Sigma$ \cite{sigma2, sigma2-gates} molecules as qubits. These schemes allow for a large number of qubits with coherence times of up to $5$ s \cite{demille}.

	Here, we aim to extend this paper to
	demonstrate the possibility of QA based on open-shell molecules by tuning an external dc electric or magnetic field. 
	To implement QA, it is necessary to realize a many-body quantum system with a transverse field Ising model (TFIM) that can be tuned, ideally by varying a single experimental parameter.  
	Although this has, to our knowledge, not been shown explicitly, transverse field Ising models can potentially be realized with ultracold molecules in combined dc electric and microwave fields with different polarizations and frequencies \cite{toolbox}. However, tuning such systems from a purely transverse field model to an Ising model requires careful and simultaneous adjustment of both the microwave  fields and the dc electric field. Encoding optimization problems into lattice systems for QA requires access to individual qubit-qubit interactions and qubit biases, potentially increasing the number of required microwave fields. With each microwave field affecting all molecules in the ensemble, tuning these interactions and biases as required for QA by simultaneously tuning multiple microwave fields is expected to be a very complex task. 	
		%	QA applications of molecules dressed by microwave fields are further complicated {\color{red} by the expected requirement to simultaneously tune multiple dressing fields. 
		
		In the present paper, we propose a many-body architecture with pairs of molecules in adjacent sites of an optical lattice as qubits that does not require microwave fields to realize a transverse field Ising model. This can be experimentally realized by placing molecules either in a 2D or 3D optical lattice with different lattice spacing along different dimensions. We demonstrate that the $XXZ$ spin model based on qubits encoded in individual molecules translates into a transverse field Ising model with qubits encoded into pairs of molecules. 
		
		In order to enhance the magnitudes of the transverse field couplings, we exploit the sensitivity of open-shell molecules to particular combinations of dc electric and magnetic fields. 
		Specifically, we consider molecular radicals in a $^2\Sigma$ electronic state. Several experiments have recently demonstrated laser-cooling of $^2\Sigma$ radicals to ultracold temperatures \cite{exp1, exp2, exp3, exp4, exp5, exp6, exp7, exp8, exp9, exp10, exp11}. As shown previously \cite{avoided}, $^2\Sigma$ molecules exhibit avoided crossings between Zeeman states of different rotational manifolds, when placed in superimposed electric and magnetic fields. These avoided crossings are sensitive to electromagnetic fields as well as intermolecular interactions, which has been previously exploited to suggest applications of $^2\Sigma$ molecules to study controlled bi-molecular collisions \cite{collisions, collisions2} and controlled Frenkel exciton dynamics \cite{romannjp} or for imaging weak rf fields \cite{imaging}. We show how the same avoided crossings can be used to tune a many-body molecular Hamiltonian with qubit encoding proposed here from a purely transverse field model to an Ising model by varying an external dc electric or magnetic field.

		\section{Theory}
			\subsection{Quantum annealing}

			QA is an optimization algorithm based on the adiabatic theorem of quantum mechanics. The adiabatic theorem ensures that one can prepare a quantum system in the ground state of a desired Hamiltonian ($\hat{H}_{\rm p}$) from the ground state of a simpler Hamiltonian ($\hat{H}_{\rm i}$) by a transformation of $\hat{H}_{\rm i}$ to $\hat{H}_{\rm p}$. This is possible if the ground state is separated from excited states by a finite energy gap throughout the transformation.
			In most current implementations of QA \cite{mcgeoch, advantage}, a discrete optimization problem is encoded into a graph $G = (V, E)$ with spin-1/2 vertices $V$ and edges $E$ representing an Ising model Hamiltonian:
			\begin{equation}\label{eq:im}
				\hat{H}_{\rm IM} = \sum_{i \in V} h_i \sigma^z_i + \sum_{(i, j) \in E} J_{ij}\sigma^z_i\sigma^z_j
			\end{equation}
			where $h_i$ is the bias applied to spin $i$, $J_{ij}$ is the Ising coupling between spins $i$ and $j$, and $\sigma^z_i$ is the Pauli \emph{Z} matrix applied to the $i$th spin. Encoding an optimization problem into an Ising model involves determining a suitable graph of spins, which is, in itself, an NP-hard problem \cite{mcgeoch}. 
			
			Given this encoding ($\hat{H}_{\rm p} = \hat{H}_{\rm IM}$), the spin system is initialized in the ground state of the following Hamiltonian $\hat{H}_{\rm i}$:
			\begin{equation}
				\hat{H}_{\rm i} = \sum_{i \in V}^{n}\sigma^x_i
				\label{sigma-X}
			\end{equation}
			where $\sigma^x_i$ is the Pauli \emph{X} matrix applied to the $i$th spin. The ground state of Hamiltonian (\ref{sigma-X}) is an equal superposition of all possible spin states. To allow adiabatic transformation of the ground state of $\hat{H}_{\rm i}$ into the ground state of $\hat{H}_{\rm p}$, a physical system must realize a TFIM with tunable parameters,
			\begin{equation}\label{eq:tfim}
				\hat{H}_{\rm TFIM}(s) = \sum_{i \in V} h_i(s) \sigma^z_i + \Delta_i(s) \sigma^x_i + \sum_{(i, j) \in E} J_{ij}(s)\sigma^z_i\sigma^z_j
			\end{equation}
			where $s$ is the annealing parameter varying from 0 to 1 during the annealing and $\Delta_i$ is the transverse field strength for spin $i$. The Hamiltonian parameters must be functions of the annealing parameter with the following limits:
			\begin{equation}
				s: 0\rightarrow 1 : \begin{cases}	
					h_i(s): 0\rightarrow h_i \\
					J_{ij}(s): 0\rightarrow J_{ij} \\
					\Delta_i(s): 1\rightarrow 0 \\
				\end{cases}.
				\label{QA}
			\end{equation}
			Hamiltonian \eqref{eq:im} is diagonal in the $\sigma^z$ basis so the ground state of  $\hat H_{\rm p}$ can be determined by measuring the spin configuration at $s=1$, which gives the solution to the optimization problem at hand. 
		
		\subsection{$^2\Sigma$ molecules in superimposed electric and magnetic fields}
		
\begin{figure}
			\includegraphics[width=\columnwidth]{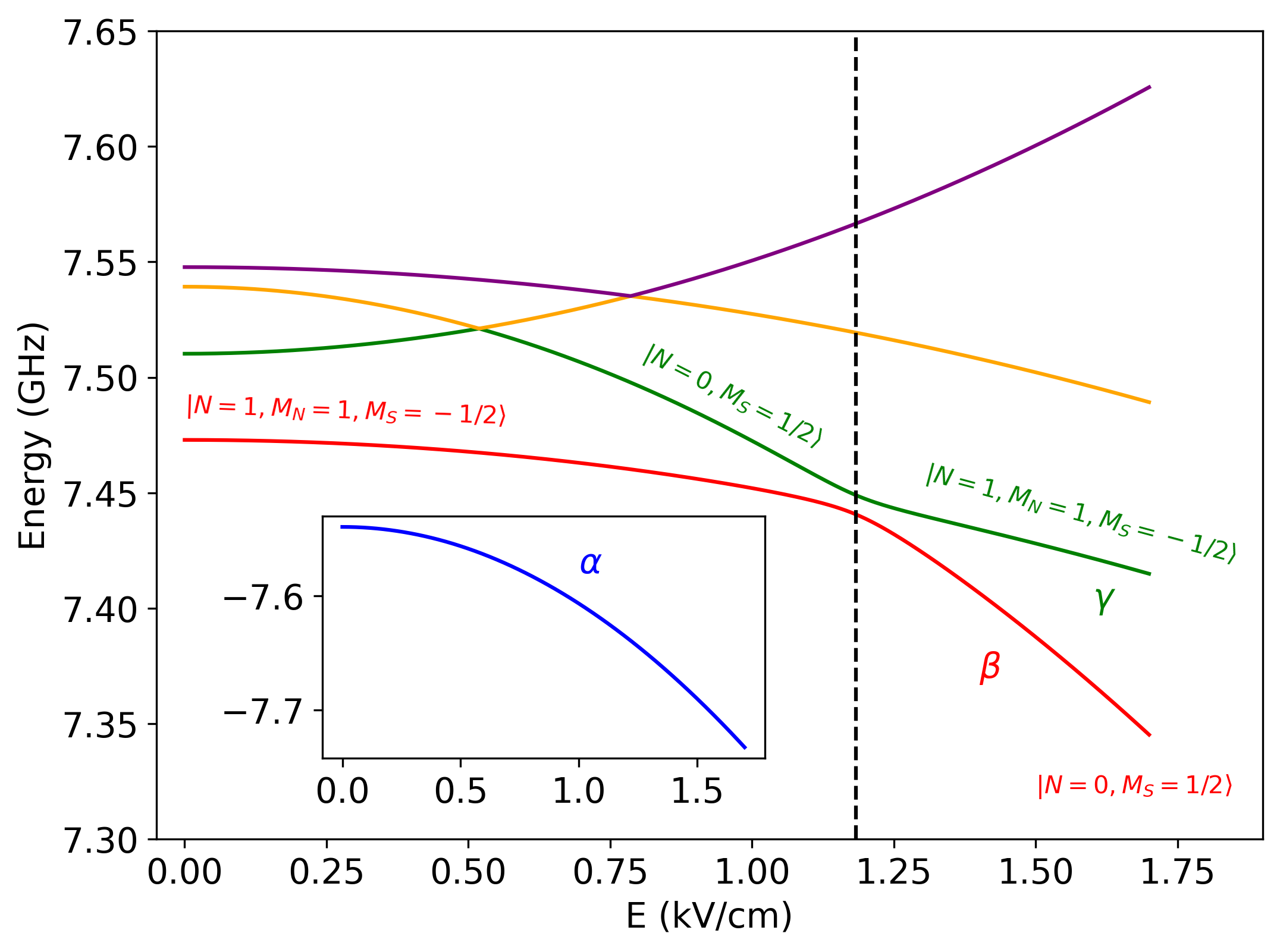} 
			\caption{Energy levels of a SrF($X^2\Sigma^+$) molecule with $B_e = 0.251 \text{ cm}^{-1}$, $\gamma_{SR}=2.49\times10^{-3} \text{ cm}^{-1}$, $d = 3.47 \text{ D}$ at $B = 538 \text{ mT}$ as functions of the strength of a dc electric field. 
				The avoided crossing between $\beta$ and $\gamma$ is indicated by the vertical dashed line at $E=1.18\text{ kV/cm}$.}
			\label{fig:energy_levels_e}
		\end{figure}

		The Hamiltonian for a $^2\Sigma$ radical in the vibrational ground state placed in a superposition of dc electric $\bm{E}$ and magnetic  $\bm{B}$ fields can be written as
		\begin{equation}\label{eq:iso_ham}
			\hat{H} = B_e \bm{N}^2 + \gamma_{SR} \bm{N}\cdot\bm{S} - \bm{E}\cdot\bm{d} + \mu_B g_S\bm{B}\cdot\bm{S}
		\end{equation}
		where $B_e$ is the rotational constant, $\gamma_{SR}$ is the spin-rotational interaction constant, $\bm{N}$ and $\bm{S}$ are the rotational and spin angular momenta of the molecule, $\bm{d}$ is the dipole moment of the molecule, $\mu_B$ is the Bohr magneton, and $g_S$ is the electron spin $g$ factor. 
		For simplicity, we assume that the magnetic- and electric-field vectors are co-aligned. Figure \ref{fig:energy_levels_e} shows the lowest five energy levels of SrF($X^2\Sigma^+$) corresponding to $N=0$ and 1. 
		The state labeled $\alpha$ is predominantly $\ket{N=0, M_S= -1/2}$, whereas the states labeled $\beta$ and $\gamma$ exhibit an avoided crossing, with $\beta$ changing from  $\ket{N=1, M_N=1, M_S=-1/2}$  to  $\ket{N=0, M_S=1/2}$ and $\gamma$ undergoing the reverse change, as the field magnitude is increased.

		%The energy levels of this Hamiltonian for SrF is displayed in figure \ref{fig:energy_levels_e} as a function of the electric field and the lowest energy levels of the rotational ground state ($N=0$) and the first rotational excited state ($N=1$) are labeled $\alpha$, $\beta$, and $\gamma$. 
		
		%The $\alpha$ states is predominantly $\ket{N=0, M_S= -1/2}$ at all field values, however the states $\beta$ and $\gamma$ go through an avoided crossing as the electric field is increased. Just before the avoided crossing, $\beta$ and $\gamma$ are principally $\ket{N=1, M_N=1, M_S=-1/2}$ and $\gamma$ is predominantly $\ket{N=0, M_S=1/2}$. After the avoided crossing, $\beta$ is essentially $\ket{N=0, M_S=1/2}$ and $\gamma$ is predominantly $\ket{N=1, M_N=1, M_S=-1/2}$.

\begin{figure}
			\begin{tabular}{c}
				\includegraphics[width=\columnwidth]{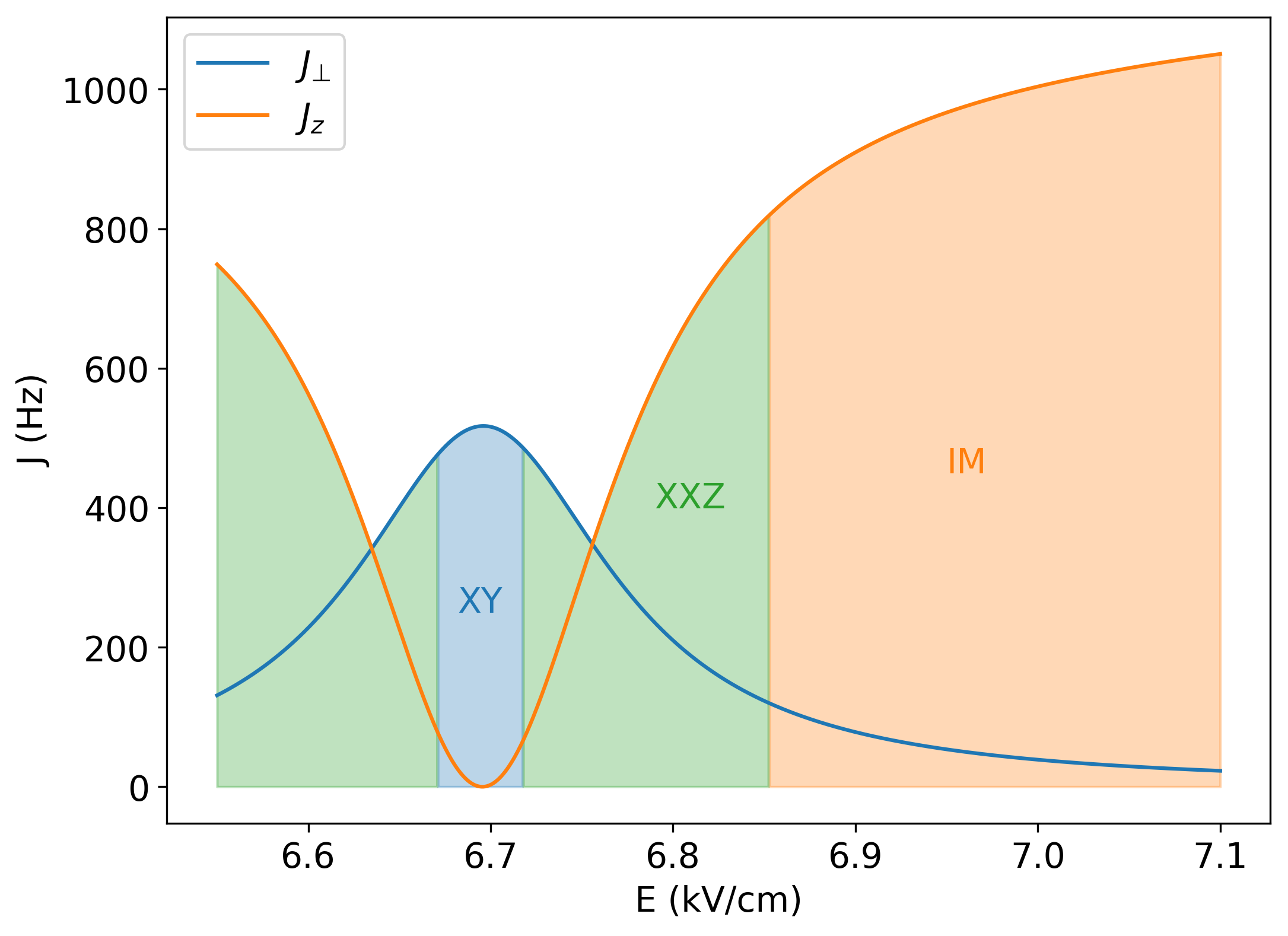}  \\
				\includegraphics[width=\columnwidth]{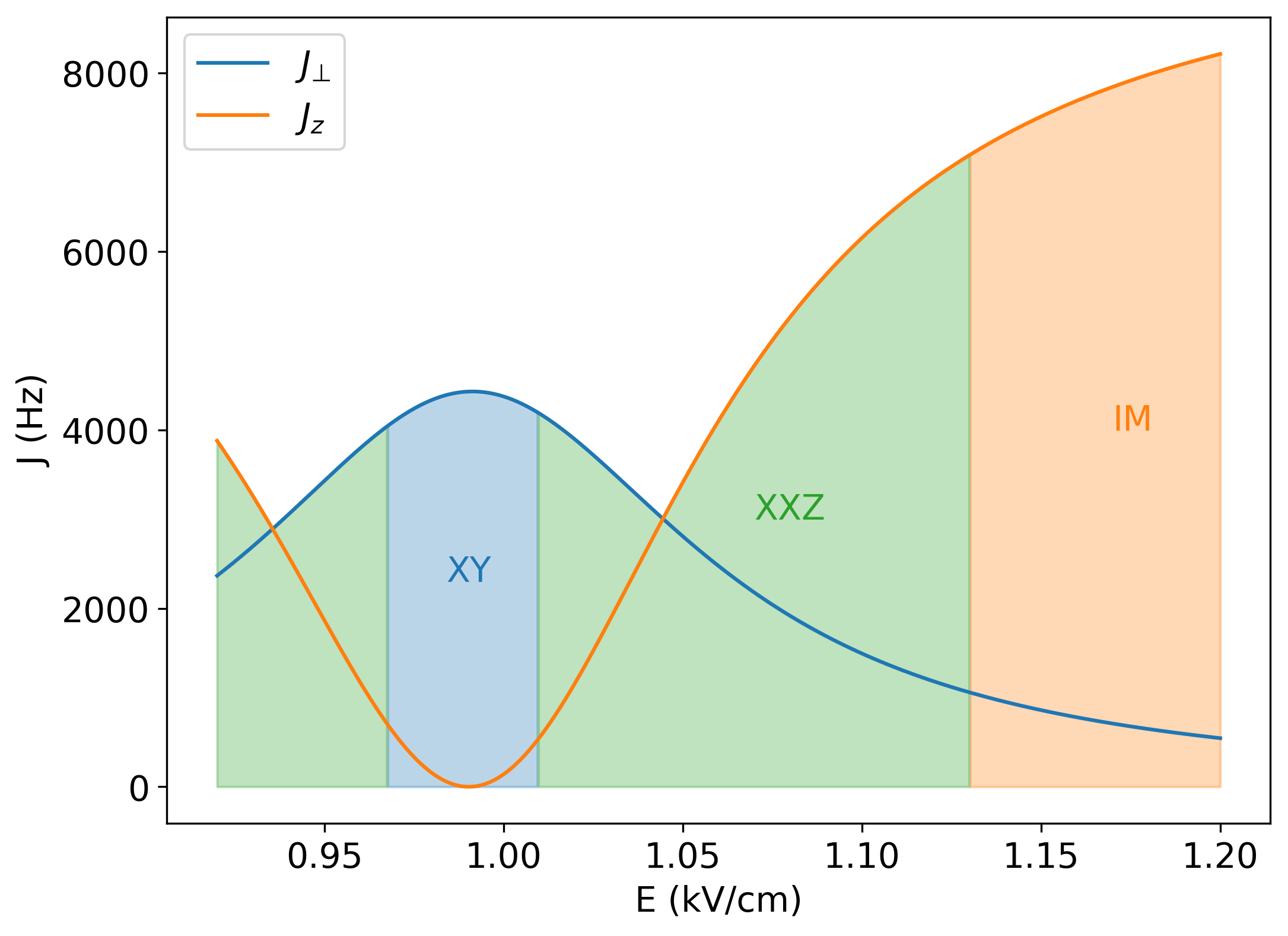}
			\end{tabular}
			\caption{Coupling constants $J_\perp$ and $J_z$ in Eq. \eqref{eq:manybody_ham} for two SrF (top) and SrI (bottom) molecules separated by 500 nm with the intermolecular axis perpendicular to $\bm E$ and $\bm B$ with $B=600$ mT (top) and $B=100$ mT (bottom). 
				The avoided crossing between $\beta$ and $\gamma$ is at $E=6.69$ kV/cm (top) and $E=0.99$ kV/cm (bottom). The shaded areas indicate the range of the parameters for the corresponding spin model, with IM denoting the Ising model.}
			\label{fig:j_bc}
		\end{figure}

		Encoding a spin-1/2 system into two isolated eigenstates of a single molecule and allowing for dipole-dipole interactions \cite{electrostatic} leads to the following many-body Hamiltonian \cite{hazzard} (assuming the energy difference between the two spin-rotational states is much larger in magnitude than the intermolecular interactions): 
		\begin{equation}\label{eq:manybody_ham}
			\hat{H} = \sum_i h_i \hat{S}_i^z + \frac{1}{2}\sum_{i\neq j} \left[\frac{J_{\perp_{ij}}}{2}\left(\hat{S}_i^+\hat{S}_j^- + {\rm h.c.}\right) + J_{z_{ij}} \hat{S}_i^z\hat{S}_j^z\right].
		\end{equation}
		Appendix \ref{ap:manybody} outlines the derivation of Eq. \eqref{eq:manybody_ham}. With $\ket{\uparrow}=\ket{\beta}$ and $\ket{\downarrow}=\ket{\gamma}$, the dependence of $J_\perp$ and $J_z$ on the electric-field magnitude produces three regimes of spin models illustrated in Fig. \ref{fig:j_bc} for two molecules (SrF and SrI) in the $X^2\Sigma^+$ state.
		%Figure \ref{fig:j_bc} shows the dependence of both couplings to the electric field magnitude for two molecules of SrF. This graph shows that the interaction between two molecules limited to the $\beta$ and $\gamma$ states can be divided into three regimes at different electric field values. 
		Near the avoided crossing, the spin-exchange interaction dominates as $J_\perp \gg J_z$, yielding a quantum $XY$ spin model ($XXZ$ model with $J_z=0$). At electric fields far detuned from the avoided crossing, the Ising coupling dominates as $J_z \gg J_\perp$, reducing the Hamiltonian to a quantum Ising model. At electric fields between these two extremes, both couplings are similar in value. 
		We define $E_\perp(B)$ to be the electric-field magnitude at which $J_\perp$ is maximized and $E_z(B)$ can be assigned to be any electric-field magnitude at which $J_z \gg J_\perp$.
		
		The coupling between molecules also depends on the angle between the intermolecular axis and the field directions following the anisotropy of dipole-dipole interactions (see Appendix \ref{ap:manybody}). While Fig. \ref{fig:j_bc} displays the couplings for two molecules with an intermolecular axis perpendicular to the electric and magnetic fields ($\theta = \pi/2$), the couplings are modulated by Eq. \eqref{eq:int_theta} for other values of $\theta$. We define ferromagnetic and anti-ferromagnetic interactions between molecules with negative and positive signs on $J_z$ respectively. Therefore, at $\theta=\pi/2$, interactions between molecules are anti-ferromagnetic ($J_z \ge 0$), and when the intermolecular axis is parallel to the fields ($\theta=0$) the interactions are ferromagnetic ($J_z \le 0$) and twice as large in magnitude as the perpendicular case.
		
		Model \eqref{eq:manybody_ham} is not suitable for QA, as it conserves the number of spin excitations. The TFIM model of Eq. \eqref{eq:tfim} includes the transverse field terms $\sigma_i^x$, which generate spin excitations during the annealing. 
			It is possible to engineer $\sigma_i^x$ terms by coupling  the $\ket{\uparrow}$ and $\ket{\downarrow}$ states of the individual molecules with near-resonant MW fields. 
			However, the biases of the qubits $h_i$ depend on the detuning of the MW field from the energy differences between the $\ket{\uparrow}$ and $\ket{\downarrow}$ states of each molecule. In order to realize an Ising model with arbitrary biases encoding an optimization problem, 
			it may be necessary to apply multiple dressing fields for the corresponding biases on each qubit. 
			The frequencies of these dressing fields must be tuned simultaneously to achieve transformation (\ref{QA}).  
			Even in simple cases with homogeneous fields and zero biases, the $\ket{\uparrow}$ -- $\ket{\downarrow}$ energy gaps are modified by the molecular interactions (Eq. \eqref{eq:xxz_int}) and the resonant frequencies vary between the qubits. QA with molecules dressed by MW fields as qubits thus appears to be a very complex task. 
		
		\subsection{Pairs of molecules as qubits}
		
		%While Figure \ref{fig:j_bc} shows how one can tune the interactions between molecules from a dynamic $XY$ model to a static Ising model, this is not sufficient to implement QA with. Because of the large energy difference between the $\beta$ and $\gamma$ states compared to the couplings, the $XXZ$ Hamiltonian \ref{eq:manybody_ham} between these two states conserves the number of excitations in the system. A global optimization of an Ising model should allow for varying number of excitations in the final optimized configuration. Here, we consider pairs of molecules exchanging an excitation with weak spin-exchange interactions with other molecules as qubits of a many-body Hamiltonian. 
		To overcome this problem, we consider pairs of molecules exchanging an excitation as qubits of a many-body Hamiltonian. This encoding scheme does not require any MW fields and allows tuning of qubit parameters and couplings by varying a single dc field, electric or magnetic. It may, however, require complicated geometry of the trapping optical lattice  and precise control of the field gradients.
		
		Within the subspace spanned by the $\ket{\uparrow\downarrow}$ and $\ket{\downarrow\uparrow}$ states, the isolated two-molecule system encodes a spin-1/2 qubit with the following Hamiltonian:
		\begin{equation}\label{eq:qubit_ham}
			\hat{H}_q = h_q\hat{S}_q^z + \Delta_q\hat{S}_q^x,
		\end{equation}
		where
		\begin{align}
			h_q &= h_1 - h_2, & \Delta_q &= J_\perp, \label{eq:bias_trans} \\ 
			S_q^x &= \frac{1}{2} \begin{pmatrix}
				0 & 1 \\
				1 & 0
			\end{pmatrix}, & S_q^z &= \frac{1}{2} \begin{pmatrix}
				1 & 0 \\
				0 & -1
			\end{pmatrix}
		\end{align}
		in the basis:
		\begin{align*}\label{eq:qubit_spin_ops}
			\ket{\uparrow\downarrow} = \ket{0} &= \begin{pmatrix}
				1 \\
				0
			\end{pmatrix} & \ket{\downarrow\uparrow} = \ket{1} &= \begin{pmatrix}
				0 \\
				1
			\end{pmatrix}
		\end{align*}
		%With this Hamiltonian, the two isolated molecules can be considered as a qubit in a quantum annealing context. 
		The bias of the qubit ($h_q$) is the difference in the energy gap of the $\ket{\uparrow}$ and $\ket{\downarrow}$ states of the two molecules, which can be tuned by applying a gradient to the electric-field magnitude. The transverse field parameter of the qubit ($\Delta_q$) depends on the magnitude of the fields  and the orientation of the fields. However, for this parameter to be relevant,  one must ensure that $h_1 - h_2 \approx \Delta_q$. We consider the following field configurations for the two molecules of each qubit: $B = B_1 = B_2$ and $E_1 = E_2 - \delta E$ with $\delta E \ll E_1, E_2$. 
		%and each qubit will be parameterized by $B, E, \delta E$.
		Qubit bias can also be set using a single molecule locked in one of the two states $\ket{\uparrow}$ or $\ket{\downarrow}$ and coupled to the qubit. This makes one of the two states more favorable energetically and has the added benefits of being of the same order of magnitude as the inter-qubit couplings, which can be easily accommodated in the annealing procedure described below. Hereafter, `qubit' refers to the two-molecule system with Hamiltonian \eqref{eq:qubit_ham}.

\begin{figure}
			\begin{tabular}{c}
				\quad\quad\includegraphics[width=0.7\columnwidth]{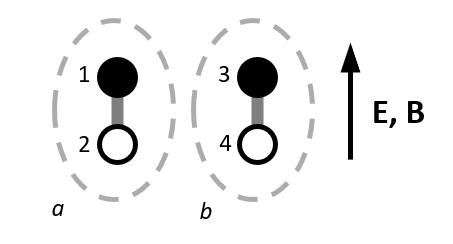} \\
				\includegraphics[width=\columnwidth]{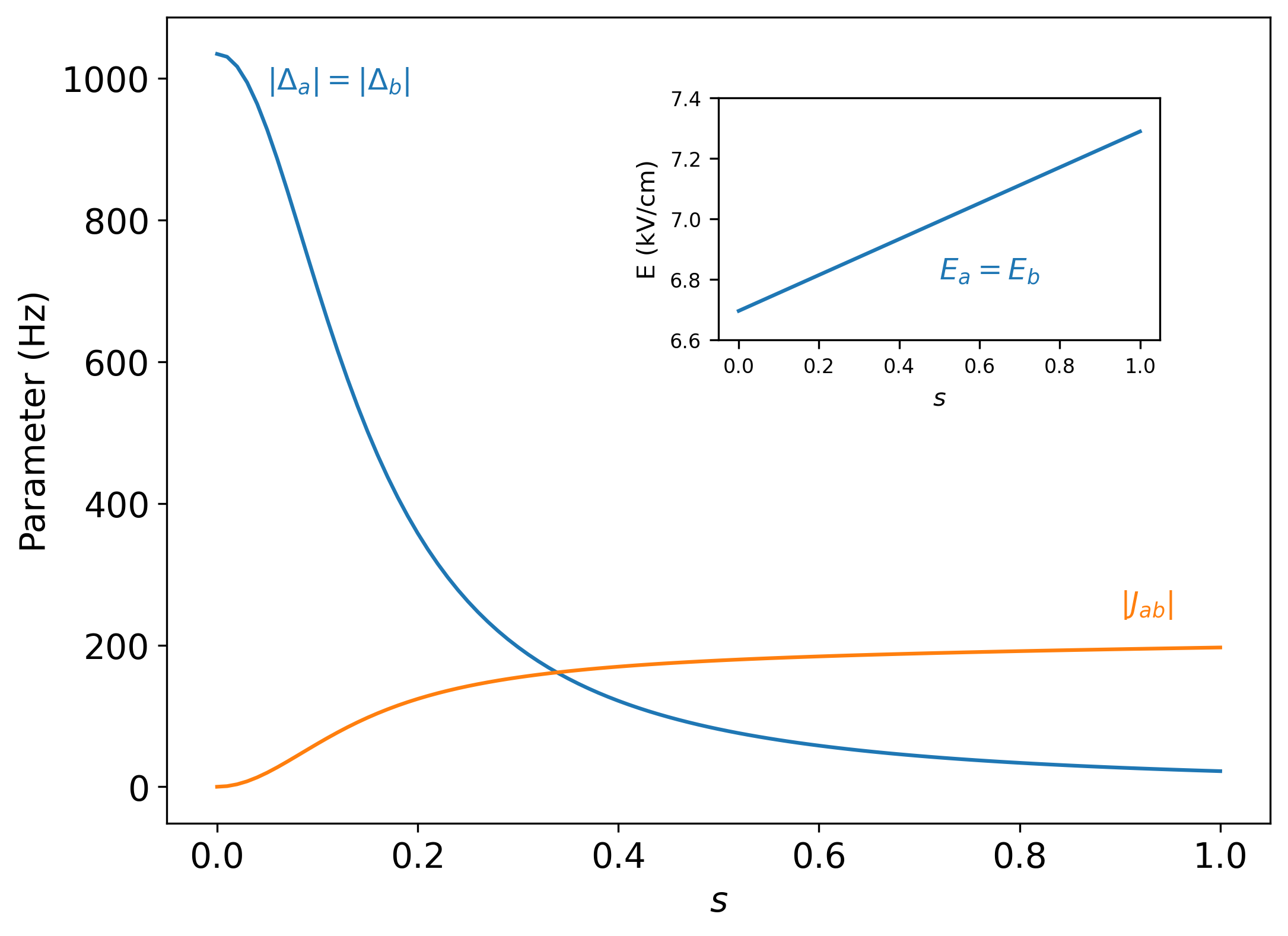}
			\end{tabular}
			\caption{
				Parameters of Hamiltonian \eqref{eq:two_qubit} during QA for two side-by-side qubits (four molecules) in a rectangular configuration in the $\ket{00}$ state (top panel). Molecules in the $\ket{\uparrow}$ state are shown as shaded circles and molecules in the $\ket{\downarrow}$ state are shown as open circles. $B_a = B_b = 600\text{ mT}$, with inter-molecular distances $r_{ij}$ set to $r_{12} = r_{34} = 500$ nm and $r_{13} = r_{24} = 1000$ nm.
				The inset shows the electric-field magnitude at both qubits during the annealing: $E(s)=6.695 + 0.594 s$ kV/cm. $h_a = h_b = 0$ in this configuration. The fields are directed along $\bm r_{12}$. 
			}
			\label{fig:qa}
		\end{figure}

		The two-qubit Hamiltonian for four molecules in a rectangular configuration (Fig. \ref{fig:qa} top) with molecules 1 and 2 forming qubit $a$ and molecules 3 and 4 forming qubit $b$ is
		%(qubit $a$ with molecules 1 and 2 and qubit $b$ with molecules 3 and 4 with couplings $J_{z_{13}}$ and $J_{\perp_{13}}$ between 1 and 3, and $J_{z_{24}}$ and $J_{\perp_{24}}$ between 2 and 4):
		%\begin{align*}
		%	\hat{H}_a &= h_a\hat{S}_a^z + \Delta_a\hat{S}_a^x & \hat{H}_b &= h_b\hat{S}_b^z + \Delta_b\hat{S}_b^x 
		%\end{align*}
		\begin{align}
			\hat{H} &= \hat{H}_a + \hat{H}_b + J_{z_{13}}S_1^zS_3^z + J_{z_{24}}S_2^zS_4^z \nonumber \\
			&+ J_{z_{14}}S_1^zS_4^z + J_{z_{23}}S_2^zS_3^z \nonumber \\
			&+ \frac{J_{\perp_{13}}}{2}\left(S_1^+ S_3^- + {\rm h.c.}\right) + \frac{J_{\perp_{24}}}{2}\left(S_2^+ S_4^- + {\rm h.c.}\right) \nonumber \\
			&+ \frac{J_{\perp_{14}}}{2}\left(S_1^+ S_4^- + {\rm h.c.}\right) + \frac{J_{\perp_{23}}}{2}\left(S_2^+ S_3^- + {\rm h.c.}\right)
		\end{align}
		where $\hat{H}_a$ and $\hat{H}_b$ are given by Eq. \eqref{eq:qubit_ham}. The spin-exchange dynamics induced by  $J_{\perp_{ij}}$ will take qubit states outside of the Hilbert space of the $\ket{0}$ and $\ket{1}$ states of each qubit (e.g. pairs of molecules within a qubit may end up in states $\ket{\uparrow\uparrow}$ and $\ket{\downarrow\downarrow}$). These interactions can be suppressed by placing the qubits further apart or detuning the qubits so that $|h_1 - h_3| \gg |J_{\perp_{13}}|$ and $|h_2 - h_4| \gg |J_{\perp_{24}}|$. This can be achieved by placing qubits in different fields by applying a gradient of an electric or magnetic field along the direction joining the qubits. 
		In this limit, the Hamiltonian reduces to 
		%All The other terms proportional to $J_{z_{13}}$ and $J_{z_{24}}$ ( also $J_{z_{14}}$ and $J_{z_{23}}$) in the Hamiltonian are all diagonal in the $\ket{0_a0_b}$, $\ket{0_a1_b}$, $\ket{1_a0_b}$, and $\ket{1_a1_b}$ basis and the Hamiltonian can be written as:
		\begin{equation}\label{eq:two_qubit}
			\hat{H} = h_a\hat{S}_a^z + \Delta_a\hat{S}_a^x + h_b\hat{S}_b^z + \Delta_b\hat{S}_b^x \pm J_{ab} S_a^z  S_b^z
		\end{equation}
		where $J_{ab} = J_{z_{13}} + J_{z_{24}} - J_{z_{14}} - J_{z_{23}}$, and the sign of the last term depends on the encoding of the qubits (positive when $\ket{0_a} = \ket{\uparrow_1\downarrow_2}$, $\ket{1_a} = \ket{\downarrow_1\uparrow_2}$ and $\ket{0_b} = \ket{\uparrow_3\downarrow_4}$, $\ket{1_b} = \ket{\downarrow_3\uparrow_4}$ and negative when one of the qubits is inverted as in $\ket{0_b} = \ket{\downarrow_3\uparrow_4}$, $\ket{1_b} = \ket{\uparrow_3\downarrow_4}$). We define ferromagnetic and anti-ferromagnetic interactions between molecules as having negative and positive signs on the $J_{ab}$ couplings respectively. The setup illustrated in the top panel of Fig. \ref{fig:qa} has anti-ferromagnetic couplings between the qubits, while also having ferromagnetic couplings between the molecules within each qubit.
		
		Qubits stacked head to head have the same Hamiltonian with half the inter-qubit interaction strength.
		Other stacking configurations can also lead to interesting coupling interplays. For example, qubits stacked perpendicularly on top of each other (in a cross shape with one qubit lying on the symmetry plane of the other qubit) would be completely decoupled from each other and configurations with asymmetric interactions on the molecules of a single qubit can lead to non-zero bias terms.
		
		Couplings between qubits are also dependent on the field magnitudes, with stronger fields resulting in stronger couplings. For SrF molecules considered here, Ising couplings between two qubits can be tuned within the range 300-2500 Hz with magnetic fields of 540-620 mT and dc electric fields of 1.6-8.5 kV/cm ($E_z(B)$) on each qubit. Electric- and magnetic-field gradients and masks could allow for this wide range of couplings between qubits in a single connected system. Combined with the anisotropy of the dipole-dipole interactions and the resulting qubit-qubit couplings, many different Ising models can be encoded into molecules. 
		
		The present method of encoding qubits can be very sensitive to inhomogeneities of external magnetic and electric fields. Field perturbations may induce significant modifications of the energy differences between the $\beta$ and $\gamma$ states, given by $h_1$ and $h_2$ in Eq. \eqref{eq:bias_trans}. Therefore, variations in the field magnitudes applied to a pair of molecules comprising a qubit could lead to significant changes in the bias of the qubit. This effect also hinders the spin-exchange interactions essential to the transverse field simulation of the system
			so care must be taken to ensure field homogeneity over length scales of individual qubits. 	
			However, small perturbations of the fields between the qubits should not be problematic as these would only slightly modify the couplings and could even be desirable, as they diminish the effect of unwanted spin-exchange interactions between qubits.
				
		\subsection{QA with $^2\Sigma$ molecules}\label{sec:qa_mols}
		
		Hamiltonian \eqref{eq:two_qubit} is the transverse field Ising model used in QA applications. 
		In order to use this system as a quantum annealer, one needs to be able to transform 
		\begin{equation}\label{eq:initial_ham}
			\hat{H}_{\rm i} = \Delta_a\hat{S}_a^x + \Delta_b\hat{S}_b^x
		\end{equation}
		to 
		\begin{equation}\label{eq:final_ham}
			\hat{H}_{\rm f} = h_a\hat{S}_a^z + h_b\hat{S}_b^z + J_z S_a^z  S_b^z
		\end{equation}
		by tuning a single external  field parameter. The sensitivity of states $\beta$ and $\gamma$ to external fields near the avoided crossing shown in Fig. \ref{fig:j_bc} makes this possible. 
		
		%Figure 4 shows the dependence of $J_{ab}$ on the magnitudes of magnetic fields $B_a$ and $B_b$ at qubits $a$ and $b$, respectively. Here, the electric fields $E_a$ and $E_b$ are chosen so that $J_z / J_\perp = 100$ and $E_z > E_\perp$.
		
		When the qubits are initialized in a superposition of $\ket{\uparrow\downarrow}$ and $\ket{\downarrow\uparrow}$ states at ($E_{\perp_a}$, $B_a$, $\delta E_a= 0$) and ($E_{\perp_b}$, $B_b$,  $\delta E_b = 0$), the Hamiltonian of the two-qubit system is given by Eq. \eqref{eq:initial_ham} with the values of $\Delta_a$ and $\Delta_b$ determined by the field magnitudes at each qubit. 
		Increasing the electric-field strength to $E_{z_i}$ decreases the magnitude of $\Delta$, while increasing $|\delta E_i|$ increases $|h_i|$. The final Hamiltonian of  two qubits at ($E_{z_a}$, $B_a$, $\delta E_a$) and ($E_{z_b}$, $B_b$, $\delta E_b$) becomes \eqref{eq:final_ham} where the values of $h_a$ and $h_b$ are determined by $\delta E$, and $J_z$ is determined by the field magnitudes at each qubit. For SrF, $\delta E \approx 10$ V/m generates a bias comparable to the Ising coupling between qubits placed 1000 nm apart.
		Alternatively, $h_a$ and $h_b$ can also be modified by coupling each qubit to a molecule locked in either $\beta$ or $\gamma$ (detuning the fields away from the avoided crossing locks the molecule in one of these two states). In this case, increasing the electric-field strength at each qubit from $E_{\perp_i}$ to $E_{z_i}$ also increases the magnitude of $h_a$ and $h_b$. The final values of $h_a$ and $h_b$ can be calculated considering the couplings between the biasing molecules and qubits.
		
		Initialization and read-out of the system can be done by resonant microwave excitation in a gradient of an electric field, as proposed by DeMille \cite{demille}. Electric-field gradients used for this purpose are similar in magnitude to the electric fields needed to apply a bias on the qubits considered here. In our examples using SrF, this can be done with field gradients $\approx 1$ kV/cm\textsuperscript{2}.
		
		In the following discussion, we define the valid qubit states as the collection of states with a single excitation in each qubit. In the subspace of valid qubit states, any measurement (in the $\ket{\uparrow}$ and $\ket{\downarrow}$ basis) leads to a qubit state of $\ket{0}$ and $\ket{1}$ for all qubits. Conversely, invalid states are measured states of the system that do not correspond to any 0-1 qubit encoding described above. These states are a result of the spin-exchange interactions of molecules in different qubits resulting in pairs of molecules in $\ket{\uparrow\uparrow}$ or $\ket{\downarrow\downarrow}$ states. The probability of obtaining such states is amplified by favorable ferromagnetic couplings between molecules in different qubits. 
		
		For QA applications, the goal is to find the minimum-energy state in the subspace of valid qubit states. In some configurations (e.g. when the intraqubit molecular interactions are ferromagnetic), the final ground state of the system (in the subspace of states with $n/2$ excitations for $n$ molecules) after annealing is an invalid state. Thus, the annealing transformation should be quasi-adiabatic and annealing times need to be carefully tuned to balance the adiabaticity of the evolution, while limiting the undesired spin-exchange interactions between qubits. Generally, in configurations where the intraqubit Ising couplings are ferromagnetic, we expect to have a larger probability of observing invalid states. 
		
		In order to test the annealing procedure with different configurations of molecular ensembles, we use quantum dynamics simulations of the corresponding spin-1/2 $XXZ$ Hamiltonian (\ref{eq:manybody_ham}) in a time-dependent electric field with the QuTip python package \cite{qutip1, qutip2}. We assume that the probability of populating any state other than $\ket{\beta}$ and $\ket{\gamma}$ is negligible. This is a reasonable assumption as these states are separated from other states by a significant energy gap. For example, in the case of SrF molecules depicted in Fig. \ref{fig:energy_levels_e}, these states are 
			separated from other states
			by $>0.1$ GHz, while the relevant coupling strengths are $<1$ MHz and the time scale of QA is $\approx10$ ms. Each system is initialized in an equal superposition of valid qubit states. Qubits with ferromagnetic couplings between molecules within qubits need to be initialized in the $\ket{0} + \ket{1}$ state and those with anti-ferromagnetic couplings need to be initialized in the $\ket{0} - \ket{1}$ state. Evolution of this initial state during annealing is then calculated by numerical integration of the time-dependent Schr\"odinger equation using the Adams-Moulton method (implicit Adams) \cite{adams} in a variable-coefficient ordinary differential equation solver \cite{vode} with a maximum order of 12, using a locally time-independent Hamiltonian in each time step. For our calculations, we used 200 time steps for the one-dimensional (1D) lattice and 100 time steps for the 2D lattice. 
		
		\section{Results}\label{sec:results}
		
		\begin{figure*}[]
			\begin{tabular}{ll}
				\quad\includegraphics[width=\columnwidth]{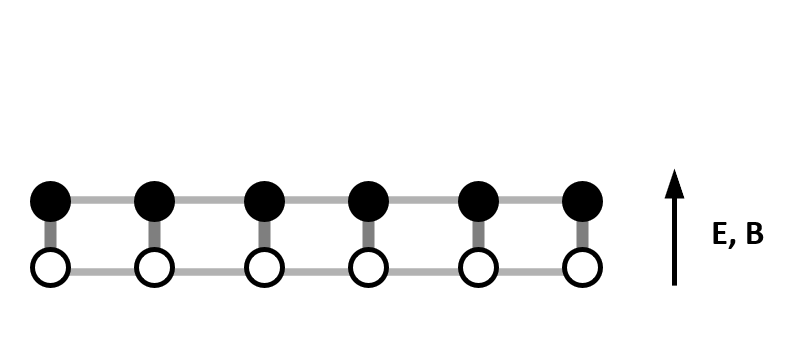} & 
				\includegraphics[width=\columnwidth]{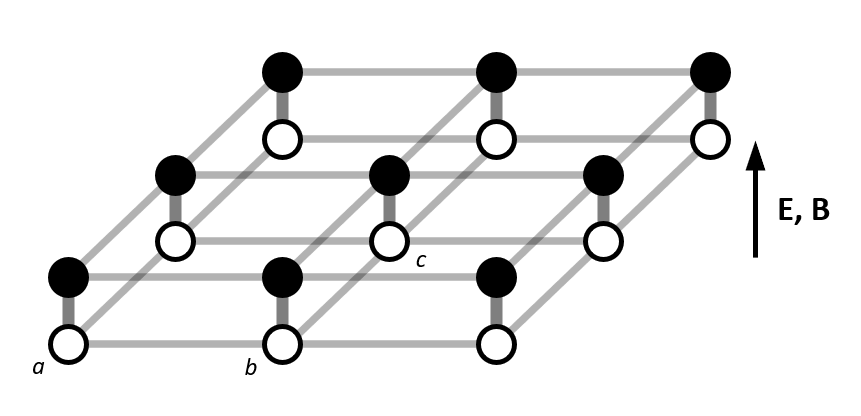} \\
				\includegraphics[width=\columnwidth]{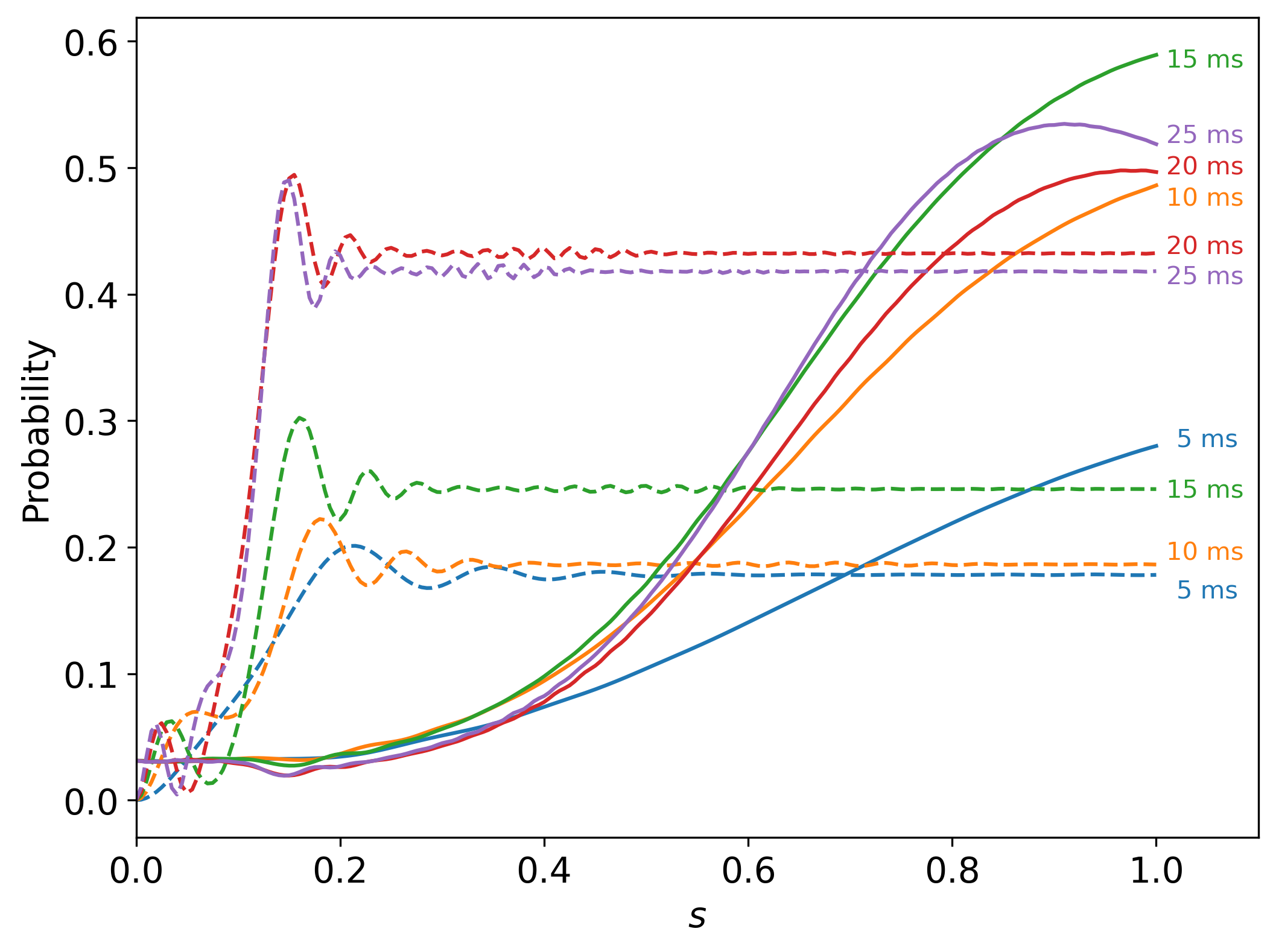} & \includegraphics[width=\columnwidth]{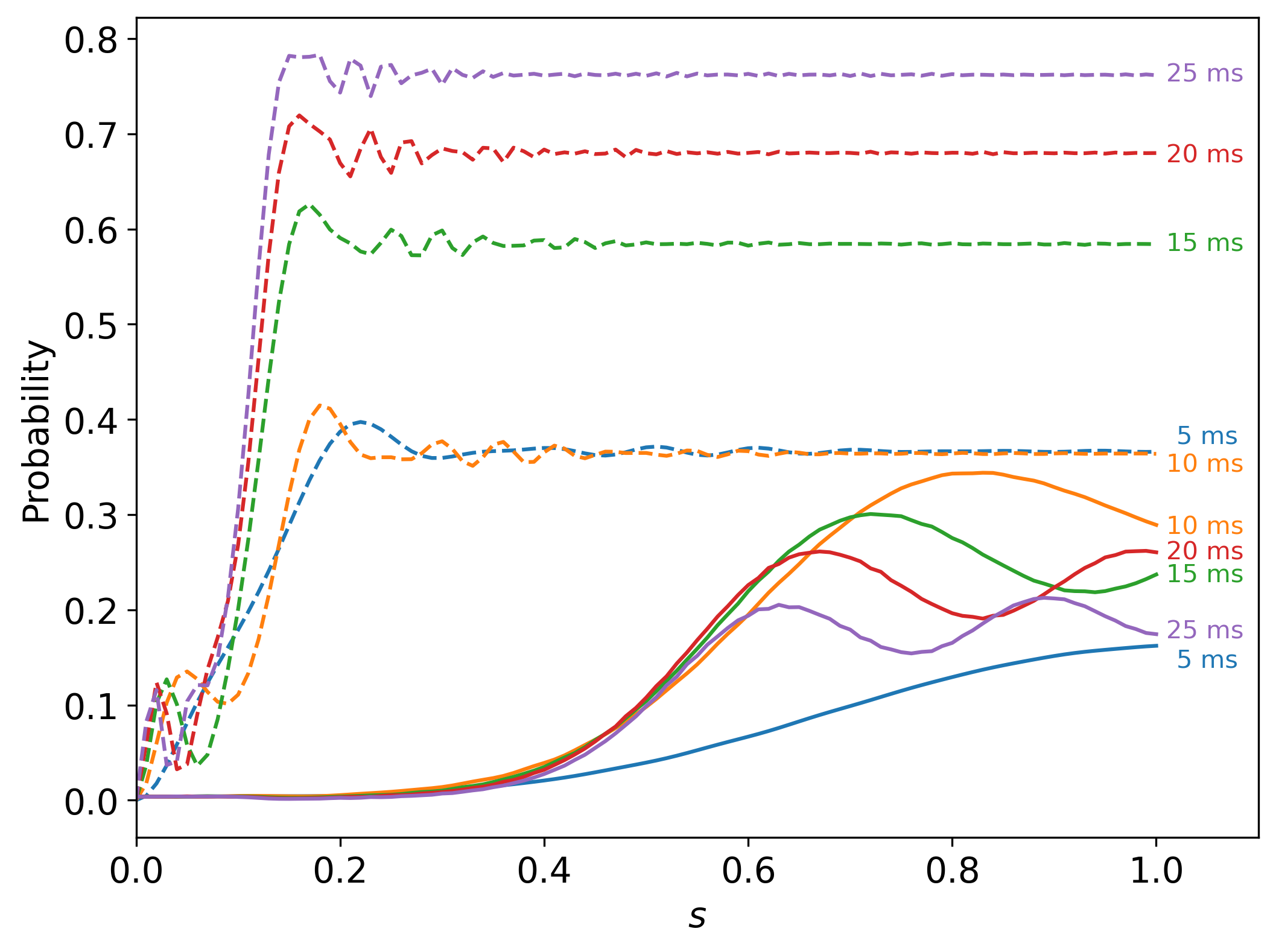} \\
				\includegraphics[width=\columnwidth]{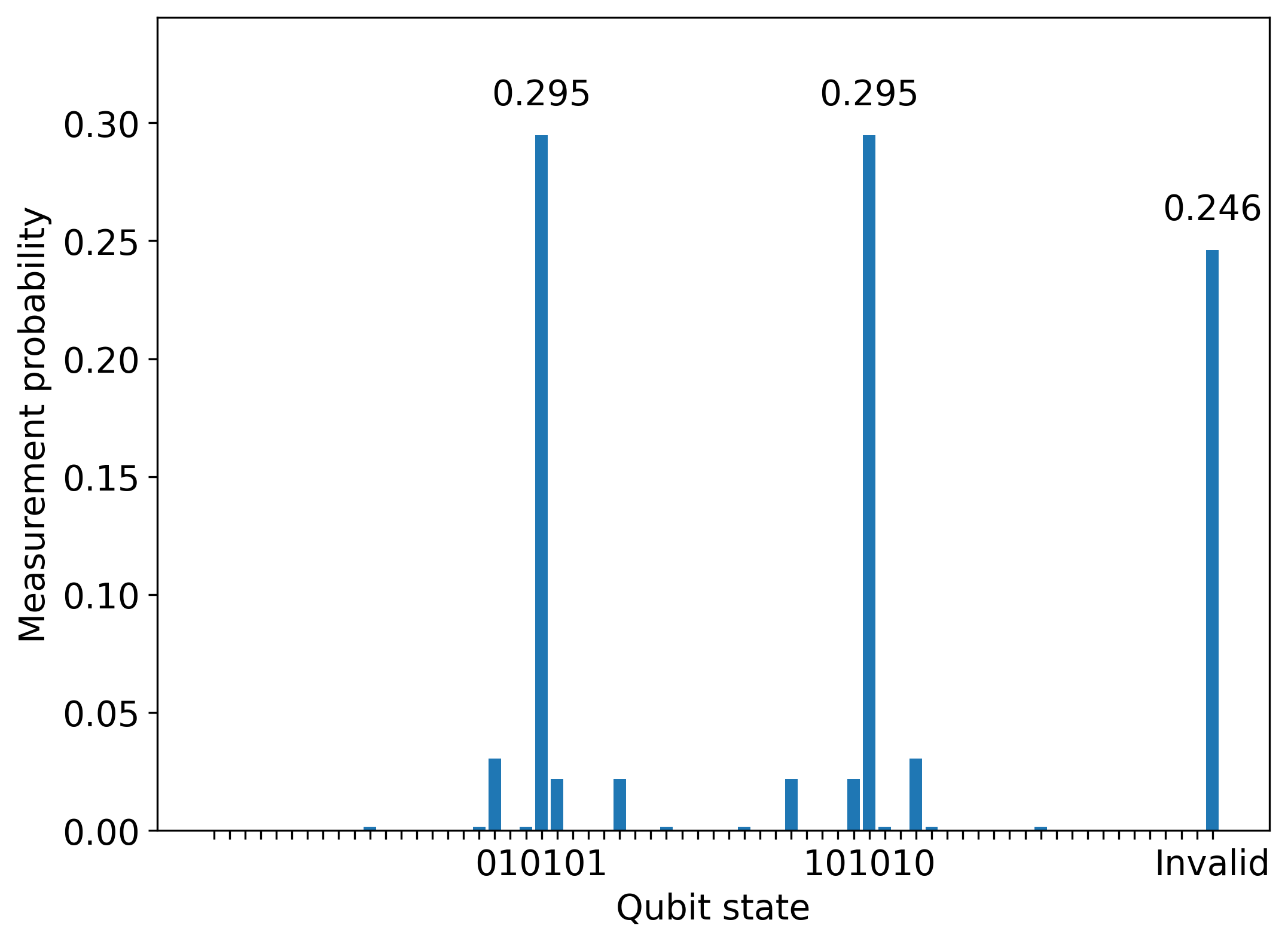} & \includegraphics[width=\columnwidth]{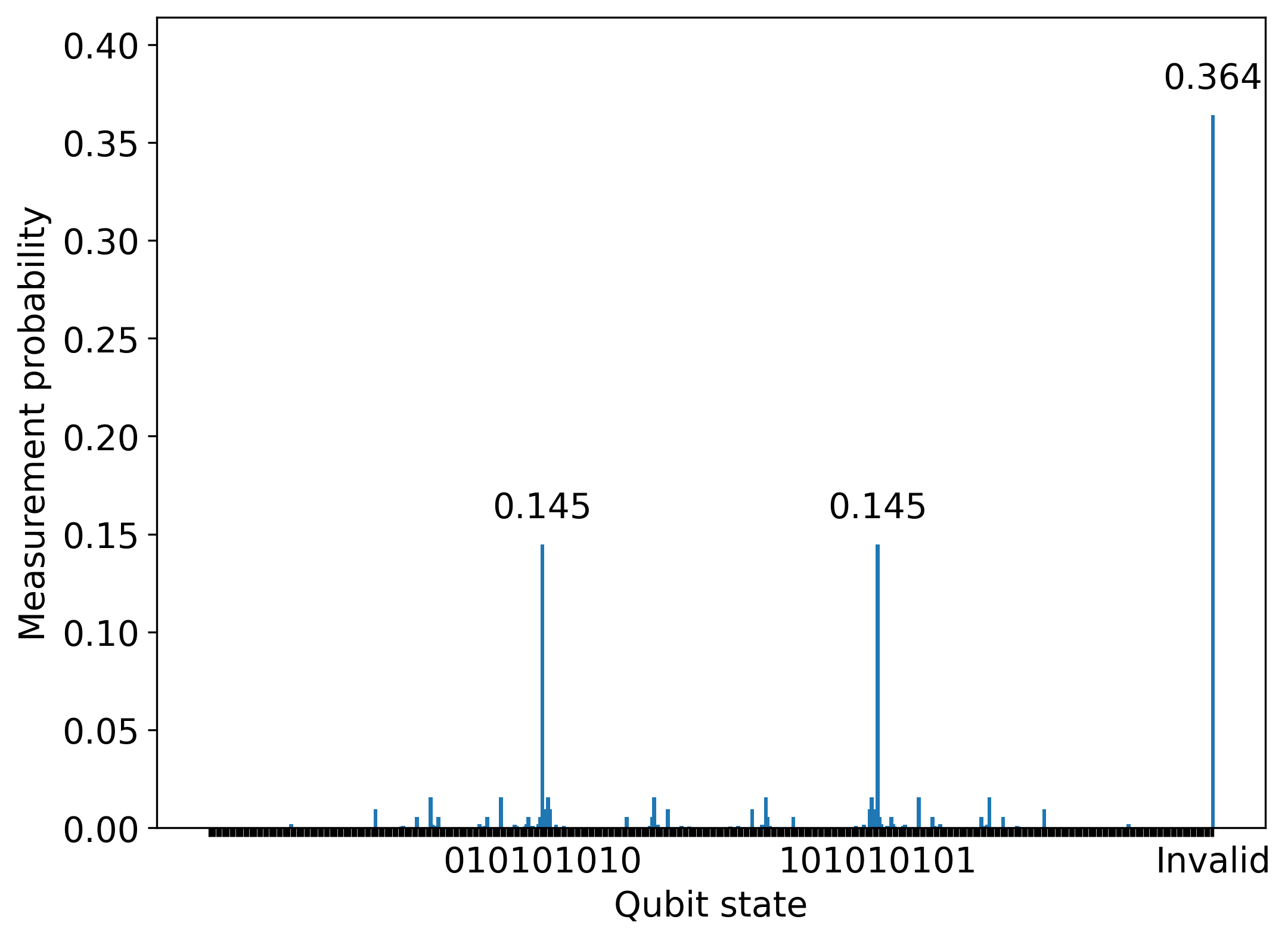} \\
			\end{tabular}
			\caption{Top panels: Examples of 1D (left) and 2D (right) anti-ferromagnetic QA setups based on pairs of molecules as qubits displayed with all qubits in the $\ket{0}$ state (anti-ferromagnetic refers to the sign of the couplings between the qubits). The circles show $^2\Sigma$ molecules trapped in an optical lattice, with shaded circles indicating molecules in the $\ket{\uparrow}$ state and open circles indicating molecules in the $\ket{\downarrow}$ state, with intra-qubit molecular spacing of $r_1=500$ nm and inter-qubit spacing of $r_2=1000$ nm in states  $\ket{\uparrow}=\ket{\beta}$ and $\ket{\downarrow}=\ket{\gamma}$ depicted in Fig. \protect\ref{fig:energy_levels_e}.
				Middle panels: Probabilities of measuring the system and observing solution states (solid lines) and invalid states (dashed lines) during annealing with homogeneous magnetic and electric fields $B = 600$ mT, $E(s)= 6.695 + 0.594 s$ kV/cm using different annealing times. The solution states for both systems are ordered anti-ferromagnetic configurations. 
				Bottom panels: Final probabilities of measuring the system after annealing in 15 ms (10 ms) of the 1D (2D) configuration. 
			}
			\label{fig:samples}
		\end{figure*}
		
		Figure \ref{fig:qa} displays a two-qubit configuration of four SrF($^2\Sigma^+$) molecules and the parameters of Hamiltonian \eqref{eq:two_qubit} during the QA procedure described above with $B_a = B_b = 600\text{ mT}$. For the present calculations, we assume that the qubits consist of two SrF molecules separated by 500 nm, while the spacing between the molecules in the direction joining the qubits is 1000 nm. We use the value 3.47 Debye for the dipole moment \cite{srf}, $2.49\times 10^{-3}\text{ cm}^{-1}$ for the spin-rotational coupling constant \cite{spinrotation} and 0.251 cm\textsuperscript{-1} for the rotational constant \cite{romannjp} of SrF($^2\Sigma^+$). The system is initialized by preparing one molecule in each qubit in the excited state. The many-body system is then relaxed to the minimum-energy state at $E_\perp = E_{\perp_a} = E_{\perp_b} = 6.695 \text{ kV/cm}$, as the electric-field strength is increased to $E_z =  E_{z_a} = E_{z_b} = 7.289 \text{ kV/cm}$, where $J_z / J_\perp = 100$. 
		
		Initially, $\Delta_a, \Delta_b \gg J_{ab}$ with $\Delta_a = \Delta_b = -1034.3\text{ Hz}$ and the spin-exchange interactions between qubits ($J_{\perp_{13}}, J_{\perp_{24}}, J_{\perp_{14}}, J_{\perp_{23}}$) are one order of magnitude weaker than the spin-exchange interaction between molecules within qubits ($J_{\perp_{13}} = J_{\perp_{24}} = 64.6\text{ Hz}$, while $J_{\perp_{12}} = \Delta_a = -1034.3\text{ Hz}$). 
		As the electric field is increased linearly from $E_\perp$ to $E_z$, $J_{ab}$ increases to its final value of 196.6 Hz and $\Delta_a$ and $\Delta_b$ decrease to about -22 Hz. In the limited subspace of valid qubit states, the lowest energy states are $\ket{01}$ and $\ket{10}$ as induced by the anti-ferromagnetic coupling $J_{ab}$. However, it should be noted that in this configuration, the dominant coupling at the end of annealing is the short-range ferromagnetic coupling between the molecules in each qubit ($J_{z_{12}} = J_{z_{34}} = -2.2\text{ kHz}$) and the ground states of the system are $\ket{\uparrow_1\uparrow_2\downarrow_3\downarrow_4}$ and $\ket{\downarrow_1\downarrow_2\uparrow_3\uparrow_4}$, which lead to invalid states.

		This setup can be extended to more complex topologies with more qubits, leading to more complex Ising models. 
		Simulating an arbitrary Ising model requires creative manipulation of qubit connectivity as the connectivities are not all independent and the signs of the couplings need to be carefully matched.
		We propose the following physically realistic setups to demonstrate the ability of the proposed QA to simulate tunable Ising models with 1D and 2D connectivities, extendable to a large number of qubits.
		
		First, we consider a chain of six qubits each arranged along the field direction as displayed in Fig. \ref{fig:samples} (top left). This configuration can optimize the following Ising model with long-range interactions:
		\begin{equation}\label{eq:2d_ham}
			\hat{H}_f = \sum_i h_i S_i^z + \sum_{i, j}J_{ij} S_i^zS_j^z
		\end{equation}
		where $h_i = 0$, $J_{ij} > 0$ with decreasing coupling magnitude as the distance between $i$ and $j$ increases. Therefore, the final state is the ground state of an anti-ferromagnetic chain of qubits.
		
		For simplicity, we consider qubits to be in a homogeneous magnetic field and a homogeneous tunable electric field: $E(s) = E_\perp + s (E_z - E_\perp)$. With all  the molecules experiencing the same magnetic field we have the same $E_\perp$ and $E_z$ for all molecules.
		The value of $E_z$ for each qubit is then determined by finding the electric-field magnitude, at which all molecules are detuned from the avoided crossing and $\Delta_i \approx 0$. With fixed magnetic fields and electric-field gradient, QA can be performed by linearly increasing the electric-field magnitude from $E_\perp$ to $E_z$.  Figure \ref{fig:samples} (middle left) shows the probabilities of measuring the system in one of the solution states or an invalid state during the annealing process with $B = 600$ mT, $E_\perp = 6.695$ kV/cm and $E_z = 7.289$ kV/cm, the intra-qubit distance $r_1=500$ nm and the inter-qubit distance $r_2=1000$ nm using different annealing times. As seen in Fig. \ref{fig:samples}, longer annealing times lead to higher probability of observing the system in one of the fully anti-ferromagnetic states at the cost of a lower probability of measuring the system in a valid qubit state. Figure \ref{fig:samples} (bottom left) displays the probabilities of observing each of the qubit states after annealing for 15 ms, showing a high probability of measuring one of the solution states. 
		
		This configuration can be generalized to an Ising model with different couplings between lattice sites using inhomogeneous fields and extended in one dimension to an arbitrary number of qubits.
		This configuration can be also extended into a 2D lattice of qubits by stacking the 1D chains of Fig. \ref{fig:samples} (top left) in the direction perpendicular to the fields. Figure \ref{fig:samples} (top right) shows one example of such configuration. With 
		homogeneous magnetic and electric fields chosen as $B = 600$ mT and $E(s)= 6.695 + 0.594 s$ kV/cm, this configuration optimizes the Ising model  (\ref{eq:2d_ham}) with $h_i = 0$ and $J_{ij} > 0$.  Figure \ref{fig:samples} (middle right) shows the evolution of the probabilities of measuring the system in solution states and invalid states for different annealing times. It also demonstrates that a linear electric-field ramp may not be optimal for the annealing procedure as the final probabilities can, for some cases, be better optimized by stopping the annealing at $s<1$ (the 10-, 15-, 20- and 25-ms cases could all benefit from this).
		
\begin{figure}
				\includegraphics[width=\columnwidth]{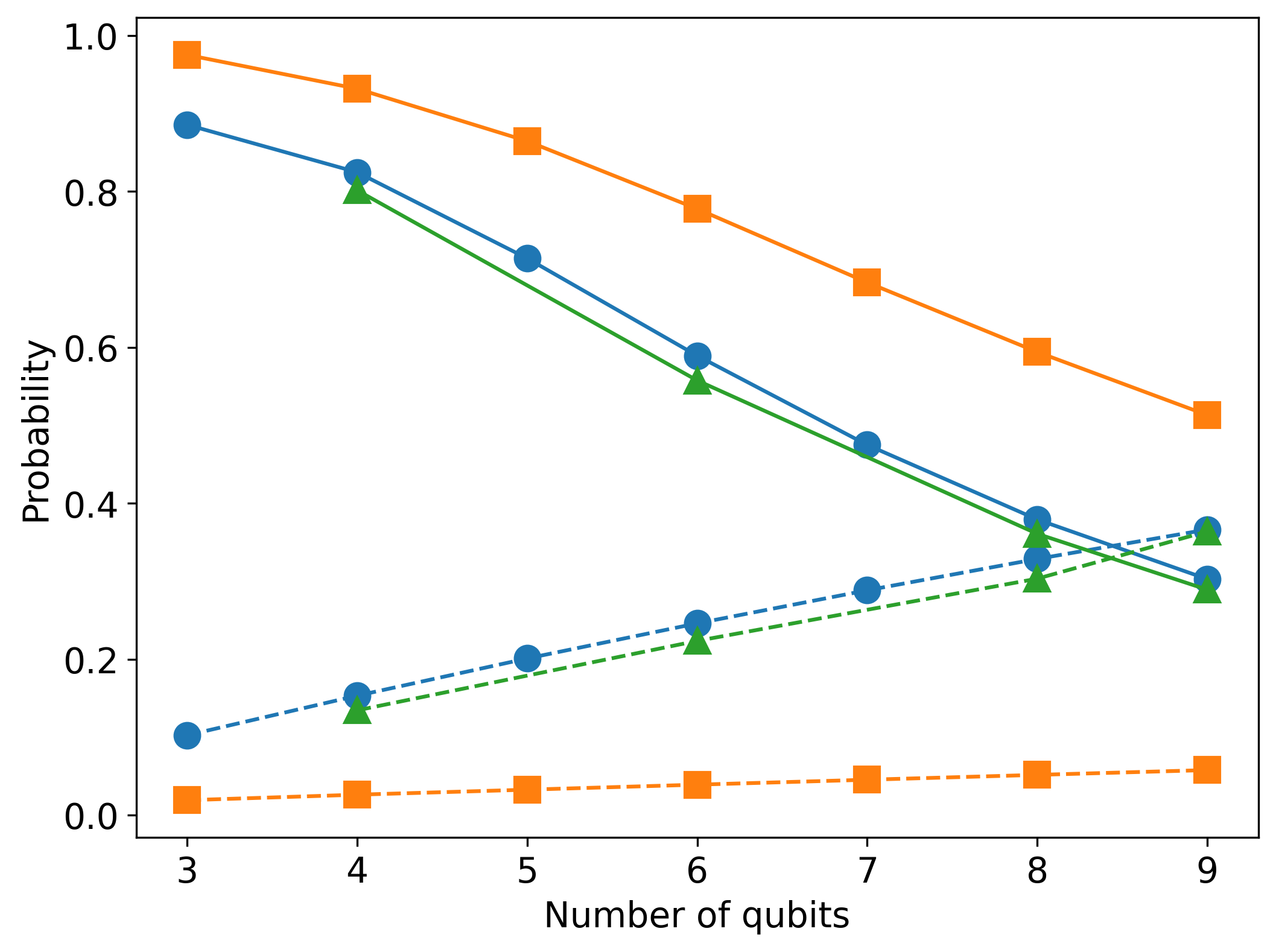} 
				\caption{Probability of obtaining the solution states (solid lines) and the invalid states (dashed lines) for three configurations of qubits. 1D anti-ferromagnetic (blue circles), 1D ferromagnetic (orange squares), and 2D anti-ferromagnetic (green triangles), where ferromagnetic and anti-ferromagnetic refer to the sign of the couplings between the qubits. For the 2D systems, we use $2\times2$, $2\times3$, $2\times4$ and $3\times3$ configurations for the corresponding number of qubits.}
				\label{fig:scale}
			\end{figure}

		Figure \ref{fig:samples} (bottom right) shows the final probabilities of measuring the system after annealing with 10 ms of annealing time. This configuration is also extendable in two dimensions (perpendicular to the fields) to an arbitrary number of qubits. It should also be noted that while both configurations considered here as examples have anti-ferromagnetic couplings between qubits, they also include ferromagnetic couplings between molecules within qubits which results in a higher probability of measuring invalid states following the adiabatic evolution of the system. In Fig. \ref{fig:scale} we show how a configuration with anti-ferromagnetic intraqubit couplings has a much lower probability of yielding invalid states. We also demonstrate the relationship between solution and invalid state probabilities and the number of qubits for these sample configurations. 
		
		For the scalability of the proposed approach, we examine the probabilities of obtaining the solution states and the invalid states for a range of system sizes (Fig. \ref{fig:scale}). The 1D and 2D anti-ferromagnetic configurations are depicted in the top panels of Fig. \ref{fig:samples}. Ferromagnetic 1D configurations can be considered as a $\pi/2$ rotation of the 1D anti-ferromagnetic system, with the intraqubit axes perpendicular to the field vectors, leading to anti-ferromagnetic intraqubit interactions and the interqubit axes parallel to the field vectors leading to ferromagnetic interqubit interactions. As described in Section \ref{sec:qa_mols}, we calculate the measurement probabilities after the QA procedure with annealing times of 5, 10, 15, 20, and 25 ms. Probabilities in Fig. \ref{fig:scale} are shown for the annealing time resulting in the highest probability of measuring the solution states. 
			
			Figure \ref{fig:scale} shows that the probabilities are highly system-dependent, as the 1D ferromagnetic system has negligible probability of yielding the invalid states while this probability becomes comparable to the solution state probabilities for the other investigated systems. This is not unexpected as the anti-ferromagnetic intraqubit interactions of the 1D ferromagnetic system protect the qubit structure throughout the annealing. 
		
\begin{figure}
			\begin{tabular}{c}
				\includegraphics[width=\columnwidth]{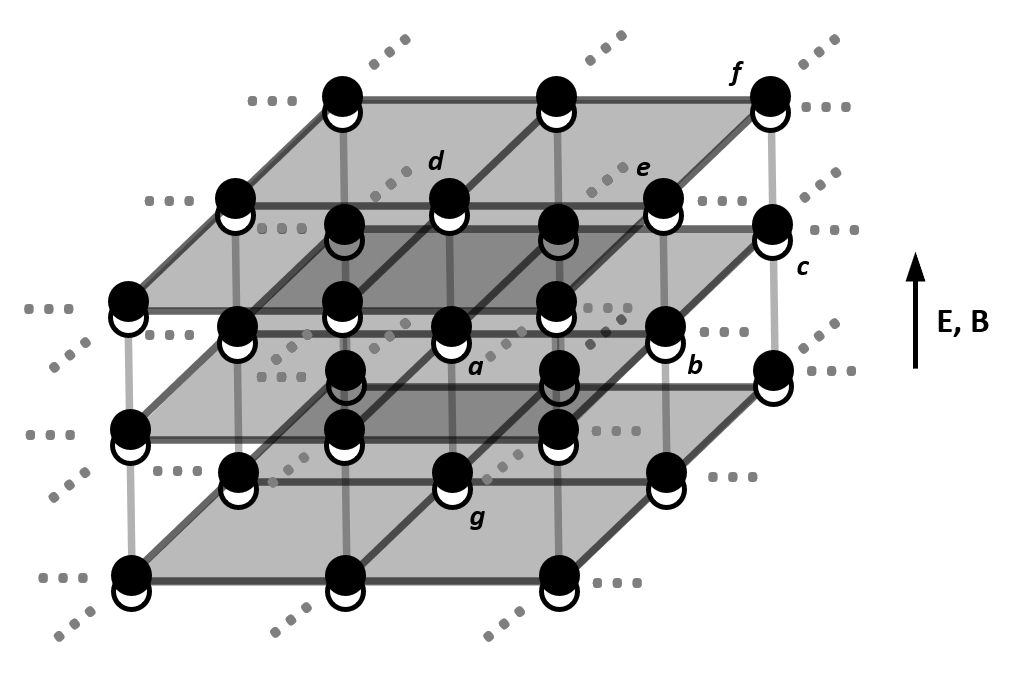} \\
				\includegraphics[width=\columnwidth]{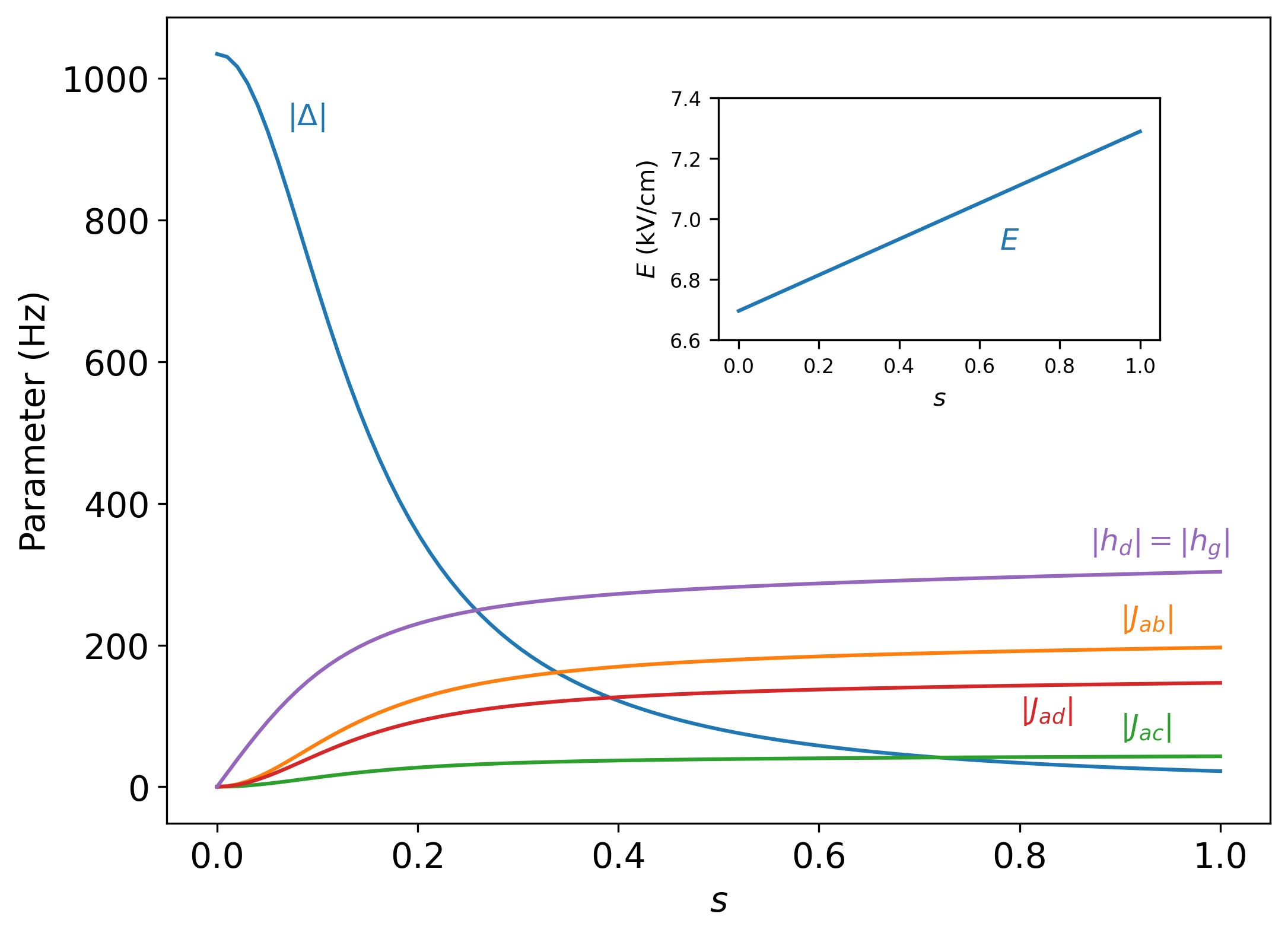}
			\end{tabular}
			\caption{(Top) A configuration of molecules leading to an Ising model with 3D connectivities. 2D layers of Fig. \protect\ref{fig:samples} (top right) are stacked on top of each other. (Bottom) Parameters of Hamiltonian \eqref{eq:2d_ham} during QA with intra-qubit molecular spacing of $r_1=500$ nm and inter-qubit spacing of $r_2=1000$ nm. $B = 600\text{ mT}$ and $E(s)= 6.695 + 0.594 s$ kV/cm.}
			\label{fig:3d_sample}
		\end{figure}

		Using more complicated optical lattice configurations, one can also implement 3D Ising lattices, as, for example, illustrated in Fig. \ref{fig:3d_sample}, which stacks 2D lattices of Fig. \ref{fig:samples} (top right) in the field direction. Fig. \ref{fig:3d_sample} (bottom) shows the parameters of Hamiltonian (\ref{eq:2d_ham}) during annealing. This configuration leads to an Ising model with anti-ferromagnetic couplings inside each 2D layer perpendicular to the fields and ferromagnetic couplings between the layers. Additionally, molecules inside each qubit of the bottom and top layers of this configuration (e.g., qubits $d$ and $g$) experience different environments and will have a non-zero bias ($h_d$ and $h_g$, where $h_d = -h_g$).
		
		\section{Conclusion}
		
		In summary, we have proposed and demonstrated by numerical calculations the possibility of quantum annealing based on open-shell polar molecules in combined dc electric and magnetic fields. 
		This paper exploits the unique structure of $^2\Sigma$ molecules exhibiting avoided crossings between states of different parity and different electron spin orientation. The structure of molecules and interactions between molecules are sensitive to external fields near these avoided crossings. 
		
		We have shown that the Ising models suitable for annealing applications can be encoded into a many-body system of $^2\Sigma$ molecules, with qubits defined by pairs of molecules sharing an excitation.
		This requires trapping molecules in an optical lattice with different lattice spacing along different directions. 
		Because molecules within qubits interact individually with molecules in different qubits,  leaking of populations outside the Hilbert space of the qubit states is possible. However, our dynamical calculations have shown that the probability of obtaining valid annealing solutions is high and can be optimized by varying the annealing times. 
		
		Given the anisotropy of the dipole - dipole interactions and sensitivity of the coupling values to the field magnitudes, a wide range of discrete optimization problems can be encoded using pairs of $^2\Sigma$ molecules in optical lattices. Here, we have demonstrated this procedure with two practical examples (anti-ferromagnetic 1D chain and 2D lattice) that require simple optical lattice configurations and homogeneous external fields. Both demonstrated examples are extendable to a large number of qubits. Our dynamics calculations for these configurations demonstrate a high probability of obtaining the annealing solution states.
		
		It is important to note that the magnetic- and electric-field requirements for the proposed QA scheme depend on the structure of molecular energy levels. For example, Fig. \ref{fig:j_bc} shows the parameters of Eq. \eqref{eq:manybody_ham} implemented using SrI, which has a significantly smaller rotational constant of 0.0367 cm\textsuperscript{-1} \cite{sri-spinrot}, a similar spin-rotational constant of $3.29\times10^{-3} \text{ cm}^{-1}$ \cite{sri-spinrot}, and a slightly larger dipole moment of 6.00 Debye \cite{sri-dipole} compared to SrF. QA with SrI molecules would require magnetic fields of about 0.1 mT compared to 0.6 mT for SrF. Molecules in the $\Sigma$ electronic state of higher spin multiplicity exhibit similar avoided crossings and can be analogously used for quantum annealing applications discussed here. Of particular interest are $^3\Sigma$ molecules that can be produced at ultracold temperatures by photo- or magneto-association of ultracold alkali-metal atoms.
		
		It remains to be seen whether non-trivial Ising models encoding useful problems can be simulated with molecule-based systems. 
		Non-trivial practical applications would require either a complex optical lattice configuration or complicated field patterns. 
		Such field patterns can potentially be identified with machine learning approaches. 
		Similar setups using optical tweezer arrays offering more control over qubit geometry may be possible. However, optical tweezer arrays have not yet been realized with interparticle distances of less than 1 \textmu m.
		Finally, we note that the approach proposed in the present paper can be extended to realize qudits of more than two states. As pairs of molecules simulate qubits, $n$ molecules arranged in a symmetrical configuration can realize $n$-state qudits. However, interactions between such qudits are more complex and a more in-depth study is needed to assess their value as quantum information processing units.
		
		\section*{Acknowledgments}
		
		This work was supported by Natural Sciences and Engineering Research Council Grant No. 543245.
		
			\appendix
			\section{Many-body Hamiltonian}\label{ap:manybody}
			
			The dipole-dipole interactions between two molecules in a dc electric field can be written as 
			\begin{align}
				V_{\rm d-d} &= R^{-3}  \left[ -\frac{3}{2} \sin^2\theta \left(d_{-1}d_{-1} + d_{1}d_{1}\right) \right. \nonumber \\ 
				&\left. - \frac{3}{\sqrt{2}} \sin \theta \cos\theta \left(d_{-1}d_{0} + d_{0}d_{-1} - d_{1}d_{0} + d_{0}d_{1}\right) \right. \nonumber\\
				&\left. - \frac{1}{2}(3\cos^2\theta - 1) \left(d_{1}d_{-1} + 2d_{0}d_{0} + d_{-1}d_{1}\right) \right] \label{eq:dipole_int}
			\end{align} 
			where $\theta$ is the angle between the quantization axis (direction of the electric field) and the inter-molecular vector ($\hat{R}$), and $d_i$ are the spherical components of the dipole moment operator. The matrix elements of the dipole moment components can be evaluated as
			\begin{align}
				\Braket{N M_N M_S}{d_{i}}{N' M'_N M'_S} &= D\delta_{M_S,M'_S}(-1)^{M_N}\sqrt{(2N+1)(2N'+1)} \nonumber \\ 
				&\times\begin{pmatrix}
					N & 1 & N' \\
					0 & 0 & 0 
				\end{pmatrix} \begin{pmatrix}
					N & 1 & N' \\
					-M_N & i & M'_N
				\end{pmatrix} 
			\end{align}
			where $D$ is the permanent dipole moment of the molecule and the brackets denote $3j$ symbols.  
			
			We encode a spin-1/2 system into two eigenstates of the Hamiltonian given by Eq. \eqref{eq:iso_ham}. We denote these two states as $\ket{\uparrow}$ and $\ket{\downarrow}$. As long as the spin-exchange interaction strengths are much smaller than the energy differences between the eigenstates, the number of molecules in each of the two states is  conserved and off-resonant matrix elements of $V_{\rm d-d}$ can be ignored \cite{agranovich}. Under this condition, only the final three terms of Eq. \eqref{eq:dipole_int}, involving $d_id_{-i}$ terms, are relevant and the interactions between molecules depend on $\theta$ as
			\begin{equation}\label{eq:int_theta}
				V_{\rm d-d} \propto \frac{1}{2}(3\cos^2\theta - 1).
			\end{equation}
			Dipole-dipole interactions between molecules $i$ and $j$ can then be calculated in the subspace spanned by $\{\ket{\uparrow}, \ket{\downarrow}\}$ as \cite{hazzard}
			\begin{align}
				V_{\rm d-d} =& J_z \hat{S}_i^z\hat{S}_j^z + \frac{J_\perp}{2} \left(\hat{S}_i^+\hat{S}_j^- + \hat{S}_i^-\hat{S}_j^+\right) \nonumber \\
				&+ W\mathbb{I}_i\hat{S}_j^z + K\hat{S}_i^z\mathbb{I}_j + V\mathbb{I}_i\mathbb{I}_j. \label{eq:xxz_int}
			\end{align}
			The full Hamiltonian for the two-molecule system becomes
			\begin{align}
				\hat H =& J_z \hat{S}_i^z\hat{S}_j^z + \frac{J_\perp}{2} \left(\hat{S}_i^+\hat{S}_j^- + \hat{S}_i^-\hat{S}_j^+\right) \nonumber \\
				&+ h_i\hat{S}_i^z + h_j\hat{S}_j^z
			\end{align}
			where $h_i = \epsilon_{i,\downarrow} - \epsilon_{i,\uparrow} + K$ and $h_j = \epsilon_{j,\downarrow} - \epsilon_{j,\uparrow} + W$. Extending this two-body Hamiltonian to a many-body system of molecules yields Eq. \eqref{eq:manybody_ham}.
			
			\section{Dependence of QA parameters on molecular constants}\label{sec:mol_params}
			
			In this section we discuss the relationship between each of the parameters in the molecular Hamiltonian given by Eq. \eqref{eq:iso_ham} and the parameters of QA (i.e. fields and couplings). To examine the effect of each molecular constant, we vary one molecular constant starting from the values relevant for the SrF molecule ($B_e = 0.251 \text{ cm}^{-1}$, $\gamma_{SR}=2.49\times10^{-3} \text{ cm}^{-1}$, $d = 3.47 \text{ D}$) while keeping all other molecular constants fixed. We set the magnetic field to 0.6 mT. This limits the range of some parameters as the $\ket{\beta}-\ket{\gamma}$ avoided crossing may not occur for this magnetic field at some electric fields for a particular set of molecular parameters. $E_z$ is defined as the smallest electric-field magnitude larger than $E_\perp$ where $J_z = 100 J_\perp$.
			
\begin{figure}
				\begin{tabular}{c}
					\includegraphics[width=\columnwidth]{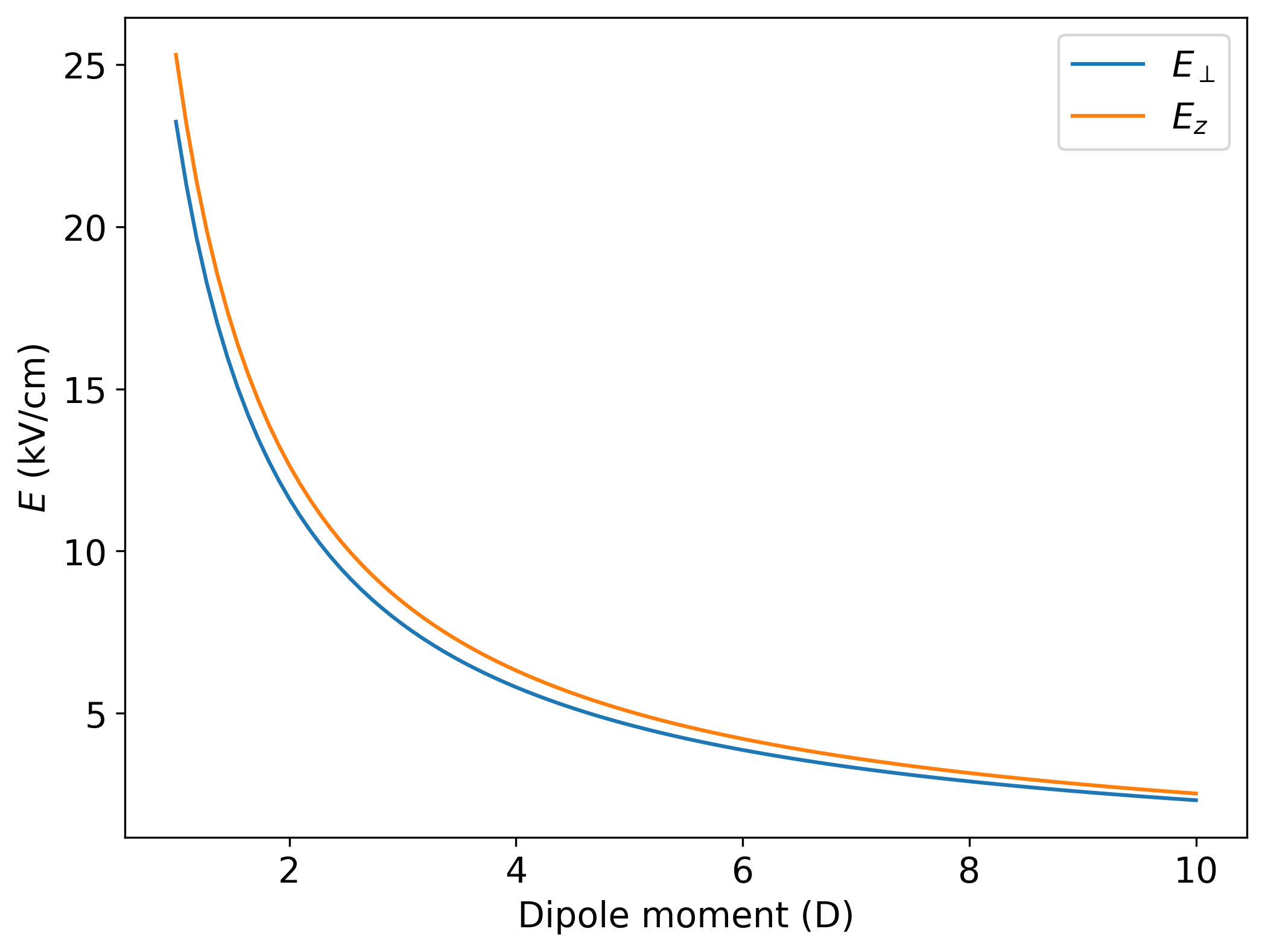} \\
					\includegraphics[width=\columnwidth]{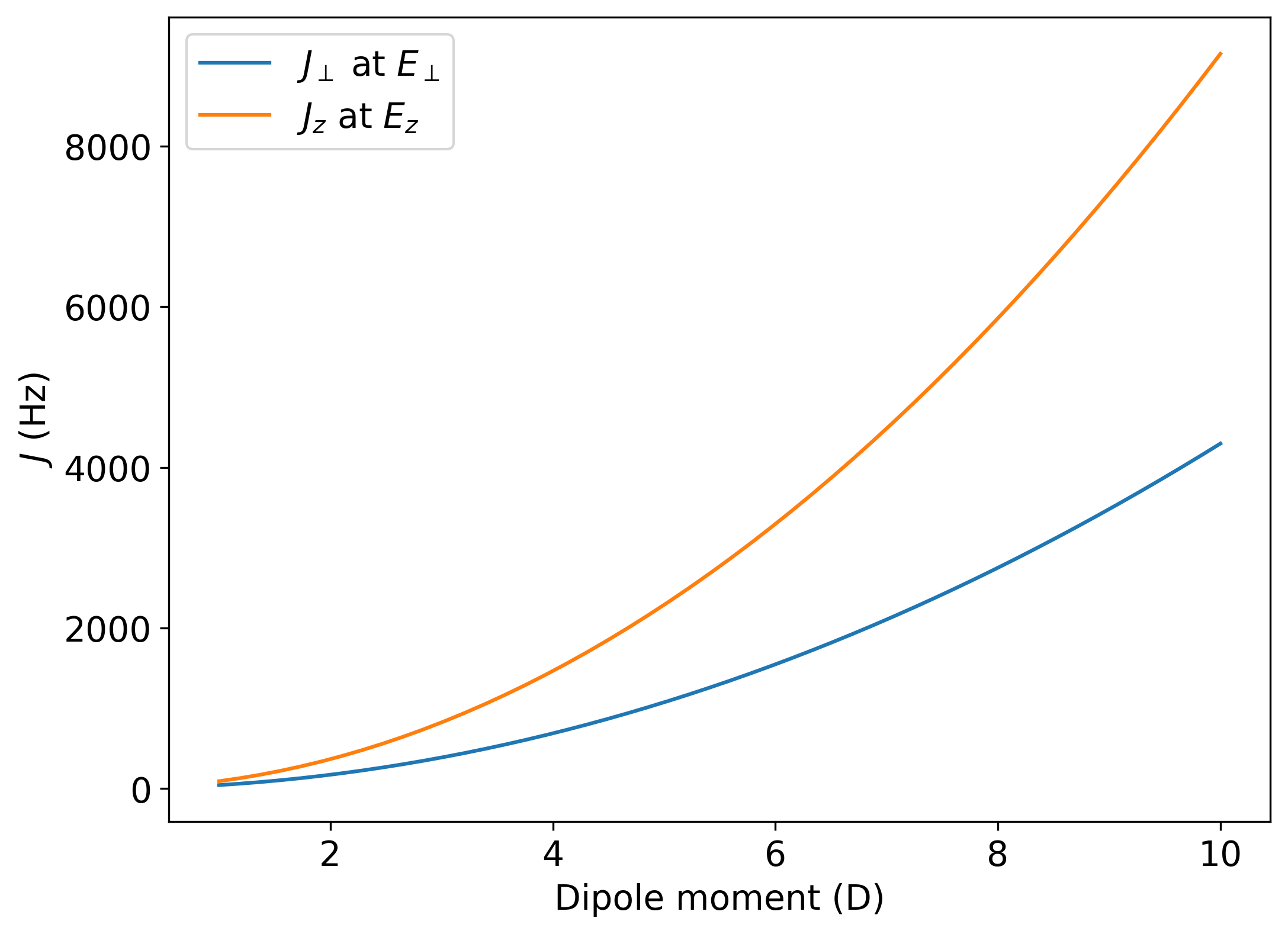}
				\end{tabular}
				\caption{Dependence of the QA field parameters and couplings on the permanent dipole of molecules. The magnetic field is fixed at 0.6 mT and other molecular parameters of the system are fixed at values for SrF.}
				\label{fig:ej_d}
			\end{figure}

			Figure \ref{fig:ej_d} shows how $E_\perp$ and $E_z$ (top panel) and $J_\perp$ and $J_z$ at these electric-field magnitudes (bottom panel) change with respect to the permanent dipole moment of the molecule. $E_\perp$ and $E_z$ are larger for molecules with a smaller dipole moment as the Stark effect is weakened. The relevant couplings at these field magnitudes increase quadratically with the permanent dipole moment magnitude. This shows that the matrix elements of the dipole moment are larger for molecules with larger dipole moments, even though the electric field used is weaker. 
			
			The spin-rotational structure of a molecule is very sensitive to changes in the rotational constant. As a result, there is a very limited range of $B_e$ values permitting the avoided crossing at 0.6 mT. Above (below) this range, the low magnetic field seeking rotational state of the $N=0$ manifold (yellow and green lines in Fig. \ref{fig:energy_levels_e}) is lower (much higher) in energy than the high magnetic field seeking rotational states (red line in Fig. \ref{fig:energy_levels_e}) of the $N=1$ manifold at zero electric field. 
			
\begin{figure}
				\begin{tabular}{c}
					\includegraphics[width=\columnwidth]{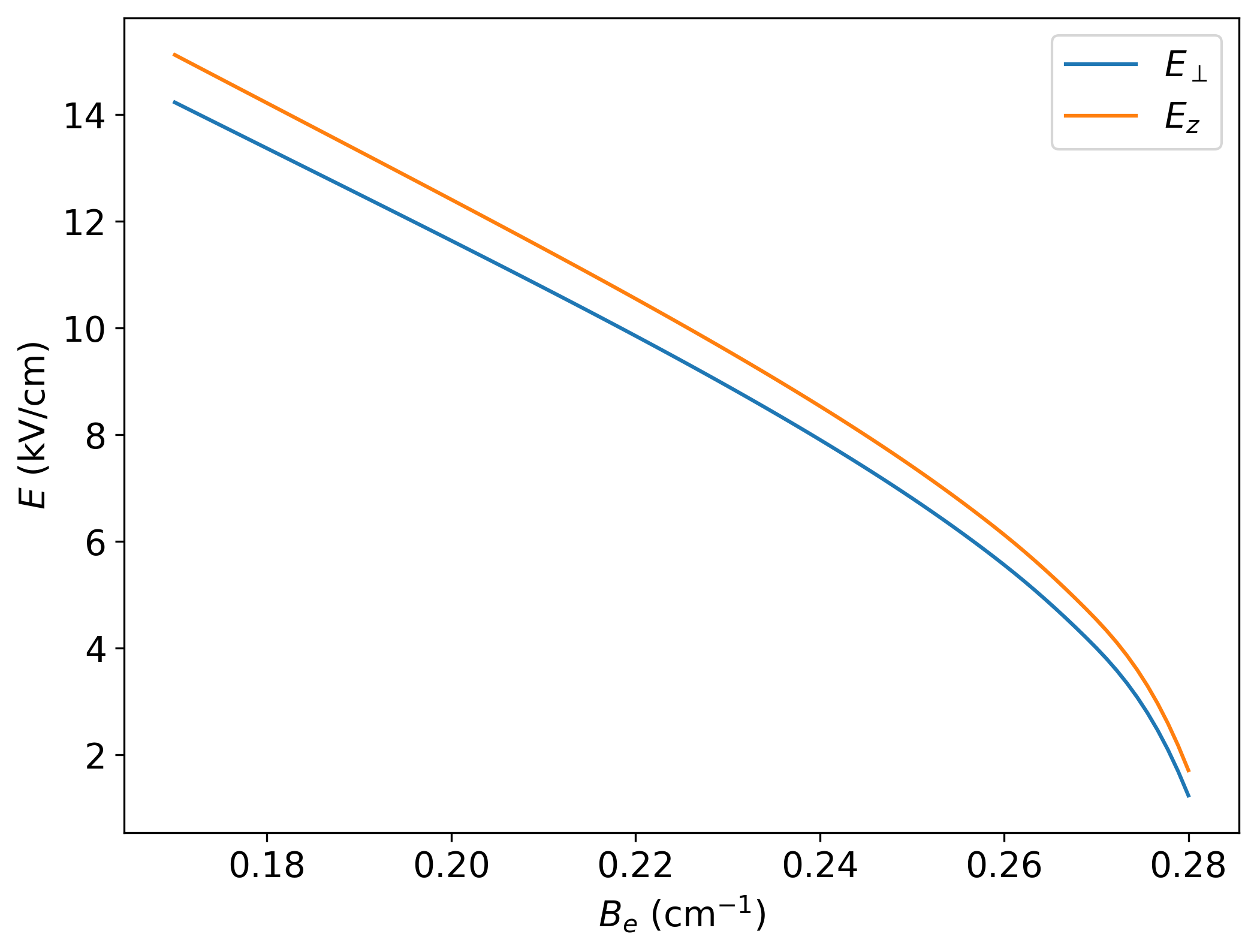} \\
					\includegraphics[width=\columnwidth]{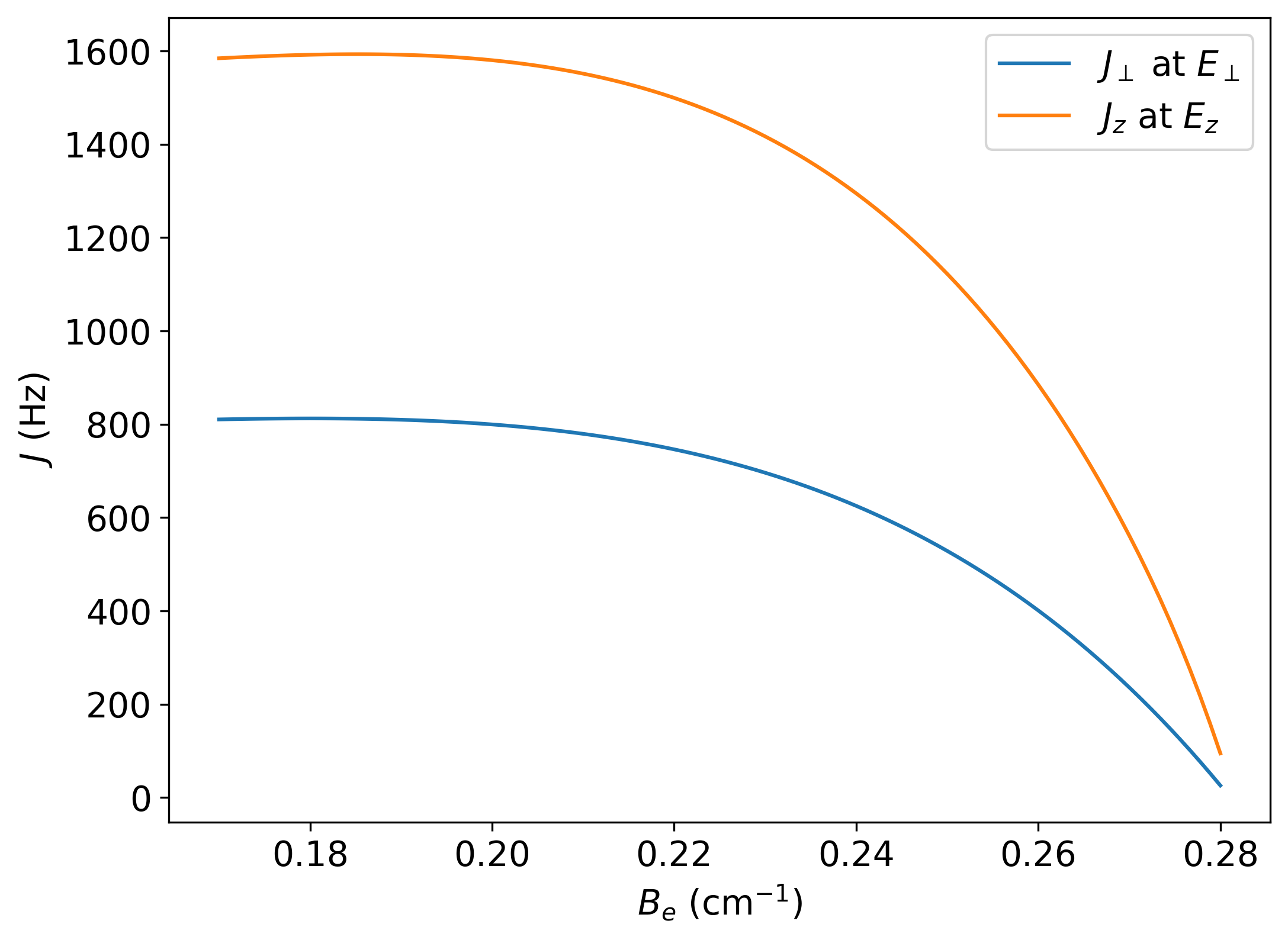}
				\end{tabular}
				\caption{Dependence of the QA field parameters and couplings on the rotational constant of molecules ($B_e$). The magnetic field is fixed at 0.6 mT and other molecular parameters of the system are fixed at values for SrF.}
				\label{fig:ej_b}
			\end{figure}

			Figure \ref{fig:ej_b} shows the variation of $E_\perp$ and $E_z$ (top panel) and $J_\perp$ and $J_z$ at these electric-field magnitudes (bottom panel) with the rotational constant ($B_e$) of the molecule. $E_\perp$ and $E_z$ are larger for molecules with a smaller rotational constant. This can be explained by observing that the avoided crossing in Fig. \ref{fig:energy_levels_e} occurs with the state from the $N=0$ manifold approaching the state from the $N=1$ manifold from above. As the rotational constant is increased, the two states become closer in energy and the electric field required to observe the crossing is lower. The relevant couplings at these field magnitudes are nearly constant at smaller $B_e$ values with a sharp decrease towards the larger $B_e$ values. This happens because smaller electric fields needed at larger rotational constants result in smaller dipole matrix elements, while larger electric fields needed for smaller rotational constants result in larger dipole matrix elements. At small rotational constants and high electric fields, the dipole matrix elements are saturated. The couplings are then derived from the dipole matrix elements according to Eq. (\ref{eq:dipole_int}).
			
\begin{figure}
				\includegraphics[width=\columnwidth]{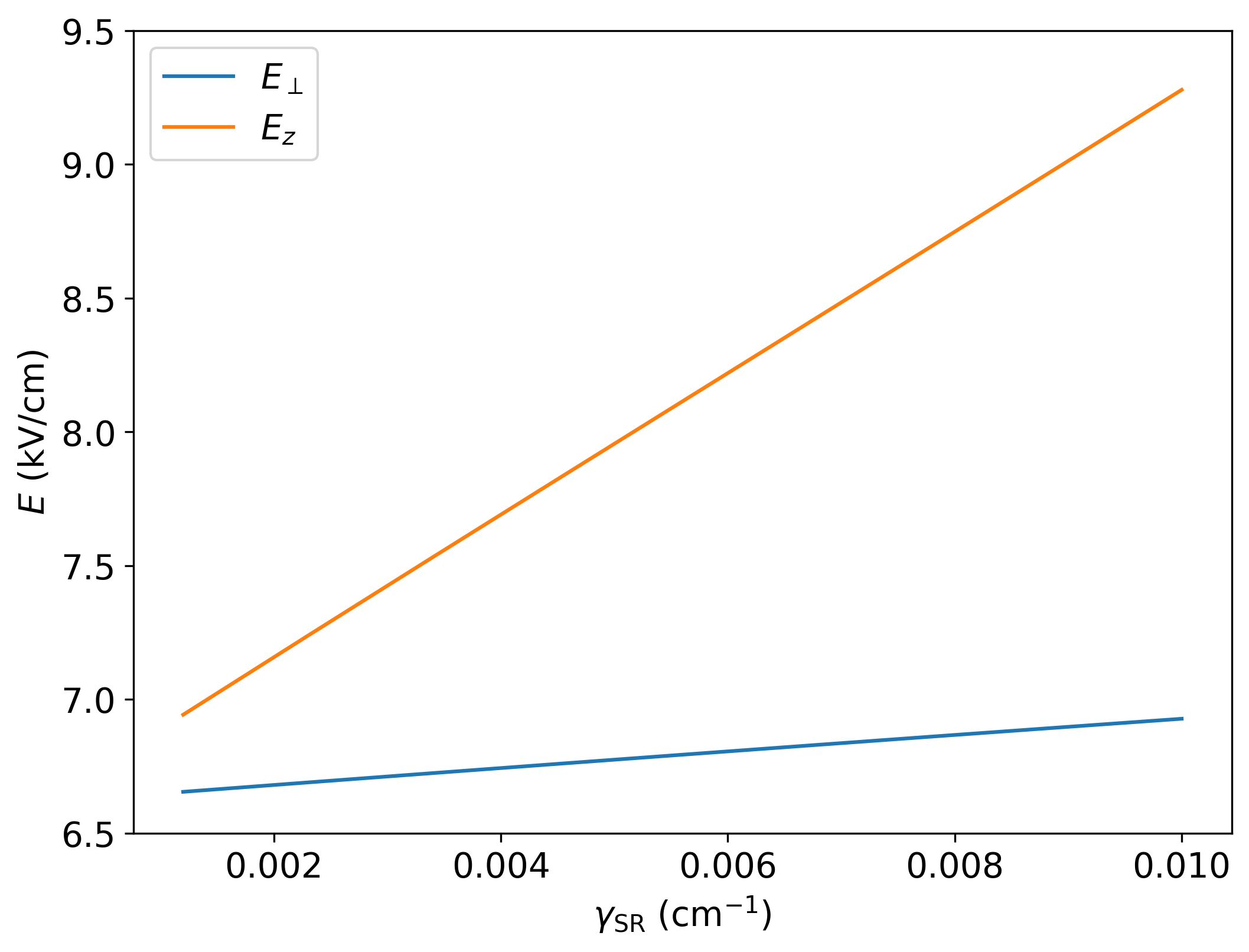} 
				\caption{Dependence of the QA field parameters on the spin-rotational constant of molecules ($\gamma_{\rm SR}$). The magnetic field is fixed at 0.6 mT and other molecular parameters of the system are fixed at SrF values.}
				\label{fig:e_gamma}
			\end{figure}

\begin{figure}
				\begin{tabular}{c}
					\includegraphics[width=\columnwidth]{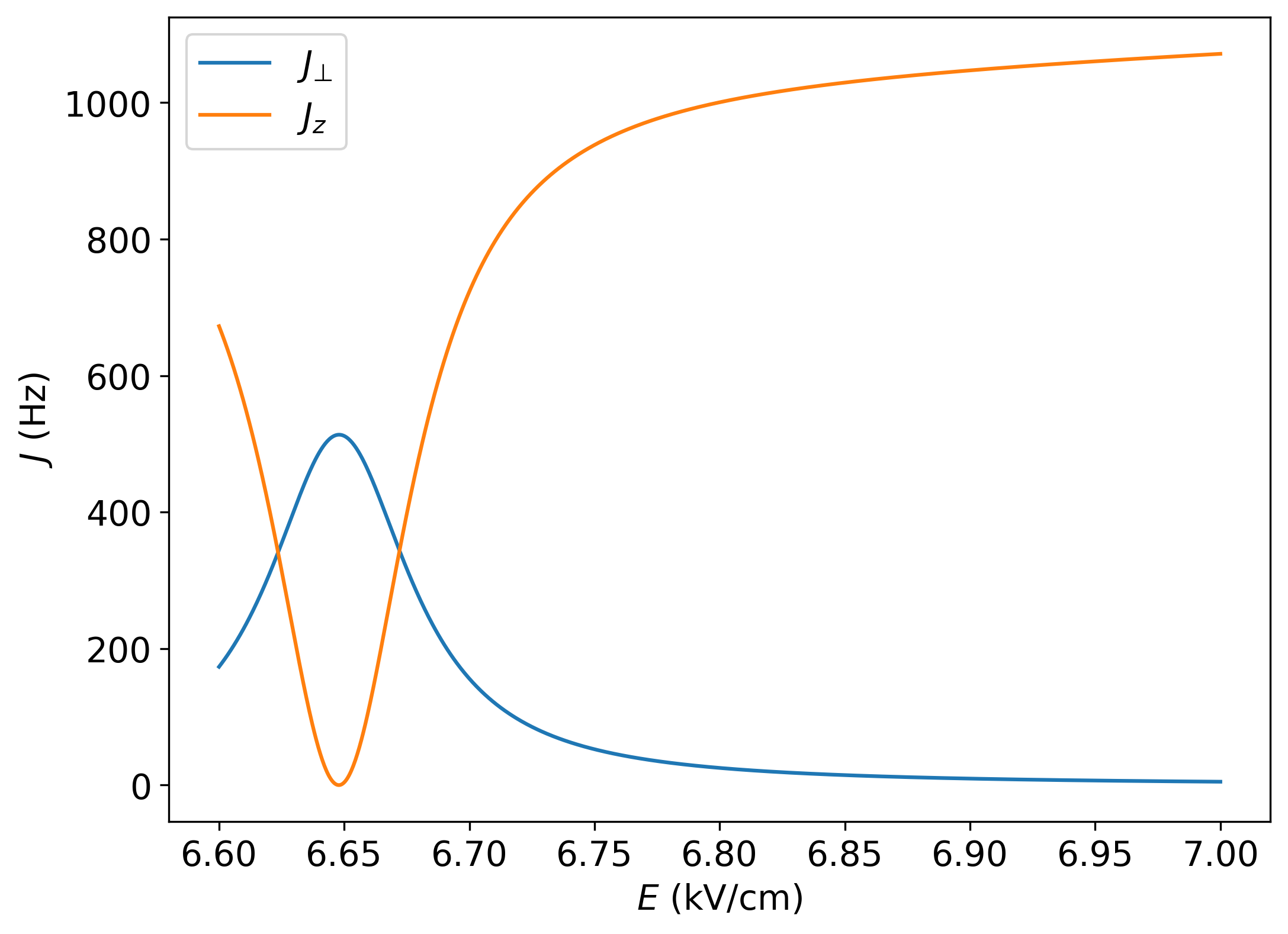} \\
					\includegraphics[width=\columnwidth]{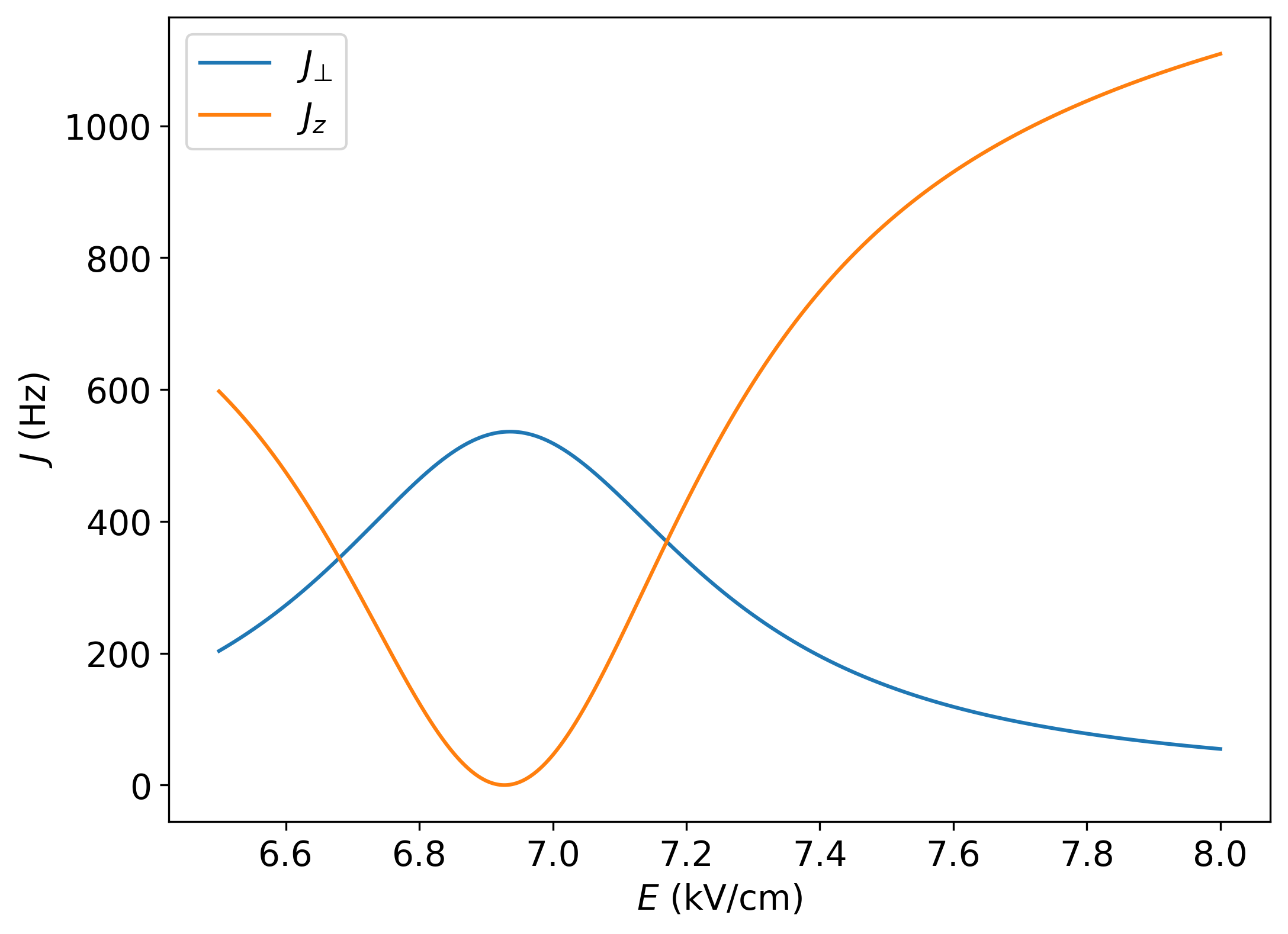}
				\end{tabular}
				\caption{Coupling constants $J_\perp$ and $J_z$ in Eq. \eqref{eq:manybody_ham} for two molecules with spin-rotational constants of 0.0001 cm\textsuperscript{-1} (top) and 0.01 cm\textsuperscript{-1} (bottom), separated by 500 nm with the intermolecular axis perpendicular to $\bm E$ and $\bm B$ with $B=600$ mT, as functions of the electric field magnitude.  }
				\label{fig:j_gamma}
			\end{figure}

			The spin-rotational constant ($\gamma_{\rm SR}$) does not significantly affect the coupling or the electric fields at $E_\perp$ and $E_z$. Figure \ref{fig:e_gamma} shows the dependence of $E_\perp$ and $E_z$ on $\gamma_{\rm SR}$ with negligible change in $E_\perp$, but significant increase in $E_z$ as $\gamma_{\rm SR}$ is increased. This can be explained by observing the couplings at and around the avoided crossing for two different $\gamma_{\rm SR}$ values. Figure \ref{fig:j_gamma} shows the dependence of couplings $J_\perp$ and $J_z$ on the electric field for a molecule with spin rotational constant of 0.0001 cm\textsuperscript{-1} (top) and 0.01 cm\textsuperscript{-1} (bottom). The spin-rotational constant is proportional to the matrix elements of the molecular Hamiltonian \eqref{eq:iso_ham} which couple the two states giving rise to the avoided crossing. At lower $\gamma_{\rm SR}$ values, the crossing is sharper. This explains the significant increase in $E_z$ for molecules with a larger spin-rotational constant, as the range of electric fields around the avoided crossing where $J_z$ is not $\gg J_\perp$ is much larger. 
			
		\clearpage
		\nocite{*}
		\bibliography{manuscript}

\begin{thebibliography}{87}%
\makeatletter
\providecommand \@ifxundefined [1]{%
 \@ifx{#1\undefined}
}%
\providecommand \@ifnum [1]{%
 \ifnum #1\expandafter \@firstoftwo
 \else \expandafter \@secondoftwo
 \fi
}%
\providecommand \@ifx [1]{%
 \ifx #1\expandafter \@firstoftwo
 \else \expandafter \@secondoftwo
 \fi
}%
\providecommand \natexlab [1]{#1}%
\providecommand \enquote  [1]{``#1''}%
\providecommand \bibnamefont  [1]{#1}%
\providecommand \bibfnamefont [1]{#1}%
\providecommand \citenamefont [1]{#1}%
\providecommand \href@noop [0]{\@secondoftwo}%
\providecommand \href [0]{\begingroup \@sanitize@url \@href}%
\providecommand \@href[1]{\@@startlink{#1}\@@href}%
\providecommand \@@href[1]{\endgroup#1\@@endlink}%
\providecommand \@sanitize@url [0]{\catcode `\\12\catcode `\$12\catcode
  `\&12\catcode `\#12\catcode `\^12\catcode `\_12\catcode `\%12\relax}%
\providecommand \@@startlink[1]{}%
\providecommand \@@endlink[0]{}%
\providecommand \url  [0]{\begingroup\@sanitize@url \@url }%
\providecommand \@url [1]{\endgroup\@href {#1}{\urlprefix }}%
\providecommand \urlprefix  [0]{URL }%
\providecommand \Eprint [0]{\href }%
\providecommand \doibase [0]{https://doi.org/}%
\providecommand \selectlanguage [0]{\@gobble}%
\providecommand \bibinfo  [0]{\@secondoftwo}%
\providecommand \bibfield  [0]{\@secondoftwo}%
\providecommand \translation [1]{[#1]}%
\providecommand \BibitemOpen [0]{}%
\providecommand \bibitemStop [0]{}%
\providecommand \bibitemNoStop [0]{.\EOS\space}%
\providecommand \EOS [0]{\spacefactor3000\relax}%
\providecommand \BibitemShut  [1]{\csname bibitem#1\endcsname}%
\let\auto@bib@innerbib\@empty
%</preamble>
\bibitem [{\citenamefont {Kadowaki}\ and\ \citenamefont
  {Nishimori}(1998)}]{kadowaki}%
  \BibitemOpen
  \bibfield  {author} {\bibinfo {author} {\bibfnamefont {T.}~\bibnamefont
  {Kadowaki}}\ and\ \bibinfo {author} {\bibfnamefont {H.}~\bibnamefont
  {Nishimori}},\ }\bibfield  {title} {\bibinfo {title} {Quantum annealing in
  the transverse ising model},\ }\href
  {https://doi.org/10.1103/PhysRevE.58.5355} {\bibfield  {journal} {\bibinfo
  {journal} {Phys. Rev. E}\ }\textbf {\bibinfo {volume} {58}},\ \bibinfo
  {pages} {5355} (\bibinfo {year} {1998})}\BibitemShut {NoStop}%
\bibitem [{\citenamefont {Harvardla}\ \emph {et~al.}(1994)\citenamefont
  {Harvardla}, \citenamefont {Gomez}, \citenamefont {Sebenik}, \citenamefont
  {Stenson},\ and\ \citenamefont {Doll}}]{qa}%
  \BibitemOpen
  \bibfield  {author} {\bibinfo {author} {\bibfnamefont {A.}~\bibnamefont
  {Harvardla}}, \bibinfo {author} {\bibfnamefont {M.}~\bibnamefont {Gomez}},
  \bibinfo {author} {\bibfnamefont {C.}~\bibnamefont {Sebenik}}, \bibinfo
  {author} {\bibfnamefont {C.}~\bibnamefont {Stenson}},\ and\ \bibinfo {author}
  {\bibfnamefont {J.}~\bibnamefont {Doll}},\ }\bibfield  {title} {\bibinfo
  {title} {Quantum annealing: A new method for minimizing multidimensional
  functions},\ }\href
  {https://doi.org/https://doi.org/10.1016/0009-2614(94)00117-0} {\bibfield
  {journal} {\bibinfo  {journal} {Chem. Phys. Lett.}\ }\textbf {\bibinfo
  {volume} {219}},\ \bibinfo {pages} {343} (\bibinfo {year}
  {1994})}\BibitemShut {NoStop}%
\bibitem [{\citenamefont {Hogg}(2000)}]{search}%
  \BibitemOpen
  \bibfield  {author} {\bibinfo {author} {\bibfnamefont {T.}~\bibnamefont
  {Hogg}},\ }\bibfield  {title} {\bibinfo {title} {Quantum search heuristics},\
  }\href {https://doi.org/10.1103/PhysRevA.61.052311} {\bibfield  {journal}
  {\bibinfo  {journal} {Phys. Rev. A}\ }\textbf {\bibinfo {volume} {61}},\
  \bibinfo {pages} {052311} (\bibinfo {year} {2000})}\BibitemShut {NoStop}%
\bibitem [{\citenamefont {Farhi}\ \emph
  {et~al.}(2001{\natexlab{a}})\citenamefont {Farhi}, \citenamefont {Goldstone},
  \citenamefont {Gutmann}, \citenamefont {Lapan}, \citenamefont {Lundgren},\
  and\ \citenamefont {Preda}}]{qa1}%
  \BibitemOpen
  \bibfield  {author} {\bibinfo {author} {\bibfnamefont {E.}~\bibnamefont
  {Farhi}}, \bibinfo {author} {\bibfnamefont {J.}~\bibnamefont {Goldstone}},
  \bibinfo {author} {\bibfnamefont {S.}~\bibnamefont {Gutmann}}, \bibinfo
  {author} {\bibfnamefont {J.}~\bibnamefont {Lapan}}, \bibinfo {author}
  {\bibfnamefont {A.}~\bibnamefont {Lundgren}},\ and\ \bibinfo {author}
  {\bibfnamefont {D.}~\bibnamefont {Preda}},\ }\bibfield  {title} {\bibinfo
  {title} {A quantum adiabatic evolution algorithm applied to random instances
  of an np-complete problem},\ }\href@noop {} {\bibfield  {journal} {\bibinfo
  {journal} {Science}\ }\textbf {\bibinfo {volume} {292}},\ \bibinfo {pages}
  {472} (\bibinfo {year} {2001}{\natexlab{a}})}\BibitemShut {NoStop}%
\bibitem [{\citenamefont {Barahona}(1982)}]{spin-glass}%
  \BibitemOpen
  \bibfield  {author} {\bibinfo {author} {\bibfnamefont {F.}~\bibnamefont
  {Barahona}},\ }\bibfield  {title} {\bibinfo {title} {On the computational
  complexity of ising spin glass models},\ }\href
  {https://doi.org/10.1088/0305-4470/15/10/028} {\bibfield  {journal} {\bibinfo
   {journal} {J. Phys. A: Math. Gen.}\ }\textbf {\bibinfo {volume} {15}},\
  \bibinfo {pages} {3241} (\bibinfo {year} {1982})}\BibitemShut {NoStop}%
\bibitem [{\citenamefont {las Cuevas}\ and\ \citenamefont
  {Cubitt}(2016)}]{universal}%
  \BibitemOpen
  \bibfield  {author} {\bibinfo {author} {\bibfnamefont {G.~D.}\ \bibnamefont
  {las Cuevas}}\ and\ \bibinfo {author} {\bibfnamefont {T.~S.}\ \bibnamefont
  {Cubitt}},\ }\bibfield  {title} {\bibinfo {title} {Simple universal models
  capture all classical spin physics},\ }\href
  {https://doi.org/10.1126/science.aab3326} {\bibfield  {journal} {\bibinfo
  {journal} {Science}\ }\textbf {\bibinfo {volume} {351}},\ \bibinfo {pages}
  {1180} (\bibinfo {year} {2016})}\BibitemShut {NoStop}%
\bibitem [{\citenamefont {Neven}\ \emph {et~al.}(2009)\citenamefont {Neven},
  \citenamefont {Denchev}, \citenamefont {Rose},\ and\ \citenamefont
  {Macready}}]{qa2}%
  \BibitemOpen
  \bibfield  {author} {\bibinfo {author} {\bibfnamefont {H.}~\bibnamefont
  {Neven}}, \bibinfo {author} {\bibfnamefont {V.~S.}\ \bibnamefont {Denchev}},
  \bibinfo {author} {\bibfnamefont {G.}~\bibnamefont {Rose}},\ and\ \bibinfo
  {author} {\bibfnamefont {W.~G.}\ \bibnamefont {Macready}},\ }\bibfield
  {title} {\bibinfo {title} {Training a large scale classifier with the quantum
  adiabatic algorithm},\ }\href@noop {} {\bibfield  {journal} {\bibinfo
  {journal} {arXiv preprint arXiv:0912.0779}\ } (\bibinfo {year}
  {2009})}\BibitemShut {NoStop}%
\bibitem [{\citenamefont {Babbush}\ \emph {et~al.}(2014)\citenamefont
  {Babbush}, \citenamefont {Love},\ and\ \citenamefont {Aspuru-Guzik}}]{qa3}%
  \BibitemOpen
  \bibfield  {author} {\bibinfo {author} {\bibfnamefont {R.}~\bibnamefont
  {Babbush}}, \bibinfo {author} {\bibfnamefont {P.~J.}\ \bibnamefont {Love}},\
  and\ \bibinfo {author} {\bibfnamefont {A.}~\bibnamefont {Aspuru-Guzik}},\
  }\bibfield  {title} {\bibinfo {title} {Adiabatic quantum simulation of
  quantum chemistry},\ }\href@noop {} {\bibfield  {journal} {\bibinfo
  {journal} {Sci. Rep.}\ }\textbf {\bibinfo {volume} {4}},\ \bibinfo {pages}
  {6603} (\bibinfo {year} {2014})}\BibitemShut {NoStop}%
\bibitem [{\citenamefont {Hernandez}\ and\ \citenamefont {Aramon}(2017)}]{qa4}%
  \BibitemOpen
  \bibfield  {author} {\bibinfo {author} {\bibfnamefont {M.}~\bibnamefont
  {Hernandez}}\ and\ \bibinfo {author} {\bibfnamefont {M.}~\bibnamefont
  {Aramon}},\ }\bibfield  {title} {\bibinfo {title} {Enhancing quantum
  annealing performance for the molecular similarity problem},\ }\href@noop {}
  {\bibfield  {journal} {\bibinfo  {journal} {Quantum Inf. Process.}\ }\textbf
  {\bibinfo {volume} {16}},\ \bibinfo {pages} {133} (\bibinfo {year}
  {2017})}\BibitemShut {NoStop}%
\bibitem [{\citenamefont {Lloyd}\ \emph {et~al.}(2013)\citenamefont {Lloyd},
  \citenamefont {Mohseni},\ and\ \citenamefont {Rebentrost}}]{qa5}%
  \BibitemOpen
  \bibfield  {author} {\bibinfo {author} {\bibfnamefont {S.}~\bibnamefont
  {Lloyd}}, \bibinfo {author} {\bibfnamefont {M.}~\bibnamefont {Mohseni}},\
  and\ \bibinfo {author} {\bibfnamefont {P.}~\bibnamefont {Rebentrost}},\
  }\bibfield  {title} {\bibinfo {title} {Quantum algorithms for supervised and
  unsupervised machine learning},\ }\href@noop {} {\bibfield  {journal}
  {\bibinfo  {journal} {arXiv preprint arXiv:1307.0411}\ } (\bibinfo {year}
  {2013})}\BibitemShut {NoStop}%
\bibitem [{\citenamefont {Benedetti}\ \emph {et~al.}(2017)\citenamefont
  {Benedetti}, \citenamefont {Realpe-G{\'o}mez}, \citenamefont {Biswas},\ and\
  \citenamefont {Perdomo-Ortiz}}]{qa6}%
  \BibitemOpen
  \bibfield  {author} {\bibinfo {author} {\bibfnamefont {M.}~\bibnamefont
  {Benedetti}}, \bibinfo {author} {\bibfnamefont {J.}~\bibnamefont
  {Realpe-G{\'o}mez}}, \bibinfo {author} {\bibfnamefont {R.}~\bibnamefont
  {Biswas}},\ and\ \bibinfo {author} {\bibfnamefont {A.}~\bibnamefont
  {Perdomo-Ortiz}},\ }\bibfield  {title} {\bibinfo {title} {Quantum-assisted
  learning of hardware-embedded probabilistic graphical models},\ }\href@noop
  {} {\bibfield  {journal} {\bibinfo  {journal} {Phys. Rev. X}\ }\textbf
  {\bibinfo {volume} {7}},\ \bibinfo {pages} {041052} (\bibinfo {year}
  {2017})}\BibitemShut {NoStop}%
\bibitem [{\citenamefont {Li}\ \emph {et~al.}(2018)\citenamefont {Li},
  \citenamefont {Di~Felice}, \citenamefont {Rohs},\ and\ \citenamefont
  {Lidar}}]{qa7}%
  \BibitemOpen
  \bibfield  {author} {\bibinfo {author} {\bibfnamefont {R.~Y.}\ \bibnamefont
  {Li}}, \bibinfo {author} {\bibfnamefont {R.}~\bibnamefont {Di~Felice}},
  \bibinfo {author} {\bibfnamefont {R.}~\bibnamefont {Rohs}},\ and\ \bibinfo
  {author} {\bibfnamefont {D.~A.}\ \bibnamefont {Lidar}},\ }\bibfield  {title}
  {\bibinfo {title} {Quantum annealing versus classical machine learning
  applied to a simplified computational biology problem},\ }\href@noop {}
  {\bibfield  {journal} {\bibinfo  {journal} {NPJ Quantum Inf.}\ }\textbf
  {\bibinfo {volume} {4}},\ \bibinfo {pages} {1} (\bibinfo {year}
  {2018})}\BibitemShut {NoStop}%
\bibitem [{\citenamefont {Garnerone}\ \emph {et~al.}(2012)\citenamefont
  {Garnerone}, \citenamefont {Zanardi},\ and\ \citenamefont {Lidar}}]{qa8}%
  \BibitemOpen
  \bibfield  {author} {\bibinfo {author} {\bibfnamefont {S.}~\bibnamefont
  {Garnerone}}, \bibinfo {author} {\bibfnamefont {P.}~\bibnamefont {Zanardi}},\
  and\ \bibinfo {author} {\bibfnamefont {D.~A.}\ \bibnamefont {Lidar}},\
  }\bibfield  {title} {\bibinfo {title} {Adiabatic quantum algorithm for search
  engine ranking},\ }\href@noop {} {\bibfield  {journal} {\bibinfo  {journal}
  {Phys. Rev, Lett.}\ }\textbf {\bibinfo {volume} {108}},\ \bibinfo {pages}
  {230506} (\bibinfo {year} {2012})}\BibitemShut {NoStop}%
\bibitem [{\citenamefont {Perdomo-Ortiz}\ \emph {et~al.}(2012)\citenamefont
  {Perdomo-Ortiz}, \citenamefont {Dickson}, \citenamefont {Drew-Brook},
  \citenamefont {Rose},\ and\ \citenamefont {Aspuru-Guzik}}]{qa9}%
  \BibitemOpen
  \bibfield  {author} {\bibinfo {author} {\bibfnamefont {A.}~\bibnamefont
  {Perdomo-Ortiz}}, \bibinfo {author} {\bibfnamefont {N.}~\bibnamefont
  {Dickson}}, \bibinfo {author} {\bibfnamefont {M.}~\bibnamefont {Drew-Brook}},
  \bibinfo {author} {\bibfnamefont {G.}~\bibnamefont {Rose}},\ and\ \bibinfo
  {author} {\bibfnamefont {A.}~\bibnamefont {Aspuru-Guzik}},\ }\bibfield
  {title} {\bibinfo {title} {Finding low-energy conformations of lattice
  protein models by quantum annealing},\ }\href@noop {} {\bibfield  {journal}
  {\bibinfo  {journal} {Sci. Rep.}\ }\textbf {\bibinfo {volume} {2}},\ \bibinfo
  {pages} {1} (\bibinfo {year} {2012})}\BibitemShut {NoStop}%
\bibitem [{\citenamefont {Rosenberg}\ \emph {et~al.}(2016)\citenamefont
  {Rosenberg}, \citenamefont {Haghnegahdar}, \citenamefont {Goddard},
  \citenamefont {Carr}, \citenamefont {Wu},\ and\ \citenamefont
  {De~Prado}}]{qa10}%
  \BibitemOpen
  \bibfield  {author} {\bibinfo {author} {\bibfnamefont {G.}~\bibnamefont
  {Rosenberg}}, \bibinfo {author} {\bibfnamefont {P.}~\bibnamefont
  {Haghnegahdar}}, \bibinfo {author} {\bibfnamefont {P.}~\bibnamefont
  {Goddard}}, \bibinfo {author} {\bibfnamefont {P.}~\bibnamefont {Carr}},
  \bibinfo {author} {\bibfnamefont {K.}~\bibnamefont {Wu}},\ and\ \bibinfo
  {author} {\bibfnamefont {M.~L.}\ \bibnamefont {De~Prado}},\ }\bibfield
  {title} {\bibinfo {title} {Solving the optimal trading trajectory problem
  using a quantum annealer},\ }\href@noop {} {\bibfield  {journal} {\bibinfo
  {journal} {IEEE Journal of Selected Topics in Signal Processing}\ }\textbf
  {\bibinfo {volume} {10}},\ \bibinfo {pages} {1053} (\bibinfo {year}
  {2016})}\BibitemShut {NoStop}%
\bibitem [{\citenamefont {Perdomo-Ortiz}\ \emph {et~al.}(2015)\citenamefont
  {Perdomo-Ortiz}, \citenamefont {Fluegemann}, \citenamefont {Narasimhan},
  \citenamefont {Biswas},\ and\ \citenamefont {Smelyanskiy}}]{qa11}%
  \BibitemOpen
  \bibfield  {author} {\bibinfo {author} {\bibfnamefont {A.}~\bibnamefont
  {Perdomo-Ortiz}}, \bibinfo {author} {\bibfnamefont {J.}~\bibnamefont
  {Fluegemann}}, \bibinfo {author} {\bibfnamefont {S.}~\bibnamefont
  {Narasimhan}}, \bibinfo {author} {\bibfnamefont {R.}~\bibnamefont {Biswas}},\
  and\ \bibinfo {author} {\bibfnamefont {V.~N.}\ \bibnamefont {Smelyanskiy}},\
  }\bibfield  {title} {\bibinfo {title} {A quantum annealing approach for fault
  detection and diagnosis of graph-based systems},\ }\href@noop {} {\bibfield
  {journal} {\bibinfo  {journal} {Eur. Phys. J.: Spec. Top.}\ }\textbf
  {\bibinfo {volume} {224}},\ \bibinfo {pages} {131} (\bibinfo {year}
  {2015})}\BibitemShut {NoStop}%
\bibitem [{\citenamefont {Perdomo-Ortiz}\ \emph {et~al.}(2019)\citenamefont
  {Perdomo-Ortiz}, \citenamefont {Feldman}, \citenamefont {Ozaeta},
  \citenamefont {Isakov}, \citenamefont {Zhu}, \citenamefont {O'Gorman},
  \citenamefont {Katzgraber}, \citenamefont {Diedrich}, \citenamefont {Neven},
  \citenamefont {de~Kleer}, \citenamefont {Lackey},\ and\ \citenamefont
  {Biswas}}]{qa12}%
  \BibitemOpen
  \bibfield  {author} {\bibinfo {author} {\bibfnamefont {A.}~\bibnamefont
  {Perdomo-Ortiz}}, \bibinfo {author} {\bibfnamefont {A.}~\bibnamefont
  {Feldman}}, \bibinfo {author} {\bibfnamefont {A.}~\bibnamefont {Ozaeta}},
  \bibinfo {author} {\bibfnamefont {S.~V.}\ \bibnamefont {Isakov}}, \bibinfo
  {author} {\bibfnamefont {Z.}~\bibnamefont {Zhu}}, \bibinfo {author}
  {\bibfnamefont {B.}~\bibnamefont {O'Gorman}}, \bibinfo {author}
  {\bibfnamefont {H.~G.}\ \bibnamefont {Katzgraber}}, \bibinfo {author}
  {\bibfnamefont {A.}~\bibnamefont {Diedrich}}, \bibinfo {author}
  {\bibfnamefont {H.}~\bibnamefont {Neven}}, \bibinfo {author} {\bibfnamefont
  {J.}~\bibnamefont {de~Kleer}}, \bibinfo {author} {\bibfnamefont
  {B.}~\bibnamefont {Lackey}},\ and\ \bibinfo {author} {\bibfnamefont
  {R.}~\bibnamefont {Biswas}},\ }\bibfield  {title} {\bibinfo {title}
  {Readiness of quantum optimization machines for industrial applications},\
  }\href@noop {} {\bibfield  {journal} {\bibinfo  {journal} {Phys. Rev. Appl.}\
  }\textbf {\bibinfo {volume} {12}},\ \bibinfo {pages} {014004} (\bibinfo
  {year} {2019})}\BibitemShut {NoStop}%
\bibitem [{\citenamefont {Rieffel}\ \emph {et~al.}(2015)\citenamefont
  {Rieffel}, \citenamefont {Venturelli}, \citenamefont {O'Gorman},
  \citenamefont {Do}, \citenamefont {Prystay},\ and\ \citenamefont
  {Smelyanskiy}}]{qa13}%
  \BibitemOpen
  \bibfield  {author} {\bibinfo {author} {\bibfnamefont {E.~G.}\ \bibnamefont
  {Rieffel}}, \bibinfo {author} {\bibfnamefont {D.}~\bibnamefont {Venturelli}},
  \bibinfo {author} {\bibfnamefont {B.}~\bibnamefont {O'Gorman}}, \bibinfo
  {author} {\bibfnamefont {M.~B.}\ \bibnamefont {Do}}, \bibinfo {author}
  {\bibfnamefont {E.~M.}\ \bibnamefont {Prystay}},\ and\ \bibinfo {author}
  {\bibfnamefont {V.~N.}\ \bibnamefont {Smelyanskiy}},\ }\bibfield  {title}
  {\bibinfo {title} {A case study in programming a quantum annealer for hard
  operational planning problems},\ }\href@noop {} {\bibfield  {journal}
  {\bibinfo  {journal} {Quantum Inf. Process.}\ }\textbf {\bibinfo {volume}
  {14}},\ \bibinfo {pages} {1} (\bibinfo {year} {2015})}\BibitemShut {NoStop}%
\bibitem [{\citenamefont {Venturelli}\ \emph {et~al.}(2015)\citenamefont
  {Venturelli}, \citenamefont {Marchand},\ and\ \citenamefont {Rojo}}]{qa14}%
  \BibitemOpen
  \bibfield  {author} {\bibinfo {author} {\bibfnamefont {D.}~\bibnamefont
  {Venturelli}}, \bibinfo {author} {\bibfnamefont {D.~J.}\ \bibnamefont
  {Marchand}},\ and\ \bibinfo {author} {\bibfnamefont {G.}~\bibnamefont
  {Rojo}},\ }\bibfield  {title} {\bibinfo {title} {Quantum annealing
  implementation of job-shop scheduling},\ }\href@noop {} {\bibfield  {journal}
  {\bibinfo  {journal} {arXiv preprint arXiv:1506.08479}\ } (\bibinfo {year}
  {2015})}\BibitemShut {NoStop}%
\bibitem [{\citenamefont {Khoshaman}\ \emph {et~al.}(2018)\citenamefont
  {Khoshaman}, \citenamefont {Vinci}, \citenamefont {Denis}, \citenamefont
  {Andriyash}, \citenamefont {Sadeghi},\ and\ \citenamefont {Amin}}]{qa15}%
  \BibitemOpen
  \bibfield  {author} {\bibinfo {author} {\bibfnamefont {A.}~\bibnamefont
  {Khoshaman}}, \bibinfo {author} {\bibfnamefont {W.}~\bibnamefont {Vinci}},
  \bibinfo {author} {\bibfnamefont {B.}~\bibnamefont {Denis}}, \bibinfo
  {author} {\bibfnamefont {E.}~\bibnamefont {Andriyash}}, \bibinfo {author}
  {\bibfnamefont {H.}~\bibnamefont {Sadeghi}},\ and\ \bibinfo {author}
  {\bibfnamefont {M.~H.}\ \bibnamefont {Amin}},\ }\bibfield  {title} {\bibinfo
  {title} {Quantum variational autoencoder},\ }\href@noop {} {\bibfield
  {journal} {\bibinfo  {journal} {Quantum Sci. Technol.}\ }\textbf {\bibinfo
  {volume} {4}},\ \bibinfo {pages} {014001} (\bibinfo {year}
  {2018})}\BibitemShut {NoStop}%
\bibitem [{\citenamefont {Wilson}\ \emph {et~al.}(2021)\citenamefont {Wilson},
  \citenamefont {Vandal}, \citenamefont {Hogg},\ and\ \citenamefont
  {Rieffel}}]{qa16}%
  \BibitemOpen
  \bibfield  {author} {\bibinfo {author} {\bibfnamefont {M.}~\bibnamefont
  {Wilson}}, \bibinfo {author} {\bibfnamefont {T.}~\bibnamefont {Vandal}},
  \bibinfo {author} {\bibfnamefont {T.}~\bibnamefont {Hogg}},\ and\ \bibinfo
  {author} {\bibfnamefont {E.~G.}\ \bibnamefont {Rieffel}},\ }\bibfield
  {title} {\bibinfo {title} {Quantum-assisted associative adversarial network:
  Applying quantum annealing in deep learning},\ }\href@noop {} {\bibfield
  {journal} {\bibinfo  {journal} {Quantum Machine Intelligence}\ }\textbf
  {\bibinfo {volume} {3}},\ \bibinfo {pages} {1} (\bibinfo {year}
  {2021})}\BibitemShut {NoStop}%
\bibitem [{\citenamefont {Mott}\ \emph {et~al.}(2017)\citenamefont {Mott},
  \citenamefont {Job}, \citenamefont {Vlimant}, \citenamefont {Lidar},\ and\
  \citenamefont {Spiropulu}}]{qa17}%
  \BibitemOpen
  \bibfield  {author} {\bibinfo {author} {\bibfnamefont {A.}~\bibnamefont
  {Mott}}, \bibinfo {author} {\bibfnamefont {J.}~\bibnamefont {Job}}, \bibinfo
  {author} {\bibfnamefont {J.-R.}\ \bibnamefont {Vlimant}}, \bibinfo {author}
  {\bibfnamefont {D.}~\bibnamefont {Lidar}},\ and\ \bibinfo {author}
  {\bibfnamefont {M.}~\bibnamefont {Spiropulu}},\ }\bibfield  {title} {\bibinfo
  {title} {Solving a higgs optimization problem with quantum annealing for
  machine learning},\ }\href@noop {} {\bibfield  {journal} {\bibinfo  {journal}
  {Nature}\ }\textbf {\bibinfo {volume} {550}},\ \bibinfo {pages} {375}
  (\bibinfo {year} {2017})}\BibitemShut {NoStop}%
\bibitem [{\citenamefont {Das}\ \emph {et~al.}(2019)\citenamefont {Das},
  \citenamefont {Wildridge}, \citenamefont {Vaidya},\ and\ \citenamefont
  {Jung}}]{qa18}%
  \BibitemOpen
  \bibfield  {author} {\bibinfo {author} {\bibfnamefont {S.}~\bibnamefont
  {Das}}, \bibinfo {author} {\bibfnamefont {A.~J.}\ \bibnamefont {Wildridge}},
  \bibinfo {author} {\bibfnamefont {S.~B.}\ \bibnamefont {Vaidya}},\ and\
  \bibinfo {author} {\bibfnamefont {A.}~\bibnamefont {Jung}},\ }\bibfield
  {title} {\bibinfo {title} {Track clustering with a quantum annealer for
  primary vertex reconstruction at hadron colliders},\ }\href@noop {}
  {\bibfield  {journal} {\bibinfo  {journal} {arXiv preprint arXiv:1903.08879}\
  } (\bibinfo {year} {2019})}\BibitemShut {NoStop}%
\bibitem [{\citenamefont {Preskill}(2018)}]{preskill}%
  \BibitemOpen
  \bibfield  {author} {\bibinfo {author} {\bibfnamefont {J.}~\bibnamefont
  {Preskill}},\ }\bibfield  {title} {\bibinfo {title} {Quantum {C}omputing in
  the {NISQ} era and beyond},\ }\href
  {https://doi.org/10.22331/q-2018-08-06-79} {\bibfield  {journal} {\bibinfo
  {journal} {{Quantum}}\ }\textbf {\bibinfo {volume} {2}},\ \bibinfo {pages}
  {79} (\bibinfo {year} {2018})}\BibitemShut {NoStop}%
\bibitem [{\citenamefont {McGeoch}(2020)}]{mcgeoch}%
  \BibitemOpen
  \bibfield  {author} {\bibinfo {author} {\bibfnamefont {C.~C.}\ \bibnamefont
  {McGeoch}},\ }\bibfield  {title} {\bibinfo {title} {Theory versus practice in
  annealing-based quantum computing},\ }\href
  {https://doi.org/https://doi.org/10.1016/j.tcs.2020.01.024} {\bibfield
  {journal} {\bibinfo  {journal} {Theor. Comput. Sci.}\ }\textbf {\bibinfo
  {volume} {816}},\ \bibinfo {pages} {169} (\bibinfo {year}
  {2020})}\BibitemShut {NoStop}%
\bibitem [{\citenamefont {Boothby}\ \emph {et~al.}(2021)\citenamefont
  {Boothby}, \citenamefont {Enderud}, \citenamefont {Lanting}, \citenamefont
  {Molavi}, \citenamefont {Tsai}, \citenamefont {Volkmann}, \citenamefont
  {Altomare}, \citenamefont {Amin}, \citenamefont {Babcock}, \citenamefont
  {Berkley}, \citenamefont {Aznar}, \citenamefont {Boschnak}, \citenamefont
  {Christiani}, \citenamefont {Ejtemaee}, \citenamefont {Evert}, \citenamefont
  {Gullen}, \citenamefont {Hager}, \citenamefont {Harris}, \citenamefont
  {Hoskinson}, \citenamefont {Hilton}, \citenamefont {Jooya}, \citenamefont
  {Huang}, \citenamefont {Johnson}, \citenamefont {King}, \citenamefont
  {Ladizinsky}, \citenamefont {Li}, \citenamefont {MacDonald}, \citenamefont
  {Fernandez}, \citenamefont {Neufeld}, \citenamefont {Norouzpour},
  \citenamefont {Oh}, \citenamefont {Ozfidan}, \citenamefont {Paddon},
  \citenamefont {Perminov}, \citenamefont {Poulin-Lamarre}, \citenamefont
  {Prescott}, \citenamefont {Raymond}, \citenamefont {Reis}, \citenamefont
  {Rich}, \citenamefont {Roy}, \citenamefont {Esfahani}, \citenamefont {Sato},
  \citenamefont {Sheldan}, \citenamefont {Smirnov}, \citenamefont {Swenson},
  \citenamefont {Whittaker}, \citenamefont {Yao}, \citenamefont {Yarovoy},\
  and\ \citenamefont {Bunyk}}]{advantage}%
  \BibitemOpen
  \bibfield  {author} {\bibinfo {author} {\bibfnamefont {K.}~\bibnamefont
  {Boothby}}, \bibinfo {author} {\bibfnamefont {C.}~\bibnamefont {Enderud}},
  \bibinfo {author} {\bibfnamefont {T.}~\bibnamefont {Lanting}}, \bibinfo
  {author} {\bibfnamefont {R.}~\bibnamefont {Molavi}}, \bibinfo {author}
  {\bibfnamefont {N.}~\bibnamefont {Tsai}}, \bibinfo {author} {\bibfnamefont
  {M.~H.}\ \bibnamefont {Volkmann}}, \bibinfo {author} {\bibfnamefont
  {F.}~\bibnamefont {Altomare}}, \bibinfo {author} {\bibfnamefont {M.~H.}\
  \bibnamefont {Amin}}, \bibinfo {author} {\bibfnamefont {M.}~\bibnamefont
  {Babcock}}, \bibinfo {author} {\bibfnamefont {A.~J.}\ \bibnamefont
  {Berkley}}, \bibinfo {author} {\bibfnamefont {C.~B.}\ \bibnamefont {Aznar}},
  \bibinfo {author} {\bibfnamefont {M.}~\bibnamefont {Boschnak}}, \bibinfo
  {author} {\bibfnamefont {H.}~\bibnamefont {Christiani}}, \bibinfo {author}
  {\bibfnamefont {S.}~\bibnamefont {Ejtemaee}}, \bibinfo {author}
  {\bibfnamefont {B.}~\bibnamefont {Evert}}, \bibinfo {author} {\bibfnamefont
  {M.}~\bibnamefont {Gullen}}, \bibinfo {author} {\bibfnamefont
  {M.}~\bibnamefont {Hager}}, \bibinfo {author} {\bibfnamefont
  {R.}~\bibnamefont {Harris}}, \bibinfo {author} {\bibfnamefont
  {E.}~\bibnamefont {Hoskinson}}, \bibinfo {author} {\bibfnamefont {J.~P.}\
  \bibnamefont {Hilton}}, \bibinfo {author} {\bibfnamefont {K.}~\bibnamefont
  {Jooya}}, \bibinfo {author} {\bibfnamefont {A.}~\bibnamefont {Huang}},
  \bibinfo {author} {\bibfnamefont {M.~W.}\ \bibnamefont {Johnson}}, \bibinfo
  {author} {\bibfnamefont {A.~D.}\ \bibnamefont {King}}, \bibinfo {author}
  {\bibfnamefont {E.}~\bibnamefont {Ladizinsky}}, \bibinfo {author}
  {\bibfnamefont {R.}~\bibnamefont {Li}}, \bibinfo {author} {\bibfnamefont
  {A.}~\bibnamefont {MacDonald}}, \bibinfo {author} {\bibfnamefont {T.~M.}\
  \bibnamefont {Fernandez}}, \bibinfo {author} {\bibfnamefont {R.}~\bibnamefont
  {Neufeld}}, \bibinfo {author} {\bibfnamefont {M.}~\bibnamefont {Norouzpour}},
  \bibinfo {author} {\bibfnamefont {T.}~\bibnamefont {Oh}}, \bibinfo {author}
  {\bibfnamefont {I.}~\bibnamefont {Ozfidan}}, \bibinfo {author} {\bibfnamefont
  {P.}~\bibnamefont {Paddon}}, \bibinfo {author} {\bibfnamefont
  {I.}~\bibnamefont {Perminov}}, \bibinfo {author} {\bibfnamefont
  {G.}~\bibnamefont {Poulin-Lamarre}}, \bibinfo {author} {\bibfnamefont
  {T.}~\bibnamefont {Prescott}}, \bibinfo {author} {\bibfnamefont
  {J.}~\bibnamefont {Raymond}}, \bibinfo {author} {\bibfnamefont
  {M.}~\bibnamefont {Reis}}, \bibinfo {author} {\bibfnamefont {C.}~\bibnamefont
  {Rich}}, \bibinfo {author} {\bibfnamefont {A.}~\bibnamefont {Roy}}, \bibinfo
  {author} {\bibfnamefont {H.~S.}\ \bibnamefont {Esfahani}}, \bibinfo {author}
  {\bibfnamefont {Y.}~\bibnamefont {Sato}}, \bibinfo {author} {\bibfnamefont
  {B.}~\bibnamefont {Sheldan}}, \bibinfo {author} {\bibfnamefont
  {A.}~\bibnamefont {Smirnov}}, \bibinfo {author} {\bibfnamefont {L.~J.}\
  \bibnamefont {Swenson}}, \bibinfo {author} {\bibfnamefont {J.}~\bibnamefont
  {Whittaker}}, \bibinfo {author} {\bibfnamefont {J.}~\bibnamefont {Yao}},
  \bibinfo {author} {\bibfnamefont {A.}~\bibnamefont {Yarovoy}},\ and\ \bibinfo
  {author} {\bibfnamefont {P.~I.}\ \bibnamefont {Bunyk}},\ }\href@noop {}
  {\bibinfo {title} {Architectural considerations in the design of a
  third-generation superconducting quantum annealing processor}} (\bibinfo
  {year} {2021}),\ \Eprint {https://arxiv.org/abs/2108.02322} {arXiv:2108.02322
  [quant-ph]} \BibitemShut {NoStop}%
\bibitem [{\citenamefont {Micheli}\ \emph {et~al.}(2006)\citenamefont
  {Micheli}, \citenamefont {Brennen},\ and\ \citenamefont {Zoller}}]{toolbox}%
  \BibitemOpen
  \bibfield  {author} {\bibinfo {author} {\bibfnamefont {A.}~\bibnamefont
  {Micheli}}, \bibinfo {author} {\bibfnamefont {G.}~\bibnamefont {Brennen}},\
  and\ \bibinfo {author} {\bibfnamefont {P.}~\bibnamefont {Zoller}},\
  }\bibfield  {title} {\bibinfo {title} {A toolbox for lattice spin models with
  polar molecules},\ }\href {https://doi.org/10.1038/nphys287} {\bibfield
  {journal} {\bibinfo  {journal} {Nat. Phys.}\ }\textbf {\bibinfo {volume}
  {2}},\ \bibinfo {pages} {341} (\bibinfo {year} {2006})}\BibitemShut {NoStop}%
\bibitem [{\citenamefont {Gorshkov}\ \emph
  {et~al.}(2011{\natexlab{a}})\citenamefont {Gorshkov}, \citenamefont
  {Manmana}, \citenamefont {Chen}, \citenamefont {Ye}, \citenamefont {Demler},
  \citenamefont {Lukin},\ and\ \citenamefont {Rey}}]{superfluid}%
  \BibitemOpen
  \bibfield  {author} {\bibinfo {author} {\bibfnamefont {A.~V.}\ \bibnamefont
  {Gorshkov}}, \bibinfo {author} {\bibfnamefont {S.~R.}\ \bibnamefont
  {Manmana}}, \bibinfo {author} {\bibfnamefont {G.}~\bibnamefont {Chen}},
  \bibinfo {author} {\bibfnamefont {J.}~\bibnamefont {Ye}}, \bibinfo {author}
  {\bibfnamefont {E.}~\bibnamefont {Demler}}, \bibinfo {author} {\bibfnamefont
  {M.~D.}\ \bibnamefont {Lukin}},\ and\ \bibinfo {author} {\bibfnamefont
  {A.~M.}\ \bibnamefont {Rey}},\ }\bibfield  {title} {\bibinfo {title} {Tunable
  superfluidity and quantum magnetism with ultracold polar molecules},\ }\href
  {https://doi.org/10.1103/PhysRevLett.107.115301} {\bibfield  {journal}
  {\bibinfo  {journal} {Phys. Rev. Lett.}\ }\textbf {\bibinfo {volume} {107}},\
  \bibinfo {pages} {115301} (\bibinfo {year} {2011}{\natexlab{a}})}\BibitemShut
  {NoStop}%
\bibitem [{\citenamefont {Gorshkov}\ \emph
  {et~al.}(2011{\natexlab{b}})\citenamefont {Gorshkov}, \citenamefont
  {Manmana}, \citenamefont {Chen}, \citenamefont {Demler}, \citenamefont
  {Lukin},\ and\ \citenamefont {Rey}}]{mag}%
  \BibitemOpen
  \bibfield  {author} {\bibinfo {author} {\bibfnamefont {A.~V.}\ \bibnamefont
  {Gorshkov}}, \bibinfo {author} {\bibfnamefont {S.~R.}\ \bibnamefont
  {Manmana}}, \bibinfo {author} {\bibfnamefont {G.}~\bibnamefont {Chen}},
  \bibinfo {author} {\bibfnamefont {E.}~\bibnamefont {Demler}}, \bibinfo
  {author} {\bibfnamefont {M.~D.}\ \bibnamefont {Lukin}},\ and\ \bibinfo
  {author} {\bibfnamefont {A.~M.}\ \bibnamefont {Rey}},\ }\bibfield  {title}
  {\bibinfo {title} {Quantum magnetism with polar alkali-metal dimers},\ }\href
  {https://doi.org/10.1103/PhysRevA.84.033619} {\bibfield  {journal} {\bibinfo
  {journal} {Phys. Rev. A}\ }\textbf {\bibinfo {volume} {84}},\ \bibinfo
  {pages} {033619} (\bibinfo {year} {2011}{\natexlab{b}})}\BibitemShut
  {NoStop}%
\bibitem [{\citenamefont {Sundar}\ \emph {et~al.}(2018)\citenamefont {Sundar},
  \citenamefont {Gadway},\ and\ \citenamefont {Hazzard}}]{synthetic}%
  \BibitemOpen
  \bibfield  {author} {\bibinfo {author} {\bibfnamefont {B.}~\bibnamefont
  {Sundar}}, \bibinfo {author} {\bibfnamefont {B.}~\bibnamefont {Gadway}},\
  and\ \bibinfo {author} {\bibfnamefont {K.}~\bibnamefont {Hazzard}},\
  }\bibfield  {title} {\bibinfo {title} {Synthetic dimensions in ultracold
  polar molecules},\ }\href {https://doi.org/10.1038/s41598-018-21699-x}
  {\bibfield  {journal} {\bibinfo  {journal} {Sci. Rep.}\ }\textbf {\bibinfo
  {volume} {8}},\ \bibinfo {pages} {3422} (\bibinfo {year} {2018})}\BibitemShut
  {NoStop}%
\bibitem [{\citenamefont {Gadway}\ and\ \citenamefont
  {Yan}(2016)}]{dd-interact}%
  \BibitemOpen
  \bibfield  {author} {\bibinfo {author} {\bibfnamefont {B.}~\bibnamefont
  {Gadway}}\ and\ \bibinfo {author} {\bibfnamefont {B.}~\bibnamefont {Yan}},\
  }\bibfield  {title} {\bibinfo {title} {Strongly interacting ultracold polar
  molecules},\ }\href {https://doi.org/10.1088/0953-4075/49/15/152002}
  {\bibfield  {journal} {\bibinfo  {journal} {J. Phys. B: At. Mol. Opt. Phys.}\
  }\textbf {\bibinfo {volume} {49}},\ \bibinfo {pages} {152002} (\bibinfo
  {year} {2016})}\BibitemShut {NoStop}%
\bibitem [{\citenamefont {B{\"u}chler}\ \emph {et~al.}(2007)\citenamefont
  {B{\"u}chler}, \citenamefont {Micheli},\ and\ \citenamefont
  {Zoller}}]{three-body}%
  \BibitemOpen
  \bibfield  {author} {\bibinfo {author} {\bibfnamefont {H.~P.}\ \bibnamefont
  {B{\"u}chler}}, \bibinfo {author} {\bibfnamefont {A.}~\bibnamefont
  {Micheli}},\ and\ \bibinfo {author} {\bibfnamefont {P.}~\bibnamefont
  {Zoller}},\ }\bibfield  {title} {\bibinfo {title} {Three-body interactions
  with cold polar molecules},\ }\href {https://doi.org/10.1038/nphys678}
  {\bibfield  {journal} {\bibinfo  {journal} {Nat. Phys.}\ }\textbf {\bibinfo
  {volume} {3}},\ \bibinfo {pages} {726} (\bibinfo {year} {2007})}\BibitemShut
  {NoStop}%
\bibitem [{\citenamefont {Weimer}(2013)}]{many-sim}%
  \BibitemOpen
  \bibfield  {author} {\bibinfo {author} {\bibfnamefont {H.}~\bibnamefont
  {Weimer}},\ }\bibfield  {title} {\bibinfo {title} {Quantum simulation of
  many-body spin interactions with ultracold polar molecules},\ }\href
  {https://doi.org/10.1080/00268976.2013.789567} {\bibfield  {journal}
  {\bibinfo  {journal} {Mol. Phys.}\ }\textbf {\bibinfo {volume} {111}},\
  \bibinfo {pages} {1753} (\bibinfo {year} {2013})}\BibitemShut {NoStop}%
\bibitem [{\citenamefont {Hazzard}\ \emph
  {et~al.}(2014{\natexlab{a}})\citenamefont {Hazzard}, \citenamefont {van~den
  Worm}, \citenamefont {Foss-Feig}, \citenamefont {Manmana}, \citenamefont
  {Dalla~Torre}, \citenamefont {Pfau}, \citenamefont {Kastner},\ and\
  \citenamefont {Rey}}]{far-equib}%
  \BibitemOpen
  \bibfield  {author} {\bibinfo {author} {\bibfnamefont {K.~R.~A.}\
  \bibnamefont {Hazzard}}, \bibinfo {author} {\bibfnamefont {M.}~\bibnamefont
  {van~den Worm}}, \bibinfo {author} {\bibfnamefont {M.}~\bibnamefont
  {Foss-Feig}}, \bibinfo {author} {\bibfnamefont {S.~R.}\ \bibnamefont
  {Manmana}}, \bibinfo {author} {\bibfnamefont {E.~G.}\ \bibnamefont
  {Dalla~Torre}}, \bibinfo {author} {\bibfnamefont {T.}~\bibnamefont {Pfau}},
  \bibinfo {author} {\bibfnamefont {M.}~\bibnamefont {Kastner}},\ and\ \bibinfo
  {author} {\bibfnamefont {A.~M.}\ \bibnamefont {Rey}},\ }\bibfield  {title}
  {\bibinfo {title} {Quantum correlations and entanglement in
  far-from-equilibrium spin systems},\ }\href
  {https://doi.org/10.1103/PhysRevA.90.063622} {\bibfield  {journal} {\bibinfo
  {journal} {Phys. Rev. A}\ }\textbf {\bibinfo {volume} {90}},\ \bibinfo
  {pages} {063622} (\bibinfo {year} {2014}{\natexlab{a}})}\BibitemShut
  {NoStop}%
\bibitem [{\citenamefont {Hazzard}\ \emph
  {et~al.}(2014{\natexlab{b}})\citenamefont {Hazzard}, \citenamefont {Gadway},
  \citenamefont {Foss-Feig}, \citenamefont {Yan}, \citenamefont {Moses},
  \citenamefont {Covey}, \citenamefont {Yao}, \citenamefont {Lukin},
  \citenamefont {Ye}, \citenamefont {Jin},\ and\ \citenamefont
  {Rey}}]{dipolar-dyn}%
  \BibitemOpen
  \bibfield  {author} {\bibinfo {author} {\bibfnamefont {K.~R.~A.}\
  \bibnamefont {Hazzard}}, \bibinfo {author} {\bibfnamefont {B.}~\bibnamefont
  {Gadway}}, \bibinfo {author} {\bibfnamefont {M.}~\bibnamefont {Foss-Feig}},
  \bibinfo {author} {\bibfnamefont {B.}~\bibnamefont {Yan}}, \bibinfo {author}
  {\bibfnamefont {S.~A.}\ \bibnamefont {Moses}}, \bibinfo {author}
  {\bibfnamefont {J.~P.}\ \bibnamefont {Covey}}, \bibinfo {author}
  {\bibfnamefont {N.~Y.}\ \bibnamefont {Yao}}, \bibinfo {author} {\bibfnamefont
  {M.~D.}\ \bibnamefont {Lukin}}, \bibinfo {author} {\bibfnamefont
  {J.}~\bibnamefont {Ye}}, \bibinfo {author} {\bibfnamefont {D.~S.}\
  \bibnamefont {Jin}},\ and\ \bibinfo {author} {\bibfnamefont {A.~M.}\
  \bibnamefont {Rey}},\ }\bibfield  {title} {\bibinfo {title} {Many-body
  dynamics of dipolar molecules in an optical lattice},\ }\href
  {https://doi.org/10.1103/PhysRevLett.113.195302} {\bibfield  {journal}
  {\bibinfo  {journal} {Phys. Rev. Lett.}\ }\textbf {\bibinfo {volume} {113}},\
  \bibinfo {pages} {195302} (\bibinfo {year} {2014}{\natexlab{b}})}\BibitemShut
  {NoStop}%
\bibitem [{\citenamefont {de~Paz}\ \emph {et~al.}(2013)\citenamefont {de~Paz},
  \citenamefont {Sharma}, \citenamefont {Chotia}, \citenamefont {Mar\'echal},
  \citenamefont {Huckans}, \citenamefont {Pedri}, \citenamefont {Santos},
  \citenamefont {Gorceix}, \citenamefont {Vernac},\ and\ \citenamefont
  {Laburthe-Tolra}}]{mag-lattice}%
  \BibitemOpen
  \bibfield  {author} {\bibinfo {author} {\bibfnamefont {A.}~\bibnamefont
  {de~Paz}}, \bibinfo {author} {\bibfnamefont {A.}~\bibnamefont {Sharma}},
  \bibinfo {author} {\bibfnamefont {A.}~\bibnamefont {Chotia}}, \bibinfo
  {author} {\bibfnamefont {E.}~\bibnamefont {Mar\'echal}}, \bibinfo {author}
  {\bibfnamefont {J.~H.}\ \bibnamefont {Huckans}}, \bibinfo {author}
  {\bibfnamefont {P.}~\bibnamefont {Pedri}}, \bibinfo {author} {\bibfnamefont
  {L.}~\bibnamefont {Santos}}, \bibinfo {author} {\bibfnamefont
  {O.}~\bibnamefont {Gorceix}}, \bibinfo {author} {\bibfnamefont
  {L.}~\bibnamefont {Vernac}},\ and\ \bibinfo {author} {\bibfnamefont
  {B.}~\bibnamefont {Laburthe-Tolra}},\ }\bibfield  {title} {\bibinfo {title}
  {Nonequilibrium quantum magnetism in a dipolar lattice gas},\ }\href
  {https://doi.org/10.1103/PhysRevLett.111.185305} {\bibfield  {journal}
  {\bibinfo  {journal} {Phys. Rev. Lett.}\ }\textbf {\bibinfo {volume} {111}},\
  \bibinfo {pages} {185305} (\bibinfo {year} {2013})}\BibitemShut {NoStop}%
\bibitem [{\citenamefont {Gorshkov}\ \emph {et~al.}(2013)\citenamefont
  {Gorshkov}, \citenamefont {Hazzard},\ and\ \citenamefont {Rey}}]{kitaev}%
  \BibitemOpen
  \bibfield  {author} {\bibinfo {author} {\bibfnamefont {A.~V.}\ \bibnamefont
  {Gorshkov}}, \bibinfo {author} {\bibfnamefont {K.~R.}\ \bibnamefont
  {Hazzard}},\ and\ \bibinfo {author} {\bibfnamefont {A.~M.}\ \bibnamefont
  {Rey}},\ }\bibfield  {title} {\bibinfo {title} {Kitaev honeycomb and other
  exotic spin models with polar molecules},\ }\href
  {https://doi.org/10.1080/00268976.2013.800604} {\bibfield  {journal}
  {\bibinfo  {journal} {Mol. Phys.}\ }\textbf {\bibinfo {volume} {111}},\
  \bibinfo {pages} {1908} (\bibinfo {year} {2013})}\BibitemShut {NoStop}%
\bibitem [{\citenamefont {Covey}\ \emph {et~al.}(2018)\citenamefont {Covey},
  \citenamefont {Moses}, \citenamefont {Ye},\ and\ \citenamefont
  {Jin}}]{control}%
  \BibitemOpen
  \bibfield  {author} {\bibinfo {author} {\bibfnamefont {J.~P.}\ \bibnamefont
  {Covey}}, \bibinfo {author} {\bibfnamefont {S.~A.}\ \bibnamefont {Moses}},
  \bibinfo {author} {\bibfnamefont {J.}~\bibnamefont {Ye}},\ and\ \bibinfo
  {author} {\bibfnamefont {D.~S.}\ \bibnamefont {Jin}},\ }\bibfield  {title}
  {\bibinfo {title} {Chapter 11 controlling a quantum gas of polar molecules in
  an optical lattice},\ }in\ \href
  {https://doi.org/10.1039/9781782626800-00537} {\emph {\bibinfo {booktitle}
  {Cold Chemistry: Molecular Scattering and Reactivity Near Absolute Zero}}}\
  (\bibinfo  {publisher} {The Royal Society of Chemistry},\ \bibinfo {address}
  {Cambridge, UK},\ \bibinfo {year} {2018})\ pp.\ \bibinfo {pages}
  {537--578}\BibitemShut {NoStop}%
\bibitem [{\citenamefont {Blackmore}\ \emph {et~al.}(2018)\citenamefont
  {Blackmore}, \citenamefont {Caldwell}, \citenamefont {Gregory}, \citenamefont
  {Bridge}, \citenamefont {Sawant}, \citenamefont {Aldegunde}, \citenamefont
  {Mur-Petit}, \citenamefont {Jaksch}, \citenamefont {Hutson}, \citenamefont
  {Sauer}, \citenamefont {Tarbutt},\ and\ \citenamefont
  {Cornish}}]{simulation}%
  \BibitemOpen
  \bibfield  {author} {\bibinfo {author} {\bibfnamefont {J.~A.}\ \bibnamefont
  {Blackmore}}, \bibinfo {author} {\bibfnamefont {L.}~\bibnamefont {Caldwell}},
  \bibinfo {author} {\bibfnamefont {P.~D.}\ \bibnamefont {Gregory}}, \bibinfo
  {author} {\bibfnamefont {E.~M.}\ \bibnamefont {Bridge}}, \bibinfo {author}
  {\bibfnamefont {R.}~\bibnamefont {Sawant}}, \bibinfo {author} {\bibfnamefont
  {J.}~\bibnamefont {Aldegunde}}, \bibinfo {author} {\bibfnamefont
  {J.}~\bibnamefont {Mur-Petit}}, \bibinfo {author} {\bibfnamefont
  {D.}~\bibnamefont {Jaksch}}, \bibinfo {author} {\bibfnamefont {J.~M.}\
  \bibnamefont {Hutson}}, \bibinfo {author} {\bibfnamefont {B.~E.}\
  \bibnamefont {Sauer}}, \bibinfo {author} {\bibfnamefont {M.~R.}\ \bibnamefont
  {Tarbutt}},\ and\ \bibinfo {author} {\bibfnamefont {S.~L.}\ \bibnamefont
  {Cornish}},\ }\bibfield  {title} {\bibinfo {title} {Ultracold molecules for
  quantum simulation: rotational coherences in {CaF} and {RbCs}},\ }\href
  {https://doi.org/10.1088/2058-9565/aaee35} {\bibfield  {journal} {\bibinfo
  {journal} {Quantum Sci. Technol.}\ }\textbf {\bibinfo {volume} {4}},\
  \bibinfo {pages} {014010} (\bibinfo {year} {2018})}\BibitemShut {NoStop}%
\bibitem [{\citenamefont {Rosson}\ \emph {et~al.}(2020)\citenamefont {Rosson},
  \citenamefont {Kiffner}, \citenamefont {Mur-Petit},\ and\ \citenamefont
  {Jaksch}}]{b-h}%
  \BibitemOpen
  \bibfield  {author} {\bibinfo {author} {\bibfnamefont {P.}~\bibnamefont
  {Rosson}}, \bibinfo {author} {\bibfnamefont {M.}~\bibnamefont {Kiffner}},
  \bibinfo {author} {\bibfnamefont {J.}~\bibnamefont {Mur-Petit}},\ and\
  \bibinfo {author} {\bibfnamefont {D.}~\bibnamefont {Jaksch}},\ }\bibfield
  {title} {\bibinfo {title} {Characterizing the phase diagram of finite-size
  dipolar bose-hubbard systems},\ }\href
  {https://doi.org/10.1103/PhysRevA.101.013616} {\bibfield  {journal} {\bibinfo
   {journal} {Phys. Rev. A}\ }\textbf {\bibinfo {volume} {101}},\ \bibinfo
  {pages} {013616} (\bibinfo {year} {2020})}\BibitemShut {NoStop}%
\bibitem [{\citenamefont {Herrera}\ and\ \citenamefont
  {Krems}(2011)}]{holstein}%
  \BibitemOpen
  \bibfield  {author} {\bibinfo {author} {\bibfnamefont {F.}~\bibnamefont
  {Herrera}}\ and\ \bibinfo {author} {\bibfnamefont {R.~V.}\ \bibnamefont
  {Krems}},\ }\bibfield  {title} {\bibinfo {title} {Tunable holstein model with
  cold polar molecules},\ }\href {https://doi.org/10.1103/PhysRevA.84.051401}
  {\bibfield  {journal} {\bibinfo  {journal} {Phys. Rev. A}\ }\textbf {\bibinfo
  {volume} {84}},\ \bibinfo {pages} {051401(R)} (\bibinfo {year}
  {2011})}\BibitemShut {NoStop}%
\bibitem [{\citenamefont {Dawid}\ \emph {et~al.}(2018)\citenamefont {Dawid},
  \citenamefont {Lewenstein},\ and\ \citenamefont {Tomza}}]{twointeract}%
  \BibitemOpen
  \bibfield  {author} {\bibinfo {author} {\bibfnamefont {A.}~\bibnamefont
  {Dawid}}, \bibinfo {author} {\bibfnamefont {M.}~\bibnamefont {Lewenstein}},\
  and\ \bibinfo {author} {\bibfnamefont {M.}~\bibnamefont {Tomza}},\ }\bibfield
   {title} {\bibinfo {title} {Two interacting ultracold molecules in a
  one-dimensional harmonic trap},\ }\href
  {https://doi.org/10.1103/PhysRevA.97.063618} {\bibfield  {journal} {\bibinfo
  {journal} {Phys. Rev. A}\ }\textbf {\bibinfo {volume} {97}},\ \bibinfo
  {pages} {063618} (\bibinfo {year} {2018})}\BibitemShut {NoStop}%
\bibitem [{\citenamefont {Xiang}\ \emph {et~al.}(2012)\citenamefont {Xiang},
  \citenamefont {Litinskaya},\ and\ \citenamefont {Krems}}]{excitons}%
  \BibitemOpen
  \bibfield  {author} {\bibinfo {author} {\bibfnamefont {P.}~\bibnamefont
  {Xiang}}, \bibinfo {author} {\bibfnamefont {M.}~\bibnamefont {Litinskaya}},\
  and\ \bibinfo {author} {\bibfnamefont {R.~V.}\ \bibnamefont {Krems}},\
  }\bibfield  {title} {\bibinfo {title} {Tunable exciton interactions in
  optical lattices with polar molecules},\ }\href
  {https://doi.org/10.1103/PhysRevA.85.061401} {\bibfield  {journal} {\bibinfo
  {journal} {Phys. Rev. A}\ }\textbf {\bibinfo {volume} {85}},\ \bibinfo
  {pages} {061401(R)} (\bibinfo {year} {2012})}\BibitemShut {NoStop}%
\bibitem [{\citenamefont {Kruckenhauser}\ \emph {et~al.}(2020)\citenamefont
  {Kruckenhauser}, \citenamefont {Sieberer}, \citenamefont {De~Marco},
  \citenamefont {Li}, \citenamefont {Matsuda}, \citenamefont {Tobias},
  \citenamefont {Valtolina}, \citenamefont {Ye}, \citenamefont {Rey},
  \citenamefont {Baranov},\ and\ \citenamefont {Zoller}}]{many-quant}%
  \BibitemOpen
  \bibfield  {author} {\bibinfo {author} {\bibfnamefont {A.}~\bibnamefont
  {Kruckenhauser}}, \bibinfo {author} {\bibfnamefont {L.~M.}\ \bibnamefont
  {Sieberer}}, \bibinfo {author} {\bibfnamefont {L.}~\bibnamefont {De~Marco}},
  \bibinfo {author} {\bibfnamefont {J.-R.}\ \bibnamefont {Li}}, \bibinfo
  {author} {\bibfnamefont {K.}~\bibnamefont {Matsuda}}, \bibinfo {author}
  {\bibfnamefont {W.~G.}\ \bibnamefont {Tobias}}, \bibinfo {author}
  {\bibfnamefont {G.}~\bibnamefont {Valtolina}}, \bibinfo {author}
  {\bibfnamefont {J.}~\bibnamefont {Ye}}, \bibinfo {author} {\bibfnamefont
  {A.~M.}\ \bibnamefont {Rey}}, \bibinfo {author} {\bibfnamefont {M.~A.}\
  \bibnamefont {Baranov}},\ and\ \bibinfo {author} {\bibfnamefont
  {P.}~\bibnamefont {Zoller}},\ }\bibfield  {title} {\bibinfo {title} {Quantum
  many-body physics with ultracold polar molecules: Nanostructured potential
  barriers and interactions},\ }\href
  {https://doi.org/10.1103/PhysRevA.102.023320} {\bibfield  {journal} {\bibinfo
   {journal} {Phys. Rev. A}\ }\textbf {\bibinfo {volume} {102}},\ \bibinfo
  {pages} {023320} (\bibinfo {year} {2020})}\BibitemShut {NoStop}%
\bibitem [{\citenamefont {B\"uchler}\ \emph {et~al.}(2007)\citenamefont
  {B\"uchler}, \citenamefont {Demler}, \citenamefont {Lukin}, \citenamefont
  {Micheli}, \citenamefont {Prokof'ev}, \citenamefont {Pupillo},\ and\
  \citenamefont {Zoller}}]{2d-phase}%
  \BibitemOpen
  \bibfield  {author} {\bibinfo {author} {\bibfnamefont {H.~P.}\ \bibnamefont
  {B\"uchler}}, \bibinfo {author} {\bibfnamefont {E.}~\bibnamefont {Demler}},
  \bibinfo {author} {\bibfnamefont {M.}~\bibnamefont {Lukin}}, \bibinfo
  {author} {\bibfnamefont {A.}~\bibnamefont {Micheli}}, \bibinfo {author}
  {\bibfnamefont {N.}~\bibnamefont {Prokof'ev}}, \bibinfo {author}
  {\bibfnamefont {G.}~\bibnamefont {Pupillo}},\ and\ \bibinfo {author}
  {\bibfnamefont {P.}~\bibnamefont {Zoller}},\ }\bibfield  {title} {\bibinfo
  {title} {Strongly correlated 2d quantum phases with cold polar molecules:
  Controlling the shape of the interaction potential},\ }\href
  {https://doi.org/10.1103/PhysRevLett.98.060404} {\bibfield  {journal}
  {\bibinfo  {journal} {Phys. Rev. Lett.}\ }\textbf {\bibinfo {volume} {98}},\
  \bibinfo {pages} {060404} (\bibinfo {year} {2007})}\BibitemShut {NoStop}%
\bibitem [{\citenamefont {Brennen}\ \emph {et~al.}(2007)\citenamefont
  {Brennen}, \citenamefont {Micheli},\ and\ \citenamefont {Zoller}}]{spin-1}%
  \BibitemOpen
  \bibfield  {author} {\bibinfo {author} {\bibfnamefont {G.~K.}\ \bibnamefont
  {Brennen}}, \bibinfo {author} {\bibfnamefont {A.}~\bibnamefont {Micheli}},\
  and\ \bibinfo {author} {\bibfnamefont {P.}~\bibnamefont {Zoller}},\
  }\bibfield  {title} {\bibinfo {title} {Designing spin-1 lattice models using
  polar molecules},\ }\href {https://doi.org/10.1088/1367-2630/9/5/138}
  {\bibfield  {journal} {\bibinfo  {journal} {New J. Phys.}\ }\textbf {\bibinfo
  {volume} {9}},\ \bibinfo {pages} {138} (\bibinfo {year} {2007})}\BibitemShut
  {NoStop}%
\bibitem [{\citenamefont {Wall}\ \emph
  {et~al.}(2015{\natexlab{a}})\citenamefont {Wall}, \citenamefont {Maeda},\
  and\ \citenamefont {Carr}}]{symm-top}%
  \BibitemOpen
  \bibfield  {author} {\bibinfo {author} {\bibfnamefont {M.~L.}\ \bibnamefont
  {Wall}}, \bibinfo {author} {\bibfnamefont {K.}~\bibnamefont {Maeda}},\ and\
  \bibinfo {author} {\bibfnamefont {L.~D.}\ \bibnamefont {Carr}},\ }\bibfield
  {title} {\bibinfo {title} {Realizing unconventional quantum magnetism with
  symmetric top molecules},\ }\href
  {https://doi.org/10.1088/1367-2630/17/2/025001} {\bibfield  {journal}
  {\bibinfo  {journal} {New J. Phys.}\ }\textbf {\bibinfo {volume} {17}},\
  \bibinfo {pages} {025001} (\bibinfo {year} {2015}{\natexlab{a}})}\BibitemShut
  {NoStop}%
\bibitem [{\citenamefont {Olaya-Agudelo}\ \emph {et~al.}(2019)\citenamefont
  {Olaya-Agudelo}, \citenamefont {Garc{\'\i}a-Negrete},\ and\ \citenamefont
  {Rodr{\'\i}guez-Ram{\'\i}rez}}]{1d-lattice}%
  \BibitemOpen
  \bibfield  {author} {\bibinfo {author} {\bibfnamefont {V.~C.}\ \bibnamefont
  {Olaya-Agudelo}}, \bibinfo {author} {\bibfnamefont {S.}~\bibnamefont
  {Garc{\'\i}a-Negrete}},\ and\ \bibinfo {author} {\bibfnamefont
  {K.}~\bibnamefont {Rodr{\'\i}guez-Ram{\'\i}rez}},\ }\bibfield  {title}
  {\bibinfo {title} {Single-excitation correlation dynamics of polar molecules
  on 1d lattices: driving by external electric fields},\ }\href
  {https://doi.org/https://doi.org/10.1016/j.matpr.2019.05.070} {\bibfield
  {journal} {\bibinfo  {journal} {Materials Today: Proceedings}\ }\textbf
  {\bibinfo {volume} {14}},\ \bibinfo {pages} {134} (\bibinfo {year}
  {2019})}\BibitemShut {NoStop}%
\bibitem [{\citenamefont {Kaufman}\ and\ \citenamefont {Ni}(2021)}]{tweezer}%
  \BibitemOpen
  \bibfield  {author} {\bibinfo {author} {\bibfnamefont {A.~M.}\ \bibnamefont
  {Kaufman}}\ and\ \bibinfo {author} {\bibfnamefont {K.-K.}\ \bibnamefont
  {Ni}},\ }\bibfield  {title} {\bibinfo {title} {Quantum science with optical
  tweezer arrays of ultracold atoms and molecules},\ }\href
  {https://doi.org/10.1038/s41567-021-01357-2} {\bibfield  {journal} {\bibinfo
  {journal} {Nat. Phys.}\ }\textbf {\bibinfo {volume} {17}},\ \bibinfo {pages}
  {1324} (\bibinfo {year} {2021})}\BibitemShut {NoStop}%
\bibitem [{\citenamefont {DeMille}(2002)}]{demille}%
  \BibitemOpen
  \bibfield  {author} {\bibinfo {author} {\bibfnamefont {D.}~\bibnamefont
  {DeMille}},\ }\bibfield  {title} {\bibinfo {title} {Quantum computation with
  trapped polar molecules},\ }\href
  {https://doi.org/10.1103/PhysRevLett.88.067901} {\bibfield  {journal}
  {\bibinfo  {journal} {Phys. Rev. Lett.}\ }\textbf {\bibinfo {volume} {88}},\
  \bibinfo {pages} {067901} (\bibinfo {year} {2002})}\BibitemShut {NoStop}%
\bibitem [{\citenamefont {Yelin}\ \emph {et~al.}(2006)\citenamefont {Yelin},
  \citenamefont {Kirby},\ and\ \citenamefont {C\^ot\'e}}]{robust}%
  \BibitemOpen
  \bibfield  {author} {\bibinfo {author} {\bibfnamefont {S.~F.}\ \bibnamefont
  {Yelin}}, \bibinfo {author} {\bibfnamefont {K.}~\bibnamefont {Kirby}},\ and\
  \bibinfo {author} {\bibfnamefont {R.}~\bibnamefont {C\^ot\'e}},\ }\bibfield
  {title} {\bibinfo {title} {Schemes for robust quantum computation with polar
  molecules},\ }\href {https://doi.org/10.1103/PhysRevA.74.050301} {\bibfield
  {journal} {\bibinfo  {journal} {Phys. Rev. A}\ }\textbf {\bibinfo {volume}
  {74}},\ \bibinfo {pages} {050301(R)} (\bibinfo {year} {2006})}\BibitemShut
  {NoStop}%
\bibitem [{\citenamefont {Hughes}\ \emph {et~al.}(2020)\citenamefont {Hughes},
  \citenamefont {Frye}, \citenamefont {Sawant}, \citenamefont {Bhole},
  \citenamefont {Jones}, \citenamefont {Cornish}, \citenamefont {Tarbutt},
  \citenamefont {Hutson}, \citenamefont {Jaksch},\ and\ \citenamefont
  {Mur-Petit}}]{gate}%
  \BibitemOpen
  \bibfield  {author} {\bibinfo {author} {\bibfnamefont {M.}~\bibnamefont
  {Hughes}}, \bibinfo {author} {\bibfnamefont {M.~D.}\ \bibnamefont {Frye}},
  \bibinfo {author} {\bibfnamefont {R.}~\bibnamefont {Sawant}}, \bibinfo
  {author} {\bibfnamefont {G.}~\bibnamefont {Bhole}}, \bibinfo {author}
  {\bibfnamefont {J.~A.}\ \bibnamefont {Jones}}, \bibinfo {author}
  {\bibfnamefont {S.~L.}\ \bibnamefont {Cornish}}, \bibinfo {author}
  {\bibfnamefont {M.~R.}\ \bibnamefont {Tarbutt}}, \bibinfo {author}
  {\bibfnamefont {J.~M.}\ \bibnamefont {Hutson}}, \bibinfo {author}
  {\bibfnamefont {D.}~\bibnamefont {Jaksch}},\ and\ \bibinfo {author}
  {\bibfnamefont {J.}~\bibnamefont {Mur-Petit}},\ }\bibfield  {title} {\bibinfo
  {title} {Robust entangling gate for polar molecules using magnetic and
  microwave fields},\ }\href {https://doi.org/10.1103/PhysRevA.101.062308}
  {\bibfield  {journal} {\bibinfo  {journal} {Phys. Rev. A}\ }\textbf {\bibinfo
  {volume} {101}},\ \bibinfo {pages} {062308} (\bibinfo {year}
  {2020})}\BibitemShut {NoStop}%
\bibitem [{\citenamefont {Mishima}\ and\ \citenamefont
  {Yamashita}(2009)}]{rotational-gates}%
  \BibitemOpen
  \bibfield  {author} {\bibinfo {author} {\bibfnamefont {K.}~\bibnamefont
  {Mishima}}\ and\ \bibinfo {author} {\bibfnamefont {K.}~\bibnamefont
  {Yamashita}},\ }\bibfield  {title} {\bibinfo {title} {Quantum computing using
  rotational modes of two polar molecules},\ }\href
  {https://doi.org/https://doi.org/10.1016/j.chemphys.2009.05.014} {\bibfield
  {journal} {\bibinfo  {journal} {Chem. Phys.}\ }\textbf {\bibinfo {volume}
  {361}},\ \bibinfo {pages} {106} (\bibinfo {year} {2009})}\BibitemShut
  {NoStop}%
\bibitem [{\citenamefont {Zhu}\ \emph {et~al.}(2013)\citenamefont {Zhu},
  \citenamefont {Kais}, \citenamefont {Wei}, \citenamefont {Herschbach},\ and\
  \citenamefont {Friedrich}}]{logic-gates}%
  \BibitemOpen
  \bibfield  {author} {\bibinfo {author} {\bibfnamefont {J.}~\bibnamefont
  {Zhu}}, \bibinfo {author} {\bibfnamefont {S.}~\bibnamefont {Kais}}, \bibinfo
  {author} {\bibfnamefont {Q.}~\bibnamefont {Wei}}, \bibinfo {author}
  {\bibfnamefont {D.}~\bibnamefont {Herschbach}},\ and\ \bibinfo {author}
  {\bibfnamefont {B.}~\bibnamefont {Friedrich}},\ }\bibfield  {title} {\bibinfo
  {title} {Implementation of quantum logic gates using polar molecules in
  pendular states},\ }\href {https://doi.org/10.1063/1.4774058} {\bibfield
  {journal} {\bibinfo  {journal} {J. Chem. Phys.}\ }\textbf {\bibinfo {volume}
  {138}},\ \bibinfo {pages} {024104} (\bibinfo {year} {2013})}\BibitemShut
  {NoStop}%
\bibitem [{\citenamefont {Mur-Petit}\ \emph {et~al.}(2013)\citenamefont
  {Mur-Petit}, \citenamefont {P{\'e}rez-R{\'\i}os}, \citenamefont
  {Campos-Mart{\'\i}nez}, \citenamefont {Hern{\'a}ndez}, \citenamefont
  {Willitsch},\ and\ \citenamefont {Garc{\'\i}a-Ripoll}}]{ion}%
  \BibitemOpen
  \bibfield  {author} {\bibinfo {author} {\bibfnamefont {J.}~\bibnamefont
  {Mur-Petit}}, \bibinfo {author} {\bibfnamefont {J.}~\bibnamefont
  {P{\'e}rez-R{\'\i}os}}, \bibinfo {author} {\bibfnamefont {J.}~\bibnamefont
  {Campos-Mart{\'\i}nez}}, \bibinfo {author} {\bibfnamefont {M.~I.}\
  \bibnamefont {Hern{\'a}ndez}}, \bibinfo {author} {\bibfnamefont
  {S.}~\bibnamefont {Willitsch}},\ and\ \bibinfo {author} {\bibfnamefont
  {J.~J.}\ \bibnamefont {Garc{\'\i}a-Ripoll}},\ }\bibfield  {title} {\bibinfo
  {title} {Toward a molecular ion qubit},\ }in\ \href@noop {} {\emph {\bibinfo
  {booktitle} {Architecture and Design of Molecule Logic Gates and Atom
  Circuits}}},\ \bibinfo {editor} {edited by\ \bibinfo {editor} {\bibfnamefont
  {N.}~\bibnamefont {Lorente}}\ and\ \bibinfo {editor} {\bibfnamefont
  {C.}~\bibnamefont {Joachim}}}\ (\bibinfo  {publisher} {Springer Berlin
  Heidelberg},\ \bibinfo {address} {Berlin, Heidelberg},\ \bibinfo {year}
  {2013})\ pp.\ \bibinfo {pages} {267--277}\BibitemShut {NoStop}%
\bibitem [{\citenamefont {Sawant}\ \emph {et~al.}(2020)\citenamefont {Sawant},
  \citenamefont {Blackmore}, \citenamefont {Gregory}, \citenamefont
  {Mur-Petit}, \citenamefont {Jaksch}, \citenamefont {Aldegunde}, \citenamefont
  {Hutson}, \citenamefont {Tarbutt},\ and\ \citenamefont {Cornish}}]{qudit}%
  \BibitemOpen
  \bibfield  {author} {\bibinfo {author} {\bibfnamefont {R.}~\bibnamefont
  {Sawant}}, \bibinfo {author} {\bibfnamefont {J.~A.}\ \bibnamefont
  {Blackmore}}, \bibinfo {author} {\bibfnamefont {P.~D.}\ \bibnamefont
  {Gregory}}, \bibinfo {author} {\bibfnamefont {J.}~\bibnamefont {Mur-Petit}},
  \bibinfo {author} {\bibfnamefont {D.}~\bibnamefont {Jaksch}}, \bibinfo
  {author} {\bibfnamefont {J.}~\bibnamefont {Aldegunde}}, \bibinfo {author}
  {\bibfnamefont {J.~M.}\ \bibnamefont {Hutson}}, \bibinfo {author}
  {\bibfnamefont {M.~R.}\ \bibnamefont {Tarbutt}},\ and\ \bibinfo {author}
  {\bibfnamefont {S.~L.}\ \bibnamefont {Cornish}},\ }\bibfield  {title}
  {\bibinfo {title} {Ultracold polar molecules as qudits},\ }\href
  {https://doi.org/10.1088/1367-2630/ab60f4} {\bibfield  {journal} {\bibinfo
  {journal} {New J. Phys.}\ }\textbf {\bibinfo {volume} {22}},\ \bibinfo
  {pages} {013027} (\bibinfo {year} {2020})}\BibitemShut {NoStop}%
\bibitem [{\citenamefont {Gregory}\ \emph {et~al.}(2021)\citenamefont
  {Gregory}, \citenamefont {Blackmore}, \citenamefont {Bromley}, \citenamefont
  {Hutson},\ and\ \citenamefont {Cornish}}]{storage}%
  \BibitemOpen
  \bibfield  {author} {\bibinfo {author} {\bibfnamefont {P.~D.}\ \bibnamefont
  {Gregory}}, \bibinfo {author} {\bibfnamefont {J.~A.}\ \bibnamefont
  {Blackmore}}, \bibinfo {author} {\bibfnamefont {S.~L.}\ \bibnamefont
  {Bromley}}, \bibinfo {author} {\bibfnamefont {J.~M.}\ \bibnamefont
  {Hutson}},\ and\ \bibinfo {author} {\bibfnamefont {S.~L.}\ \bibnamefont
  {Cornish}},\ }\bibfield  {title} {\bibinfo {title} {Robust storage qubits in
  ultracold polar molecules},\ }\href
  {https://doi.org/10.1038/s41567-021-01328-7} {\bibfield  {journal} {\bibinfo
  {journal} {Nat. Phys.}\ }\textbf {\bibinfo {volume} {17}},\ \bibinfo {pages}
  {1149} (\bibinfo {year} {2021})}\BibitemShut {NoStop}%
\bibitem [{\citenamefont {Karra}\ \emph {et~al.}(2016)\citenamefont {Karra},
  \citenamefont {Sharma}, \citenamefont {Friedrich}, \citenamefont {Kais},\
  and\ \citenamefont {Herschbach}}]{sigma2}%
  \BibitemOpen
  \bibfield  {author} {\bibinfo {author} {\bibfnamefont {M.}~\bibnamefont
  {Karra}}, \bibinfo {author} {\bibfnamefont {K.}~\bibnamefont {Sharma}},
  \bibinfo {author} {\bibfnamefont {B.}~\bibnamefont {Friedrich}}, \bibinfo
  {author} {\bibfnamefont {S.}~\bibnamefont {Kais}},\ and\ \bibinfo {author}
  {\bibfnamefont {D.}~\bibnamefont {Herschbach}},\ }\bibfield  {title}
  {\bibinfo {title} {Prospects for quantum computing with an array of ultracold
  polar paramagnetic molecules},\ }\href {https://doi.org/10.1063/1.4942928}
  {\bibfield  {journal} {\bibinfo  {journal} {J. Chem. Phys.}\ }\textbf
  {\bibinfo {volume} {144}},\ \bibinfo {pages} {094301} (\bibinfo {year}
  {2016})}\BibitemShut {NoStop}%
\bibitem [{\citenamefont {M\"uller}\ \emph {et~al.}(2011)\citenamefont
  {M\"uller}, \citenamefont {Reich}, \citenamefont {Murphy}, \citenamefont
  {Yuan}, \citenamefont {Vala}, \citenamefont {Whaley}, \citenamefont
  {Calarco},\ and\ \citenamefont {Koch}}]{sigma2-gates}%
  \BibitemOpen
  \bibfield  {author} {\bibinfo {author} {\bibfnamefont {M.~M.}\ \bibnamefont
  {M\"uller}}, \bibinfo {author} {\bibfnamefont {D.~M.}\ \bibnamefont {Reich}},
  \bibinfo {author} {\bibfnamefont {M.}~\bibnamefont {Murphy}}, \bibinfo
  {author} {\bibfnamefont {H.}~\bibnamefont {Yuan}}, \bibinfo {author}
  {\bibfnamefont {J.}~\bibnamefont {Vala}}, \bibinfo {author} {\bibfnamefont
  {K.~B.}\ \bibnamefont {Whaley}}, \bibinfo {author} {\bibfnamefont
  {T.}~\bibnamefont {Calarco}},\ and\ \bibinfo {author} {\bibfnamefont {C.~P.}\
  \bibnamefont {Koch}},\ }\bibfield  {title} {\bibinfo {title} {Optimizing
  entangling quantum gates for physical systems},\ }\href
  {https://doi.org/10.1103/PhysRevA.84.042315} {\bibfield  {journal} {\bibinfo
  {journal} {Phys. Rev. A}\ }\textbf {\bibinfo {volume} {84}},\ \bibinfo
  {pages} {042315} (\bibinfo {year} {2011})}\BibitemShut {NoStop}%
\bibitem [{\citenamefont {Anderegg}\ \emph {et~al.}(2019)\citenamefont
  {Anderegg}, \citenamefont {Cheuk}, \citenamefont {Bao}, \citenamefont
  {Burchesky}, \citenamefont {Ketterle}, \citenamefont {Ni},\ and\
  \citenamefont {Doyle}}]{exp1}%
  \BibitemOpen
  \bibfield  {author} {\bibinfo {author} {\bibfnamefont {L.}~\bibnamefont
  {Anderegg}}, \bibinfo {author} {\bibfnamefont {L.~W.}\ \bibnamefont {Cheuk}},
  \bibinfo {author} {\bibfnamefont {Y.}~\bibnamefont {Bao}}, \bibinfo {author}
  {\bibfnamefont {S.}~\bibnamefont {Burchesky}}, \bibinfo {author}
  {\bibfnamefont {W.}~\bibnamefont {Ketterle}}, \bibinfo {author}
  {\bibfnamefont {K.-K.}\ \bibnamefont {Ni}},\ and\ \bibinfo {author}
  {\bibfnamefont {J.~M.}\ \bibnamefont {Doyle}},\ }\bibfield  {title} {\bibinfo
  {title} {An optical tweezer array of ultracold molecules},\ }\href
  {https://doi.org/10.1126/science.aax1265} {\bibfield  {journal} {\bibinfo
  {journal} {Science}\ }\textbf {\bibinfo {volume} {365}},\ \bibinfo {pages}
  {1156} (\bibinfo {year} {2019})}\BibitemShut {NoStop}%
\bibitem [{\citenamefont {Barry}\ \emph {et~al.}(2014)\citenamefont {Barry},
  \citenamefont {McCarron}, \citenamefont {Norrgard}, \citenamefont
  {Steinecker},\ and\ \citenamefont {DeMille}}]{exp2}%
  \BibitemOpen
  \bibfield  {author} {\bibinfo {author} {\bibfnamefont {J.~F.}\ \bibnamefont
  {Barry}}, \bibinfo {author} {\bibfnamefont {D.~J.}\ \bibnamefont {McCarron}},
  \bibinfo {author} {\bibfnamefont {E.~B.}\ \bibnamefont {Norrgard}}, \bibinfo
  {author} {\bibfnamefont {M.~H.}\ \bibnamefont {Steinecker}},\ and\ \bibinfo
  {author} {\bibfnamefont {D.}~\bibnamefont {DeMille}},\ }\bibfield  {title}
  {\bibinfo {title} {Magneto-optical trapping of a diatomic molecule},\ }\href
  {https://doi.org/10.1038/nature13634} {\bibfield  {journal} {\bibinfo
  {journal} {Nature}\ }\textbf {\bibinfo {volume} {512}},\ \bibinfo {pages}
  {286} (\bibinfo {year} {2014})}\BibitemShut {NoStop}%
\bibitem [{\citenamefont {Truppe}\ \emph {et~al.}(2017)\citenamefont {Truppe},
  \citenamefont {Williams}, \citenamefont {Hambach}, \citenamefont {Caldwell},
  \citenamefont {Fitch}, \citenamefont {Hinds}, \citenamefont {Sauer},\ and\
  \citenamefont {Tarbutt}}]{exp3}%
  \BibitemOpen
  \bibfield  {author} {\bibinfo {author} {\bibfnamefont {S.}~\bibnamefont
  {Truppe}}, \bibinfo {author} {\bibfnamefont {H.~J.}\ \bibnamefont
  {Williams}}, \bibinfo {author} {\bibfnamefont {M.}~\bibnamefont {Hambach}},
  \bibinfo {author} {\bibfnamefont {L.}~\bibnamefont {Caldwell}}, \bibinfo
  {author} {\bibfnamefont {N.~J.}\ \bibnamefont {Fitch}}, \bibinfo {author}
  {\bibfnamefont {E.~A.}\ \bibnamefont {Hinds}}, \bibinfo {author}
  {\bibfnamefont {B.~E.}\ \bibnamefont {Sauer}},\ and\ \bibinfo {author}
  {\bibfnamefont {M.~R.}\ \bibnamefont {Tarbutt}},\ }\bibfield  {title}
  {\bibinfo {title} {Molecules cooled below the doppler limit},\ }\href
  {https://doi.org/10.1038/nphys4241} {\bibfield  {journal} {\bibinfo
  {journal} {Nat. Phys.}\ }\textbf {\bibinfo {volume} {13}},\ \bibinfo {pages}
  {1173} (\bibinfo {year} {2017})}\BibitemShut {NoStop}%
\bibitem [{\citenamefont {Brue}\ and\ \citenamefont {Hutson}(2013)}]{exp4}%
  \BibitemOpen
  \bibfield  {author} {\bibinfo {author} {\bibfnamefont {D.~A.}\ \bibnamefont
  {Brue}}\ and\ \bibinfo {author} {\bibfnamefont {J.~M.}\ \bibnamefont
  {Hutson}},\ }\bibfield  {title} {\bibinfo {title} {Prospects of forming
  ultracold molecules in ${}^{2}\ensuremath{\Sigma}$ states by
  magnetoassociation of alkali-metal atoms with yb},\ }\href
  {https://doi.org/10.1103/PhysRevA.87.052709} {\bibfield  {journal} {\bibinfo
  {journal} {Phys. Rev. A}\ }\textbf {\bibinfo {volume} {87}},\ \bibinfo
  {pages} {052709} (\bibinfo {year} {2013})}\BibitemShut {NoStop}%
\bibitem [{\citenamefont {Jones}\ \emph {et~al.}(2006)\citenamefont {Jones},
  \citenamefont {Tiesinga}, \citenamefont {Lett},\ and\ \citenamefont
  {Julienne}}]{exp5}%
  \BibitemOpen
  \bibfield  {author} {\bibinfo {author} {\bibfnamefont {K.~M.}\ \bibnamefont
  {Jones}}, \bibinfo {author} {\bibfnamefont {E.}~\bibnamefont {Tiesinga}},
  \bibinfo {author} {\bibfnamefont {P.~D.}\ \bibnamefont {Lett}},\ and\
  \bibinfo {author} {\bibfnamefont {P.~S.}\ \bibnamefont {Julienne}},\
  }\bibfield  {title} {\bibinfo {title} {Ultracold photoassociation
  spectroscopy: Long-range molecules and atomic scattering},\ }\href
  {https://doi.org/10.1103/RevModPhys.78.483} {\bibfield  {journal} {\bibinfo
  {journal} {Rev. Mod. Phys.}\ }\textbf {\bibinfo {volume} {78}},\ \bibinfo
  {pages} {483} (\bibinfo {year} {2006})}\BibitemShut {NoStop}%
\bibitem [{\citenamefont {Okano}\ \emph {et~al.}(2009)\citenamefont {Okano},
  \citenamefont {Hara}, \citenamefont {Muramatsu}, \citenamefont {Doi},
  \citenamefont {Uetake}, \citenamefont {Takasu},\ and\ \citenamefont
  {Takahashi}}]{exp6}%
  \BibitemOpen
  \bibfield  {author} {\bibinfo {author} {\bibfnamefont {M.}~\bibnamefont
  {Okano}}, \bibinfo {author} {\bibfnamefont {H.}~\bibnamefont {Hara}},
  \bibinfo {author} {\bibfnamefont {M.}~\bibnamefont {Muramatsu}}, \bibinfo
  {author} {\bibfnamefont {K.}~\bibnamefont {Doi}}, \bibinfo {author}
  {\bibfnamefont {S.}~\bibnamefont {Uetake}}, \bibinfo {author} {\bibfnamefont
  {Y.}~\bibnamefont {Takasu}},\ and\ \bibinfo {author} {\bibfnamefont
  {Y.}~\bibnamefont {Takahashi}},\ }\bibfield  {title} {\bibinfo {title}
  {Simultaneous magneto-optical trapping of lithium and ytterbium atoms towards
  production of ultracold polar molecules},\ }\href
  {https://doi.org/10.1007/s00340-009-3728-0} {\bibfield  {journal} {\bibinfo
  {journal} {Appl. Phys. B}\ }\textbf {\bibinfo {volume} {98}},\ \bibinfo
  {pages} {691} (\bibinfo {year} {2009})}\BibitemShut {NoStop}%
\bibitem [{\citenamefont {Nemitz}\ \emph {et~al.}(2010)\citenamefont {Nemitz},
  \citenamefont {Baumer}, \citenamefont {M\"unchow}, \citenamefont {Tassy},\
  and\ \citenamefont {G\"orlitz}}]{exp7}%
  \BibitemOpen
  \bibfield  {author} {\bibinfo {author} {\bibfnamefont {N.}~\bibnamefont
  {Nemitz}}, \bibinfo {author} {\bibfnamefont {F.}~\bibnamefont {Baumer}},
  \bibinfo {author} {\bibfnamefont {F.}~\bibnamefont {M\"unchow}}, \bibinfo
  {author} {\bibfnamefont {S.}~\bibnamefont {Tassy}},\ and\ \bibinfo {author}
  {\bibfnamefont {A.}~\bibnamefont {G\"orlitz}},\ }\bibfield  {title} {\bibinfo
  {title} {Production of heteronuclear molecules in an electronically excited
  state by photoassociation in a mixture of ultracold yb and rb},\ }\href
  {https://doi.org/10.1103/PhysRevA.79.061403} {\bibfield  {journal} {\bibinfo
  {journal} {Phys. Rev. A}\ }\textbf {\bibinfo {volume} {79}},\ \bibinfo
  {pages} {061403(R)} (\bibinfo {year} {2010})}\BibitemShut {NoStop}%
\bibitem [{\citenamefont {Weinstein}\ \emph {et~al.}(1998)\citenamefont
  {Weinstein}, \citenamefont {DeCarvalho}, \citenamefont {Guillet},
  \citenamefont {Friedrich},\ and\ \citenamefont {Doyle}}]{exp8}%
  \BibitemOpen
  \bibfield  {author} {\bibinfo {author} {\bibfnamefont {J.~D.}\ \bibnamefont
  {Weinstein}}, \bibinfo {author} {\bibfnamefont {R.}~\bibnamefont
  {DeCarvalho}}, \bibinfo {author} {\bibfnamefont {T.}~\bibnamefont {Guillet}},
  \bibinfo {author} {\bibfnamefont {B.}~\bibnamefont {Friedrich}},\ and\
  \bibinfo {author} {\bibfnamefont {J.~M.}\ \bibnamefont {Doyle}},\ }\bibfield
  {title} {\bibinfo {title} {Magnetic trapping of calcium monohydride molecules
  at millikelvin temperatures},\ }\href@noop {} {\bibfield  {journal} {\bibinfo
   {journal} {Nature}\ }\textbf {\bibinfo {volume} {395}},\ \bibinfo {pages}
  {148} (\bibinfo {year} {1998})}\BibitemShut {NoStop}%
\bibitem [{\citenamefont {Maussang}\ \emph {et~al.}(2005)\citenamefont
  {Maussang}, \citenamefont {Egorov}, \citenamefont {Helton}, \citenamefont
  {Nguyen},\ and\ \citenamefont {Doyle}}]{exp9}%
  \BibitemOpen
  \bibfield  {author} {\bibinfo {author} {\bibfnamefont {K.}~\bibnamefont
  {Maussang}}, \bibinfo {author} {\bibfnamefont {D.}~\bibnamefont {Egorov}},
  \bibinfo {author} {\bibfnamefont {J.~S.}\ \bibnamefont {Helton}}, \bibinfo
  {author} {\bibfnamefont {S.~V.}\ \bibnamefont {Nguyen}},\ and\ \bibinfo
  {author} {\bibfnamefont {J.~M.}\ \bibnamefont {Doyle}},\ }\bibfield  {title}
  {\bibinfo {title} {Zeeman relaxation of caf in low-temperature collisions
  with helium},\ }\href {https://doi.org/10.1103/PhysRevLett.94.123002}
  {\bibfield  {journal} {\bibinfo  {journal} {Phys. Rev. Lett.}\ }\textbf
  {\bibinfo {volume} {94}},\ \bibinfo {pages} {123002} (\bibinfo {year}
  {2005})}\BibitemShut {NoStop}%
\bibitem [{\citenamefont {Di~Rosa}(2004)}]{exp10}%
  \BibitemOpen
  \bibfield  {author} {\bibinfo {author} {\bibfnamefont {M.}~\bibnamefont
  {Di~Rosa}},\ }\bibfield  {title} {\bibinfo {title} {Laser-cooling
  molecules},\ }\href@noop {} {\bibfield  {journal} {\bibinfo  {journal} {Eur.
  Phys. J. D}\ }\textbf {\bibinfo {volume} {31}},\ \bibinfo {pages} {395}
  (\bibinfo {year} {2004})}\BibitemShut {NoStop}%
\bibitem [{\citenamefont {Shuman}\ \emph {et~al.}(2009)\citenamefont {Shuman},
  \citenamefont {Barry}, \citenamefont {Glenn},\ and\ \citenamefont
  {DeMille}}]{exp11}%
  \BibitemOpen
  \bibfield  {author} {\bibinfo {author} {\bibfnamefont {E.~S.}\ \bibnamefont
  {Shuman}}, \bibinfo {author} {\bibfnamefont {J.~F.}\ \bibnamefont {Barry}},
  \bibinfo {author} {\bibfnamefont {D.~R.}\ \bibnamefont {Glenn}},\ and\
  \bibinfo {author} {\bibfnamefont {D.}~\bibnamefont {DeMille}},\ }\bibfield
  {title} {\bibinfo {title} {Radiative force from optical cycling on a diatomic
  molecule},\ }\href {https://doi.org/10.1103/PhysRevLett.103.223001}
  {\bibfield  {journal} {\bibinfo  {journal} {Phys. Rev. Lett.}\ }\textbf
  {\bibinfo {volume} {103}},\ \bibinfo {pages} {223001} (\bibinfo {year}
  {2009})}\BibitemShut {NoStop}%
\bibitem [{\citenamefont {Friedrich}\ and\ \citenamefont
  {Herschbach}(2000)}]{avoided}%
  \BibitemOpen
  \bibfield  {author} {\bibinfo {author} {\bibfnamefont {B.}~\bibnamefont
  {Friedrich}}\ and\ \bibinfo {author} {\bibfnamefont {D.}~\bibnamefont
  {Herschbach}},\ }\bibfield  {title} {\bibinfo {title} {Steric proficiency of
  polar ${}^{2}\ensuremath{\Sigma}$ molecules in congruent electric and
  magnetic fields},\ }\href {https://doi.org/10.1039/A908876H} {\bibfield
  {journal} {\bibinfo  {journal} {Phys. Chem. Chem. Phys.}\ }\textbf {\bibinfo
  {volume} {2}},\ \bibinfo {pages} {419} (\bibinfo {year} {2000})}\BibitemShut
  {NoStop}%
\bibitem [{\citenamefont {Abrahamsson}\ \emph {et~al.}(2007)\citenamefont
  {Abrahamsson}, \citenamefont {Tscherbul},\ and\ \citenamefont
  {Krems}}]{collisions}%
  \BibitemOpen
  \bibfield  {author} {\bibinfo {author} {\bibfnamefont {E.}~\bibnamefont
  {Abrahamsson}}, \bibinfo {author} {\bibfnamefont {T.~V.}\ \bibnamefont
  {Tscherbul}},\ and\ \bibinfo {author} {\bibfnamefont {R.~V.}\ \bibnamefont
  {Krems}},\ }\bibfield  {title} {\bibinfo {title} {Inelastic collisions of
  cold polar molecules in nonparallel electric and magnetic fields},\ }\href
  {https://doi.org/10.1063/1.2748770} {\bibfield  {journal} {\bibinfo
  {journal} {J. Chem. Phys.}\ }\textbf {\bibinfo {volume} {127}},\ \bibinfo
  {pages} {044302} (\bibinfo {year} {2007})}\BibitemShut {NoStop}%
\bibitem [{\citenamefont {Tscherbul}\ and\ \citenamefont
  {Krems}(2006)}]{collisions2}%
  \BibitemOpen
  \bibfield  {author} {\bibinfo {author} {\bibfnamefont {T.~V.}\ \bibnamefont
  {Tscherbul}}\ and\ \bibinfo {author} {\bibfnamefont {R.~V.}\ \bibnamefont
  {Krems}},\ }\bibfield  {title} {\bibinfo {title} {Controlling electronic spin
  relaxation of cold molecules with electric fields},\ }\href
  {https://doi.org/10.1103/PhysRevLett.97.083201} {\bibfield  {journal}
  {\bibinfo  {journal} {Phys. Rev. Lett.}\ }\textbf {\bibinfo {volume} {97}},\
  \bibinfo {pages} {083201} (\bibinfo {year} {2006})}\BibitemShut {NoStop}%
\bibitem [{\citenamefont {P{\'{e}}rez-R{\'{\i}}os}\ \emph
  {et~al.}(2010)\citenamefont {P{\'{e}}rez-R{\'{\i}}os}, \citenamefont
  {Herrera},\ and\ \citenamefont {Krems}}]{romannjp}%
  \BibitemOpen
  \bibfield  {author} {\bibinfo {author} {\bibfnamefont {J.}~\bibnamefont
  {P{\'{e}}rez-R{\'{\i}}os}}, \bibinfo {author} {\bibfnamefont
  {F.}~\bibnamefont {Herrera}},\ and\ \bibinfo {author} {\bibfnamefont {R.~V.}\
  \bibnamefont {Krems}},\ }\bibfield  {title} {\bibinfo {title} {External field
  control of collective spin excitations in an optical lattice of {$2\upSigma$}
  molecules},\ }\href {https://doi.org/10.1088/1367-2630/12/10/103007}
  {\bibfield  {journal} {\bibinfo  {journal} {New J. Phys.}\ }\textbf {\bibinfo
  {volume} {12}},\ \bibinfo {pages} {103007} (\bibinfo {year}
  {2010})}\BibitemShut {NoStop}%
\bibitem [{\citenamefont {Alyabyshev}\ \emph {et~al.}(2012)\citenamefont
  {Alyabyshev}, \citenamefont {Lemeshko},\ and\ \citenamefont
  {Krems}}]{imaging}%
  \BibitemOpen
  \bibfield  {author} {\bibinfo {author} {\bibfnamefont {S.~V.}\ \bibnamefont
  {Alyabyshev}}, \bibinfo {author} {\bibfnamefont {M.}~\bibnamefont
  {Lemeshko}},\ and\ \bibinfo {author} {\bibfnamefont {R.~V.}\ \bibnamefont
  {Krems}},\ }\bibfield  {title} {\bibinfo {title} {Sensitive imaging of
  electromagnetic fields with paramagnetic polar molecules},\ }\href
  {https://doi.org/10.1103/physreva.86.013409} {\bibfield  {journal} {\bibinfo
  {journal} {Phys. Rev. A}\ }\textbf {\bibinfo {volume} {86}},\ \bibinfo
  {pages} {013409} (\bibinfo {year} {2012})}\BibitemShut {NoStop}%
\bibitem [{\citenamefont {Gray}(1976)}]{electrostatic}%
  \BibitemOpen
  \bibfield  {author} {\bibinfo {author} {\bibfnamefont {C.~G.}\ \bibnamefont
  {Gray}},\ }\bibfield  {title} {\bibinfo {title} {Spherical tensor approach to
  multipole expansions. i. electrostatic interactions},\ }\href
  {https://doi.org/10.1139/p76-057} {\bibfield  {journal} {\bibinfo  {journal}
  {Canadian Journal of Physics}\ }\textbf {\bibinfo {volume} {54}},\ \bibinfo
  {pages} {505} (\bibinfo {year} {1976})}\BibitemShut {NoStop}%
\bibitem [{\citenamefont {Wall}\ \emph
  {et~al.}(2015{\natexlab{b}})\citenamefont {Wall}, \citenamefont {Hazzard},\
  and\ \citenamefont {Rey}}]{hazzard}%
  \BibitemOpen
  \bibfield  {author} {\bibinfo {author} {\bibfnamefont {M.~L.}\ \bibnamefont
  {Wall}}, \bibinfo {author} {\bibfnamefont {K.~R.~A.}\ \bibnamefont
  {Hazzard}},\ and\ \bibinfo {author} {\bibfnamefont {A.~M.}\ \bibnamefont
  {Rey}},\ }\bibinfo {title} {Quantum magnetism with ultracold molecules},\ in\
  \href {https://doi.org/10.1142/9789814678704_0001} {\emph {\bibinfo
  {booktitle} {From Atomic to Mesoscale}}}\ (\bibinfo  {publisher} {World
  Scientific Publishing},\ \bibinfo {address} {Singapore},\ \bibinfo {year}
  {2015})\ Chap.~\bibinfo {chapter} {1}, pp.\ \bibinfo {pages}
  {3--37}\BibitemShut {NoStop}%
\bibitem [{\citenamefont {Johansson}\ \emph {et~al.}(2012)\citenamefont
  {Johansson}, \citenamefont {Nation},\ and\ \citenamefont {Nori}}]{qutip1}%
  \BibitemOpen
  \bibfield  {author} {\bibinfo {author} {\bibfnamefont {J.}~\bibnamefont
  {Johansson}}, \bibinfo {author} {\bibfnamefont {P.}~\bibnamefont {Nation}},\
  and\ \bibinfo {author} {\bibfnamefont {F.}~\bibnamefont {Nori}},\ }\bibfield
  {title} {\bibinfo {title} {Qutip: An open-source python framework for the
  dynamics of open quantum systems},\ }\href
  {https://doi.org/https://doi.org/10.1016/j.cpc.2012.02.021} {\bibfield
  {journal} {\bibinfo  {journal} {Comput. Phys. Commun.}\ }\textbf {\bibinfo
  {volume} {183}},\ \bibinfo {pages} {1760} (\bibinfo {year}
  {2012})}\BibitemShut {NoStop}%
\bibitem [{\citenamefont {Johansson}\ \emph {et~al.}(2013)\citenamefont
  {Johansson}, \citenamefont {Nation},\ and\ \citenamefont {Nori}}]{qutip2}%
  \BibitemOpen
  \bibfield  {author} {\bibinfo {author} {\bibfnamefont {J.}~\bibnamefont
  {Johansson}}, \bibinfo {author} {\bibfnamefont {P.}~\bibnamefont {Nation}},\
  and\ \bibinfo {author} {\bibfnamefont {F.}~\bibnamefont {Nori}},\ }\bibfield
  {title} {\bibinfo {title} {Qutip 2: A python framework for the dynamics of
  open quantum systems},\ }\href
  {https://doi.org/https://doi.org/10.1016/j.cpc.2012.11.019} {\bibfield
  {journal} {\bibinfo  {journal} {Comput. Phys. Commun.}\ }\textbf {\bibinfo
  {volume} {184}},\ \bibinfo {pages} {1234} (\bibinfo {year}
  {2013})}\BibitemShut {NoStop}%
\bibitem [{\citenamefont {Moulton}(1926)}]{adams}%
  \BibitemOpen
  \bibfield  {author} {\bibinfo {author} {\bibfnamefont {F.~R.}\ \bibnamefont
  {Moulton}},\ }\href@noop {} {\emph {\bibinfo {title} {New methods in exterior
  ballistics}}}\ (\bibinfo  {publisher} {University of Chicago Press},\
  \bibinfo {year} {1926})\BibitemShut {NoStop}%
\bibitem [{\citenamefont {Brown}\ \emph {et~al.}(1989)\citenamefont {Brown},
  \citenamefont {Byrne},\ and\ \citenamefont {Hindmarsh}}]{vode}%
  \BibitemOpen
  \bibfield  {author} {\bibinfo {author} {\bibfnamefont {P.~N.}\ \bibnamefont
  {Brown}}, \bibinfo {author} {\bibfnamefont {G.~D.}\ \bibnamefont {Byrne}},\
  and\ \bibinfo {author} {\bibfnamefont {A.~C.}\ \bibnamefont {Hindmarsh}},\
  }\bibfield  {title} {\bibinfo {title} {Vode: A variable-coefficient ode
  solver},\ }\href {https://doi.org/10.1137/0910062} {\bibfield  {journal}
  {\bibinfo  {journal} {SIAM Journal on Scientific and Statistical Computing}\
  }\textbf {\bibinfo {volume} {10}},\ \bibinfo {pages} {1038} (\bibinfo {year}
  {1989})}\BibitemShut {NoStop}%
\bibitem [{\citenamefont {Ernst}\ \emph {et~al.}(1985)\citenamefont {Ernst},
  \citenamefont {K{\"a}ndler}, \citenamefont {Kindt},\ and\ \citenamefont
  {T{\"o}rring}}]{srf}%
  \BibitemOpen
  \bibfield  {author} {\bibinfo {author} {\bibfnamefont {W.~E.}\ \bibnamefont
  {Ernst}}, \bibinfo {author} {\bibfnamefont {J.}~\bibnamefont {K{\"a}ndler}},
  \bibinfo {author} {\bibfnamefont {S.}~\bibnamefont {Kindt}},\ and\ \bibinfo
  {author} {\bibfnamefont {T.}~\bibnamefont {T{\"o}rring}},\ }\bibfield
  {title} {\bibinfo {title} {Electric dipole moment of {SrF}
  {X}${}^{2}\ensuremath{\Sigma}^{+}$ from high-precision stark effect
  measurements},\ }\href
  {https://doi.org/https://doi.org/10.1016/0009-2614(85)80379-1} {\bibfield
  {journal} {\bibinfo  {journal} {Chem. Phys. Lett.}\ }\textbf {\bibinfo
  {volume} {113}},\ \bibinfo {pages} {351} (\bibinfo {year}
  {1985})}\BibitemShut {NoStop}%
\bibitem [{\citenamefont {Childs}\ \emph {et~al.}(1981)\citenamefont {Childs},
  \citenamefont {Goodman},\ and\ \citenamefont {Renhorn}}]{spinrotation}%
  \BibitemOpen
  \bibfield  {author} {\bibinfo {author} {\bibfnamefont {W.}~\bibnamefont
  {Childs}}, \bibinfo {author} {\bibfnamefont {L.}~\bibnamefont {Goodman}},\
  and\ \bibinfo {author} {\bibfnamefont {I.}~\bibnamefont {Renhorn}},\
  }\bibfield  {title} {\bibinfo {title} {Radio-frequency optical
  double-resonance spectrum of {SrF}: The {X}${}^{2}\ensuremath{\Sigma}^{+}$
  state},\ }\href
  {https://doi.org/https://doi.org/10.1016/0022-2852(81)90422-7} {\bibfield
  {journal} {\bibinfo  {journal} {Journal of Molecular Spectroscopy}\ }\textbf
  {\bibinfo {volume} {87}},\ \bibinfo {pages} {522} (\bibinfo {year}
  {1981})}\BibitemShut {NoStop}%
\bibitem [{\citenamefont {Schr{\"o}der}\ \emph {et~al.}(1988)\citenamefont
  {Schr{\"o}der}, \citenamefont {Nitsch},\ and\ \citenamefont
  {Ernst}}]{sri-spinrot}%
  \BibitemOpen
  \bibfield  {author} {\bibinfo {author} {\bibfnamefont {J.~O.}\ \bibnamefont
  {Schr{\"o}der}}, \bibinfo {author} {\bibfnamefont {C.}~\bibnamefont
  {Nitsch}},\ and\ \bibinfo {author} {\bibfnamefont {W.~E.}\ \bibnamefont
  {Ernst}},\ }\bibfield  {title} {\bibinfo {title} {Polarization spectroscopy
  of {Srl} in a heat pipe: The
  {B}${}^{2}\ensuremath{\Sigma}^{+}$-{X}${}^{2}\ensuremath{\Sigma}^{+}$ (0,0)
  system},\ }\href
  {https://doi.org/https://doi.org/10.1016/0022-2852(88)90066-5} {\bibfield
  {journal} {\bibinfo  {journal} {J. Mol. Spectrosc.}\ }\textbf {\bibinfo
  {volume} {132}},\ \bibinfo {pages} {166} (\bibinfo {year}
  {1988})}\BibitemShut {NoStop}%
\bibitem [{\citenamefont {T{\"o}rring}\ \emph {et~al.}(1989)\citenamefont
  {T{\"o}rring}, \citenamefont {Ernst},\ and\ \citenamefont
  {K{\"a}ndler}}]{sri-dipole}%
  \BibitemOpen
  \bibfield  {author} {\bibinfo {author} {\bibfnamefont {T.}~\bibnamefont
  {T{\"o}rring}}, \bibinfo {author} {\bibfnamefont {W.~E.}\ \bibnamefont
  {Ernst}},\ and\ \bibinfo {author} {\bibfnamefont {J.}~\bibnamefont
  {K{\"a}ndler}},\ }\bibfield  {title} {\bibinfo {title} {Energies and electric
  dipole moments of the low lying electronic states of the alkaline earth
  monohalides from an electrostatic polarization model},\ }\href
  {https://doi.org/10.1063/1.456589} {\bibfield  {journal} {\bibinfo  {journal}
  {J. Chem. Phys.}\ }\textbf {\bibinfo {volume} {90}},\ \bibinfo {pages} {4927}
  (\bibinfo {year} {1989})}\BibitemShut {NoStop}%
\bibitem [{\citenamefont {Agranovich}(2009)}]{agranovich}%
  \BibitemOpen
  \bibfield  {author} {\bibinfo {author} {\bibfnamefont {V.~M.}\ \bibnamefont
  {Agranovich}},\ }\href@noop {} {\emph {\bibinfo {title} {Excitations in
  organic solids}}},\ Vol.\ \bibinfo {volume} {142}\ (\bibinfo  {publisher}
  {OUP Oxford},\ \bibinfo {year} {2009})\BibitemShut {NoStop}%
\bibitem [{\citenamefont {Farhi}\ \emph
  {et~al.}(2001{\natexlab{b}})\citenamefont {Farhi}, \citenamefont {Goldstone},
  \citenamefont {Gutmann}, \citenamefont {Lapan}, \citenamefont {Lundgren},\
  and\ \citenamefont {Preda}}]{farhi}%
  \BibitemOpen
  \bibfield  {author} {\bibinfo {author} {\bibfnamefont {E.}~\bibnamefont
  {Farhi}}, \bibinfo {author} {\bibfnamefont {J.}~\bibnamefont {Goldstone}},
  \bibinfo {author} {\bibfnamefont {S.}~\bibnamefont {Gutmann}}, \bibinfo
  {author} {\bibfnamefont {J.}~\bibnamefont {Lapan}}, \bibinfo {author}
  {\bibfnamefont {A.}~\bibnamefont {Lundgren}},\ and\ \bibinfo {author}
  {\bibfnamefont {D.}~\bibnamefont {Preda}},\ }\bibfield  {title} {\bibinfo
  {title} {A quantum adiabatic evolution algorithm applied to random instances
  of an np-complete problem},\ }\href {https://doi.org/10.1126/science.1057726}
  {\bibfield  {journal} {\bibinfo  {journal} {Science}\ }\textbf {\bibinfo
  {volume} {292}},\ \bibinfo {pages} {472} (\bibinfo {year}
  {2001}{\natexlab{b}})}\BibitemShut {NoStop}%
\end{thebibliography}%
		
	\end{document}